\documentclass[twocolumn,floatfix]{revtex4}
\pdfoutput=1

\usepackage{dcolumn}
\usepackage{longtable}
\usepackage{amsmath}
\usepackage{amssymb}

\usepackage{mathtools}
\usepackage{adjustbox}
\usepackage{graphicx}
\usepackage{epstopdf}
\usepackage[version=3] {mhchem}
\usepackage{amssymb}
\usepackage{braket}
\usepackage{latexsym}
\usepackage{subfigure}
\usepackage{young}
\usepackage[dvipsnames]{xcolor}
\usepackage[toc,page]{appendix}
\usepackage{bbm}
\usepackage{dsfont}
\usepackage{lscape}
\usepackage{float}
\usepackage{tikz}

\newcommand{\abs}[1]{\left\lvert#1\right\rvert}

\begin{document}
\title{The islands of shape coexistence within the Elliott and the proxy-SU(3) Models 
}

\author
{Andriana Martinou$^1$, Dennis Bonatsos$^1$, T. J. Mertzimekis $^2$, K. E. Karakatsanis$^{1,3}$, I.E. Assimakis$^1$, S. K. Peroulis$^1$,  S. Sarantopoulou$^1$, and N. Minkov$^4$ }

\affiliation
{$^1$Institute of Nuclear and Particle Physics, National Centre for Scientific Research  ``Demokritos'', GR-15310 Aghia Paraskevi, Attiki, Greece}

\affiliation
{$^2$Department of Physics, National and Kapodistrian University of Athens, Zografou Campus, GR-15784 Athens, Greece}

\affiliation
{$^3$Department of Physics, Faculty of Science, University of Zagreb, HR-10000 Zagreb, Croatia}

\affiliation
{$^4$Institute of Nuclear Research and Nuclear Energy, Bulgarian Academy of Sciences, 72 Tzarigrad Road, 1784 Sofia, Bulgaria}

\begin{abstract}

A novel dual--shell mechanism for the phenomenon of shape coexistence in nuclei within the Elliott SU(3) and the proxy-SU(3) symmetry is proposed for all mass regions. It is supposed, that shape coexistence is activated by large quadrupole-quadrupole interaction and involves the interchange among the spin--orbit (SO) like shells within nucleon numbers 6-14, 14-28, 28-50, 50-82, 82-126, 126-184, which are being described by the proxy-SU(3) symmetry, and the harmonic oscillator (HO) shells within nucleon numbers 2-8, 8-20, 20-40, 40-70, 70-112, 112-168 of the Elliott SU(3) symmetry. The outcome is, that shape coexistence may occur in certain islands on the nuclear map. The dual--shell mechanism predicts without any free parameters, that nuclei with proton number ($Z$) or neutron number ($N$) between 7-8, 17-20, 34-40, 59-70, 96-112, 146-168 are possible candidates for shape coexistence. In the light nuclei the nucleons flip from the HO shell to the neighboring SO--like shell, which means, that particle excitations occur. For this mass region, the predicted islands of shape coexistence, coincide with the islands of inversion. But in medium mass and heavy nuclei, in which the nucleons inhabit the SO--like shells, shape coexistence is accompanied by a merging of the SO--like shell with the open HO shell. The shell merging can be accomplished by the outer product of the SU(3) irreps of the two shells and represents the unificaton of the HO shell with the SO--like shell.

\end{abstract}

\maketitle

\section{Introduction}
\label{intro}

Ever since the beginning of Nuclear Physics after Rutherford's discovery
of the atomic nucleus several questions have arisen about its composition and the
structure. Some questions have been tantalizing nuclear scientists for
decades, while some still do; especially those related to the fundamental
aspects of nuclear forces and the dynamics governing the coexistence of
protons and neutrons in the tiny volume of the nucleus.

The groundbreaking work by Mayer \cite{1964_Mayer,Mayer1} revealed the existence of
energy shells as the main feature of nuclear structure. Particle
excitations across such shells obey quantum rules defined by the nuclear
Hamiltonian. Very soon after the formulation of the Shell Model it became
clear, that collective nuclear phenomena can have a direct correspondence
to the single particle character, reflected on proton and neutrons
interactions providing the means for a firm explanation of rotations
and vibrations, recorded in experimental data in studies throughout
the nuclear chart.

In this framework the shape of the nucleus has a central role in understanding
the interplay of nuclear constituents at fundamental level. The presence
of symmetries, but also any violations of them, can be studied experimentally
by measuring the shape of nuclei in the ground and excited states and
taking advantage of the information becoming available through shape evolution
among different structures in the same nucleus or in different nuclei.
The correlation of shape evolution to proton and neutron numbers is not yet fully
understood. The advent of radioactive beams has drastically expanded the
limits of the known nuclear chart, posing various new and important questions
about nuclear structure at the extremes.

The behavior of the nuclear forces is responsible for the occurrence of phenomena like shape coexistence, in which an excited band with a shape different
from that of the ground band resides close to the latter. Experimentally, shape coexistence was observed early in various types of reactions, see e.g. \cite{1969_McLatchie}, but experimental studies have been intensified significantly in recent years.

The experimental interest was initially focused on rare earth isotopes,
at masses $A\approx 150$,
such as Gd and Sm \cite{1973_Burke,1974_Kleinheinz,1979_Smith,1980_Lanier},
but also in heavier masses near the magic lead nuclei \cite{1976_Hamilton,1977_Cole}, an effort, that escalated significantly
later on.

The discovery of the island of inversion around the neutron rich Mg isotopes,
where the nuclear interaction drives strong deformations, shuttering the
``traditional'' predictions of the shell model, has created new excitement
on shape coexistence at the experimental level. Early spectroscopic studies
(measurements of $\alpha$ spectroscopic factors and $E0$ values) in
Ar \cite{1977_Bohne} and
Ca isotopes \cite{1972_Middleton,1979_Fortune,1979_Fortune2}
provided strong hints for shape coexistence around the upper end of the $sd$ shell.
The search for similar occurrences has been intense in the $fp$ shell,
as well, where the protons and neutrons occupy orbitals with similar Fermi
energies, giving rise to strong proton--neutron correlations. Strong
deformations, that correlate to shape coexistence phenomena in
neutron--deficient Se \cite{2008_Ljungvall} and Kr \cite{2001_Mertzimekis}
nuclei, have been examined experimentally in line with similar studies in neighboring nuclei,
such as Ge \cite{1978_Kumar},
Se \cite{1984_Becker,1984_Eberth} and
Sr \cite{1980_Schussler} nuclei.

This mid--weight mass regime around $A\sim 80$ has been central in a series
of experiments ``east'' and ``west'' of the valley of stability. Due to the 
increasing use of radioactive beams, neutron--rich  krypton isotopes have
been examined thoroughly \cite{2004_Korten,2005_Gorgen,2007_Clement} to
understand any potential evolution of shape coexistence when neutron number crosses the magic number $N=50$. Such studies have relied on Coulomb
excitation reactions, which are known to favor strongly $E2$ transitions,
however they suffer rather low statistics when unstable beams are employed.

Shape coexistence is now recognized as a rather common phenomenon potentially
occurring in all nuclei \cite{Heyde2011}. Despite the fact that several
theoretical models have attempted to explain its dynamics (see the
discussion later in the present work), no explicit mechanism is known to
describe it fully at a fundamental level. The implicit strong correlations
in the $fp$ shell, as well as the prediction of islands of shape coexistence
in the heavier region of the nuclear chart \cite{Heyde2011} has expanded
the experimental efforts towards consolidating the spectroscopic signatures
of shape coexistence in nuclei with $A\sim120-180$. Experimental data in
Mo \cite{2016_Thomas}, 
Ag \cite{1977_Glascock},
Cd \cite{2011_Garrett},
La \cite{2020_Ionescu},
and recent measurements of lifetimes in neutron--deficient nuclei in the lead region \cite{2019_Olaizola,2020_Siciliano} provide strong evidence of the
onset of shape coexistence at neutron numbers near to 108. Given the large
gap of knowledge about characteristic experimental signatures of shape
coexistence, such as lifetimes and reduced matrix elements, in neutron--rich
Ytterbium and Hafnium nuclei, there remain open questions on the universality
of shape coexistence across the large shell forming above $N=82$. The existence
of odd parity bands next to the ground state band, as well as the presence
of a few low--lying $0^+$ states, hint the onset of shape coexistence.

According to the presently recognized understanding of the shape
coexistence at certain proton or
neutron number, a multi particle--hole (mp--mh) configuration across the
closed shell or sub--shell appears
to possess similar binding energy, compared to the 0p--0h
configuration. This can be explained
by a specific correlation of energy contributions involving pairing,
quadrupole and other interactions,
providing the ``intruder" mp--mh state with a deformation different from
that of the ground state.

In the present work a novel dual--shell mechanism for shape coexistence is proposed. It is supposed that  two types of open valence shells are active: the HO shell and the SO--like shell. The coexistence of these two types of shells is allowed at certain nucleon numbers, where the quadrupole-quadrupole interaction in the SO--like shell is greater than the one of the HO shell.

As we shall see below, the new mechanism is fully compatible with the particle--hole mechanism in the regions of the nuclear chart, in which the latter has been applied. The novel suggestion of the present work is, that shape coexistence can only occur within certain islands of the nuclear chart, fully compatible with the regions, in which shape coexistence has been seen experimentally \cite{Wood1992,Heyde2011}. In contrast to the common belief, that shape coexistence can appear at any place of the nuclear chart \cite{Heyde2011}, we prove, that this is possible only within certain well defined islands of the nuclear chart.

In order to determine the shores of these islands, we will need the SU(3) symmetry of the three-dimensional (3D) isotropic harmonic oscillator \cite{Wybourne,Smirnov}, used within the Elliott model \cite{Elliott1,Elliott2,Elliott3,Elliott4}, which is applicable in light nuclei up to the $sd$ shell, as well as the recently introduced proxy-SU(3) symmetry \cite{proxy1,proxy2,proxy4}, which is applicable beyond the $sd$ shell in medium-mass and heavy nuclei. Shape coexistence is recognized to be related to single--particle energy gaps, which are affected by the deformation \cite{Heyde2011}. Since it is a ``deformation--driving" phenomenon (see section 6 of Ref. \cite{Heyde2011}), one should do the calculations in a deformed basis and after that, projected the overall wave function on states with good angular momentum. This task is easily done within the Elliott and the proxy-SU(3) symmetry, in which the cartesian single particle basis \cite{proxy4} is initially used for the intrinsic frame and as a second step the projection onto a rotational invariant wave function  follows \cite{Elliott3,Elliott4,Wilsdon,Vergados1968}. The Elliott and proxy-SU(3) symmetry and the will be reviewed in sections \ref{Elliott} and \ref{proxy}. An application of the projection technique will be demonstrated in section \ref{O} for the calculation of the energy of the $0_2^+$ of $\ce{^{16}O}$.

The basic element of the new mechanism, namely particle excitations among the HO shells defined by the 3D isotropic HO magic numbers 2, 8, 20, 40, 70, 112, 168, and the SO--like shells defined by the SO magic numbers 6, 14, 28, 50, 82, 126, 184 \cite{Sorlin} are described in section \ref{excitations} and a detailed example for the halo nucleus $\ce{^{11}Be}$ is given. In section \ref{mn} it is explained, how the the dissolution of magic numbers emerges within certain areas of the nuclear chart, resulting in the merging of the HO and SO--like shells. The compatibility between the particle--hole mechanism and the present new mechanism is also discussed in section \ref{mn}, before summarizing the main features of the new mechanism in section \ref{mechanism}.

 The two low--lying nuclear bands being candidates for shape coexistence are discussed in terms of their relevant symmetry features (the SU(3) irreducible representations (irreps) in mathematical language) in section \ref{two}, while in section \ref{islands} the shores of the islands of shape coexistence are determined through a simple Hamiltonian bearing the SU(3) symmetry. The sudden onset of deformation in certain regions of the nuclear chart and its role in clarifying the relation between shape coexistence and shape phase transitions \cite{Jolie2004,Garcia2018,Garcia2020} is discussed in section \ref{onset}, while in section \ref{paradigm} specific examples of shape coexistence appearing in seven different regions of the nuclear chart are presented. The conclusions based on the present findings and plans for further work are given in section \ref{discussion}. At last a nuclear chart with the possible islands of shape coexistence, predicted by the dual--shell mechanism, is presented in Fig. 25.

\section{The Elliott SU(3) Model}\label{Elliott}

The Elliott SU(3) symmetry \cite{Elliott1,Elliott2} can be considered as a beyond-Shell-Model theory, in which the rotational spectrum of deformed nuclei emerges naturally from the coupling of the valence particles \cite{Elliott2}. The Elliott SU(3) symmetry is valid in a valence proton or neutron shell, constituted by orbitals  of the 3D-HO Hamiltonian with common number of quanta $\mathcal{N}$. Such shells lie among proton or neutron numbers 2, 8, 20, 40, 70, 112, 168, namely the HO magic numbers. 

The Elliott SU(3) model uses the cartesian states $\ket{n_z,n_x,n_y,m_s}$ as building blocks \cite{Elliott3,Elliott4,Harvey,proxy4}, where $m_s=\pm{1\over 2}$ is the spin projection of the nucleon and $n_z,n_x,n_y$ are the oscillator quanta in the three cartesian axes of the single particle 3D-HO Hamiltonian:
\begin{eqnarray}\label{h0}
h_0={\mathbf{p}^2\over 2M}+{1\over 2}M\omega^2 \mathbf{r}^2,
\end{eqnarray}
with $M,\mathbf{p},\mathbf{r},\omega$ being the nucleon mass, the momentum, the spatial coordinate and the oscillator frequency respectively. The eigenvalues of the Hamiltonian (\ref{h0}) are \cite{Cohen}:
\begin{equation}\label{e0}
\epsilon_0=\left(\mathcal{N}+{3\over 2}\right)\hbar\omega,
\end{equation}
where
\begin{equation}\label{Nc}
\mathcal{N}=n_z+n_x+n_y.
\end{equation}

The eight generators of the SU(3) algebra are the  $l_0$, $l_{\pm1}$, $q_{\pm 2}$, $q_{\pm 1}$, $q_0$ \cite{Lipkin,Harvey,Elliott3}, where $l_0,$ is the operator of the projection of the orbital angular momentum, $l_{\pm1}$ are the ladder operators of the algebra of the orbital angular momentum and $q_{\pm 2},q_{\pm 1},q_0$ are the five components of the algebraic quadrupole operator \cite{Kota}. These generators can be expressed in terms of the well known creation and annihilation operators $a^\dagger_z$ ,$a^\dagger_x$, $a^\dagger_y$, $a_z$, $a_x$, $a_y$ of a quantum in each cartesian direction \cite{Lipkin,Harvey,Elliott3}. Of utmost importance is the $q_0$, which measures the elongation of the $z$ axis \cite{Elliott2}:
\begin{equation}\label{q0}
q_0=2n_z-n_x-n_y.
\end{equation}

On the one hand the cartesian states $\ket{n_z,n_x,n_y,m_s}$ are eigenstates of the single particle 3D-HO Hamiltonian $h_0$ and of $q_0$. The cartesian basis is suitable for the description of the single--particle states of deformed nuclei \cite{Elliott3}, in which the quadrupole--quadrupole, two body interaction of the many nucleon system \cite{Lipkin} :
\begin{equation}\label{QQQ}
QQ=\sum_{i,i',m=\pm 2,\pm 1,0}(-1)^mq_{mi}q_{-mi'}
\end{equation}
prevails. In the above the summation is for every valence $(i,i')$ nucleon pair and for every $m$ component of the quadrupole operator.

On the other hand the cartesian states are not eigenstates of a) the single particle spin-orbit interaction $\mathbf{l}\cdot \mathbf{s}$, where $\mathbf{l,s}$ are the orbital angular momentum and the spin of the nucleon respectively, and of b) the $\mathbf{l}^2$ interaction. These interactions are included in the Shell Model \cite{Mayer1,1964_Mayer} single particle Hamiltonian \cite{Nilssonbook}:
\begin{equation}\label{h}
h=h_0+\upsilon_{ls}\hbar\omega \mathbf{l}\cdot \mathbf{s}+\upsilon_{ll}\hbar\omega(\mathbf{l}^2-\langle\mathbf{l}^2\rangle_\mathcal{N}),
\end{equation}
where $\upsilon_{ls},\upsilon_{ll}$ are strength parameters \cite{Nilssonbook} (see Table I of Ref. \cite{proxy1} for their values) and the term $\mathbf{l}^2-\langle\mathbf{l}^2\rangle_\mathcal{N}$ is usually used for the flattening of the HO potential.

The eigenstates of the Hamiltonian of Eq. (\ref{h}) are expressed in the spherical coordinate system as $\ket{n,l,j,m_{j}}$, where $n$ is the radial quantum number $n=0,1,2$ with \cite{Davies}:
\begin{equation}\label{Ns}
\mathcal{N}=2n+l,
\end{equation}
$j$ is the total angular momentum and $m_{j}$ is its projection on the $z$ axis \cite{Heyde}. The $\ket{n,l,j,m_j}$ states are the usual Shell Model orbitals, if one adds one unit in the radial quantum number $n$ and represents the orbital angular momentum $l=0,1,2,...$ by the small latin characters $s,p,d,...$. For instance the orbitals $\ket{n,l,j,m_j}$: $\ket{0,1,{3\over2},{1\over 2}}$, $\ket{1,2,{5\over 2},{3\over 2}}$ are labeled as $1p^{j=3/2}_{m_j=1/2}$, $2d^{j=5/2}_{m_j=3/2}$ in the Shell Model notation respectively.

Despite the fact that the $\ket{n_z,n_x,n_y,m_s}$ are not eigenstates of the Hamiltonian (\ref{h}), they can be transformed into them \cite{proxy4}. In Tables 1, 3, 5 of Ref. \cite{proxy4} explicit expressions were given for the transformation of the $\ket{n_z,n_x,n_y,m_s}$ states to the $\ket{n,l,j,m_j}$ orbitals. As such the cartesian orbital $\ket{1,0,0,{1\over 2}}$ transforms as:
\begin{eqnarray}\label{ex}
\ket{1,0,0,\pm{1\over 2}}=\mp{1\over \sqrt{3}}\ket{1p^{1/2}_{\pm 1/2}}+\sqrt{2\over 3}\ket{1p^{3/2}_{\pm 1/2}}.
\end{eqnarray} 
The $\ket{1p^{1/2}_{1/2}}$ orbital is the $\ket{n,l,j,m_j}=\ket{0,1,{1\over 2},{1\over 2}}$, while the $\ket{1p^{3/2}_{1/2}}$ is the $\ket{n,l,j,m_j}=\ket{0,1,{3\over 2},{1\over 2}}$, both of them possessing $\mathcal{N}=1$ number of oscillator quanta.

Nevertheless both types of bases $\ket{n_z,n_x,n_y,m_s}$, $\ket{n,l,j,m_j}$ are eigenstates of the $h_0$ of Eq. (\ref{h0}):
\begin{gather}
h_0\ket{n_z,n_x,n_y,m_s}=\epsilon_0\ket{n_z,n_x,n_y,m_s},\label{h0c}\\
h_0\ket{n,l,j,m_j}=\epsilon_0\ket{n,l,j,m_j},\label{h0s}
\end{gather}
where $\epsilon_0$ is given by Eq. (\ref{e0}) and $\mathcal{N}$ is given by Eqs. (\ref{Nc}), (\ref{Ns}).

Since the nucleons are fermions and thus they obey the Pauli exclusion principle \cite{Fermi, Pauli}, the many--body nuclear wave functions are Slater determinants \cite{Slater} of the single--particle states \cite{Ring,Elliott3,Wilsdon}. Initially in the Elliott SU(3) symmetry at Ref. \cite{Elliott2} the many particle, intrinsic nuclear wave functions $\mathcal{X}$ in the cylindrical coordinate system had been projected into the $\Psi$ many particle wave functions in the spherical coordinates, which possess the orbital angular momentum $L$ as a good quantum number. Notice, that throughout the present work we use capital letters for angular momentum quantum numbers referring to the many--body system, while we use lower case letters for angular momentum quantum numbers regarding the single--particle states. It has been proven in Ref. \cite{Elliott2}, that a single $\mathcal{X}$ wave function suffices, to generate all the states of the rotational bands, which belong to the same SU(3) irrep.

It was later realized \cite{Elliott3,Wilsdon}, that the description of the rotational features of deformed nuclei becomes even simpler in the cartesian coordinate system, in which the many--particle nuclear wave functions are labeled by $\Phi$. The $\Phi$ many particle wave functions are being derived by the single particle orbitals $\ket{n_z,n_x,n_y,m_s}$ in the cartesian coordinate system. The $L$ or $J$-projection of a single $\Phi$ wave function to wave functions, which possess the anugular momentum or the total angular momentum respectively as a good quantum number, derives all the nuclear rotational states, which belong to the same SU(3) irrep $(\lambda,\mu)$ \cite{Elliott3,Elliott4,Wilsdon}. 

To conclude, the single--particle basis $\ket{n_z,n_x,n_y,m_s}$ and the many--body wave function $\Phi$, which is derived from them, is the intrinsic basis, which describes {\it deformed} nuclei in the Elliott SU(3) symmetry. An $L$ or $J$ projection of the cartesian intrinsic wave function reveals the rotational energy bands, which belong to the the same SU(3) irrep \cite{Elliott3,Elliott4,Wilsdon}.

The cartesian states are used to derive the SU(3) irreps $(\lambda,\mu)$. The subject of the calculation of $(\lambda,\mu)$ has been exhausted in Refs. \cite{Elliott1,Elliott2,code,Kota2018,Rila2018,Book,proxy5}. The important thing for the dual--shell mechanism for shape coexistence is, that the low-lying nuclear properties are being derived by the highest weight SU(3) irreps \cite{proxy2,proxy5}. This special irrep emerges, when the valence nucleons fill the spatial, cartesian orbitals with the following specific order:
\begin{gather}
\ket{n_z,n_x,n_y}:\ket{\mathcal{N},0,0}, \ket{\mathcal{N}-1,1,0}, \ket{\mathcal{N}-1,0,1},\nonumber\\ \ket{\mathcal{N}-2,2,0},\ket{\mathcal{N}-2,1,1},\ket{\mathcal{N}-2,0,2},...,\ket{0,0,\mathcal{N}}.\nonumber\\\label{order}
\end{gather}
If this filling order is followed, then \cite{Elliott2}:
\begin{gather}
\lambda=\sum_in_{z,i}-\sum_in_{x,i},\label{lambda}\\
\mu=\sum_in_{x,i}-\sum_in_{y,i},\label{mu}
\end{gather}
where the summations are over every valence nucleon.

For the many--body problem the relevant 3D-HO Hamiltonian is the summation over every nucleon:
\begin{equation}\label{H0}
H_{0}=\sum_{i=1}^Ah_{0,i},
\end{equation}
where $A$ is the mass number. The eigenvalues of $H_0$ are labeled as $N_0$ \cite{Rowe2016}:
\begin{equation}\label{N0}
N_0=\sum_{i=1}^{A}\left(\mathcal{N}_i+{3\over 2}\right)\hbar\omega,
\end{equation}
with $\mathcal{N}$ given by Eqs. (\ref{Nc}), (\ref{Ns}).

The overall algebraic quadrupole-quadrupole ($QQ$) interaction in the Elliott SU(3) model is calculated as in \cite{Draayer1984,Draayerbook}:
\begin{equation}\label{QQ}
QQ=4C_2-3L(L+1),
\end{equation}
where $L(L+1)$ is the eigenvalue of the square of the orbital angular momentum operator for all the valence particles \cite{Elliott1,Lipkin}:
\begin{equation}
\mathbf{L}=\sum_i\mathbf{l}_i
\end{equation}
and $\hat C_2$ is the second order Casimir operator of SU(3) with eigenvalues \cite{Elliott4,Lipas}:
\begin{equation}\label{C2}
C_2=\lambda^2+\mu^2+\lambda\mu+3(\lambda+\mu).
\end{equation}
The $C_2$ operator is linked to the nuclear quadrupole deformation parameter $\beta$ of the Bohr and Mottelson Collective Model \cite{BohrII} through the equation \cite{Castanos1988}:
\begin{equation}\label{beta}
\beta^2={4\pi\over 5(A\bar r^2)^2}(C_2+3),
\end{equation}
with $\bar r^2=0.87^2A^{1/3}$ is the dimensionless mean square radius (rms). Since usually $C_2\gg 3$, $\beta^2$ is proportional to $C_2$, therefore one may treat the $C_2$ as a measure of deformation. Due to fact that the Elliott SU(3) Model treats the algebraic $QQ$ interaction, which has non zero matrix elements among single--particle states with the same number of quanta $\mathcal{N}$, while the measured $QQ$ interaction is the collective one, which has non zero matrix elements among single--particle orbitals with $\mathcal{N},\mathcal{N}\pm 2$ quanta \cite{Kota,Draayerbook}, the formula of Eq. (\ref{beta}) needs to be multiplied by a scaling factor as in Ref. \cite{proxy2}:
\begin{equation}\label{scale}
{A^2\over (S_\varpi+S_\nu)^2}
\end{equation}
where $S_\varpi,S_\nu$ is the size of the valence proton and neutron shell respectively.

A very simple SU(3) Hamiltonian, which includes only the 3D-HO and the $QQ$ interaction and adequately describes the $L=0$ states, is the following \cite{Rowe2006}:
\begin{equation}\label{H0QQ}
H=H_0-{\kappa\over 2}QQ,
\end{equation}
where the $H_0$ was defined in Eq. (\ref{H0}), while the $QQ$ is being calculated by Eqs. (\ref{QQ}) and (\ref{C2}). The parameter $\kappa$ is \cite{Rowe2006}:
\begin{equation}\label{kappa}
\kappa={\hbar\omega\over 2N_0},
\end{equation}
where in the numerator \cite{Ring}:
\begin{equation}\label{hbar}
\hbar\omega={41\over A^{1/3}}\ce{MeV}, 
\end{equation}
while in the denominator the $N_0$ is calculated with Eq. (\ref{N0}) by setting $\hbar\omega=1$. 

The Elliott SU(3) symmetry is valid in a 3D-HO valence shell, which is constituted by orbitals with common number of quanta $\mathcal{N}$ and lies among the 3D-HO magic numbers 2, 8, 20, 40, 70, 112, 168. Since a significant spin--orbit interaction leads to the spin--orbit like (SO) magic numbers 6, 14, 28, 50, 82, 126, 184 one would expect, that in the Elliott scheme either there is no spin--orbit interaction, or that this interaction is small. Nevertheless the spin--orbit interaction can and has to be applied in the Elliott SU(3) scheme with the techniques applied in Refs. \cite{Elliott3,Elliott4,Wilsdon,Bouten1967}. 

It has also been proved, that the spin--orbit interaction \cite{Elliott3,Elliott4}:
\begin{equation}\label{LS}
V_{s.o.}=\xi\sum_i\mathbf{l}_i\cdot\mathbf{s}_i
\end{equation}
(with $\xi$ being a parameter, while $i$ counts every valence nucleon) becomes stronger as more valence nucleons are added in the valence 3D-HO shell \cite{Wilsdon,Bouten1967}. At the second half of a HO shell the spin--orbit interaction has grown so much, that one has to change either to a $jj$ coupling scheme among the nucleons of the HO shell \cite{Wilsdon,Bouten1967}, or to a spin--orbit like shell \cite{Mayer1,1964_Mayer}. In rotational nuclei the 3D-HO magic numbers are valid as long as the $\ket{n,l,j,m_j}$ orbitals with common $\mathcal{N}$ mix together. For instance as long as the $2s^{1/2},1d^{3/2}$ mix with the $1d^{5/2}$,  the $sd$ 3D-HO shell among magic numbers 8, 20 is being created. But when a strong overall spin--orbit interaction, as defined in Eq. (\ref{LS}), applies in this shell, then the normal parity orbitals with $\mathcal{N}$ number of quanta: $2s^{1/2},1d^{3/2}$ join together with the intruder parity orbitals with $\mathcal{N}+1$ number of quanta: $1f^{7/2}$ and they create the SO--like shell among magic numbers 14, 28.

\section{The proxy-SU(3) Model}\label{proxy}

In the SO--like shells \cite{Sorlin} among magic numbers  6, 14, 28, 50, 82, 126, 182 (see Table 7 of \cite{proxy4}) the normal parity orbitals contain $\mathcal{N}$ quanta, while the intruders $\mathcal{N}+1$ and as a result the Elliott SU(3) Model cannot have a straightforward application. In such SO--like shells the proxy-SU(3) symmetry can be applied instead \cite{proxy1,proxy2,proxy3,proxy4}. The proxy-SU(3) symmetry had been introduced in Ref. \cite{proxy1} and delivered at once parameter-free predictions for the nuclear deformation in Ref. \cite{proxy2}. 

The proxy-SU(3) Model restores the SU(3) symmetry in the notation of the asymptotic basis of the Nilsson Model \cite{Nilsson1} by replacing the intruder Nilsson orbitals $K[\mathcal{N}n_z\Lambda]$, where $K, \Lambda$ are the projections of the total and the orbital angular momentum respectively, by their $\Delta K[\Delta\mathcal{N}\Delta n_z\Delta \Lambda]$ =$0[110]$ counterparts \cite{proxy1}. This replacement is not altering the $K, \Lambda$, while it reduces by one unit the $\mathcal{N}$ and the $n_z$ of the intruder orbitals, which in other words means, that this replacement is not affecting the quanta in the $x-y$ plane. The inspiration for this replacement was the observation of an enhanced proton--neutron interaction between the $\Delta K[\Delta\mathcal{N}\Delta n_z\Delta \Lambda]$ =$0[110]$ counterparts \cite{Burcu2010,Sofia2013}.

The proxy-SU(3) replacement of orbitals can also take place in the Shell Model basis \cite{Mayer1,1964_Mayer}, where the intruder Shell Model orbitals $\ket{n,l,j,m_j}$ possessing $\mathcal{N}+1$ quanta are replaced by their de Shalit--Goldhaber counterparts \cite{deShalit}, identified by $\ket{n,l-1,j-1,m_j}$ \cite{proxy4,Bonatsos2020}. The transformation among the de Shalit-- Goldhaber orbitals proved to be a unitary transformation, similar in nature to the one used in the pseudo-SU(3) symmetry \cite{AnnArbor}. However, in the proxy-SU(3) case, the unitary transformation affects only the quanta in the $z$ cartesian direction and leaves the $x-y$ plane intact \cite{proxy4}. As a result the proxy-SU(3) model restores the SU(3) symmetry by leaving the normal parity orbitals intact and performing a unitary transformation in the intruder parity orbitals, which affects the quantum number $n_z$.

It is interesting, to understand, what the consequences of the unitary transformation are \cite{proxy4}, which applies in the proxy-SU(3) symmetry, on the single particle energies of a SO--like shell. The orbitals of the SO--like shells are presented in Table 7 of Ref. \cite{proxy4}. For instance the 6-14 SO shell consists of the $1p^{1/2}_{m_j},1d^{5/2}_{m_j}$ spherical Shell Model orbitals. Such mixed parity shells do not possess SU(3) symmetry. But the SU(3) symmetry can be restored in this shell, by applying on the intruder $1d^{5/2}_{m_j}$ orbitals the unitary transformation \cite{proxy4}:
\begin{equation}\label{unitary}
U=a_z<a_z^\dagger a_z>^{-1/2},
\end{equation}
which reduces the quanta in the $z$ cartesian axis by one unit and normalizes the result. Such an action maps all the $1d^{5/2}_{m_j}$ orbitals (except $1d^{5/2}_{\pm 5/2}$) to the $1p^{3/2}_{m_j}$ and so the proxy shell consists by the $1p^{1/2}_{m_j},1p^{3/2}_{m_j}$ orbitals and possesses an SU(3) symmetry \cite{proxy4}.

This loss of one quantum in the $z$ cartesian direction of the intruder orbitals results to a modification of the Hamiltonian, which describes a proxy shell, as discussed in Appendix A of Ref. \cite{proxy4}. If the single particle Hamiltonian before the unitary transformation was that of Eq. (\ref{h0}), then the Hamiltonian, which describes the proxy shell after the unitary transformation must be:
\begin{equation}\label{h0pr}
h_{0,SO}=h_{0}+\epsilon_{proxy}\delta_{j,\mathcal{N}+1/2},
\end{equation}
where the Kronecker $\delta_{j,\mathcal{N}+1/2}$ is affecting only the orbitals with $j=j_{max}=\mathcal{N}+1/2$, within the proxy shell. Thus from the orbitals $1p^{1/2}_{m_j},1p^{3/2}_{m_j}$ of the 6-12 proxy shell (see Table 7 of Ref. \cite{proxy4}) only the $1p^{3/2}_{m_j}$ orbitals are affected by the Kronecker delta function, which restores the loss of the energy from the replacement of the intruder orbitals by their proxies \cite{proxy1,proxy4}. The $\epsilon_{proxy}$ is the single--particle energy difference among the intruder orbital of the SO--like shell and the proxy orbital, which replaces it (see Table 7 of Ref. \cite{proxy4}). For instance the $\epsilon_{proxy}$ in the 6-14 SO--like shell is the single--particle energy difference between the $1d^{5/2}_{m_j}$ and the $1p^{3/2}_{m_j}$ orbitals. 

The value of the $\epsilon_{proxy}$ has been given in Ref. \cite{proxy1} to be:
\begin{equation}\label{eproxy}
\epsilon_{proxy}=\left( 1-{2\varepsilon\over3}\right) \hbar\omega,
\end{equation}
where $\varepsilon$ is the deformation parameter in the asymptotic basis of the Nilsson Model \cite{Nilsson1}, connected with the deformation parameter $\delta$ of the Nilsson Model \cite{Nilsson1} through:
\begin{gather}\label{def1}
\varepsilon=\delta+{1\over 6}\delta^2,
\end{gather}
with the deformation variable $\beta$ of the Bohr and Mottelson Model \cite{BohrII} being related to $\delta$ by \cite{Nilsson1}:
\begin{equation}\label{def2}
\delta=0.95\beta.
\end{equation}
The unit in Eq. (\ref{eproxy}) restores the loss of one quantum from the unitary transformation, which applies in the proxy scheme \cite{proxy4}, while the ${2\varepsilon\over 3}$ restores the damage in the value of $q_0$ of Eq. (\ref{q0}), due to the loss of one quantum in the $z$ direction \cite{proxy4} (see section 1.3 of Ref. \cite{Book} and section 7.3.1 of Ref. \cite{Draayerbook} for deeper understanding).

It is well known from Quantum Mechanics, that the addition of a constant in the Hamiltonian is not altering the eigenstates, but simply lifts the eigenvalues by this constant, which means, that Eq. (\ref{h0s}) is modified as:
\begin{equation}
h_{0,SO}\ket{n,l,j,m_j}=\epsilon_0(\ket{n,l,j,m_j})_{SO}\ket{n,l,j,m_j},
\end{equation}
with 
\begin{equation}
\epsilon_0(\ket{n,l,j,m_j})_{SO}=\epsilon_0+\epsilon_{proxy}\delta_{j,\mathcal{N}+1/2}\label{e0pr}
\end{equation}
where $\epsilon_0$ is given by Eqs. (\ref{e0}), (\ref{Ns}).
The Hamiltonian of Eq. (\ref{h0pr}) includes the $j$ quantum number. As a consequence the $\ket{n_z,n_x,n_y,m_s}$ states are not eigenstates of the Hamiltonian of Eq. (\ref{h0pr}), while the $\ket{n,l,j,m_j}$ are.

The change in energy of the spherical Shell Model states $\ket{n,l,j,m_j}$ inevitably alters the expectation value of the energy of the cartesian states $\ket{n_z,n_x,n_y,m_s}$. In general using the Hamiltonian of Eq. (\ref{h0pr}) the expectation values of the cartesian orbitals of the proxy shells are given by:
\begin{eqnarray}\label{singlecpr}
\langle\epsilon_0(\ket{n_z,n_x,n_y,m_s})_{SO}\rangle=\nonumber\\
\braket{n_z,n_x,n_y,m_s|h_{0,SO}|n_z,n_x,n_y,m_s}=\nonumber\\
\sum_{n,l,j,m_j} \abs{\braket{n,l,j,m_{j}|n_{z},n_{x},n_{y},m_{s}}}^2
\cdot  \epsilon_0(\ket{n,l,j,m_j})_{SO},\nonumber\\
\label{Eprc}
\end{eqnarray}
where the summation is over every non--vanishing matrix element $\braket{n,l,j,m_{j}|n_{z},n_{x},n_{y},m_{s}}$ of the transformation presented in Ref. \cite{proxy4}.

The effect of Eq. (\ref{h0pr}) on the many--particle Hamiltonian is the following:
\begin{gather}\label{H0SO}
H_{0,SO}=H_0+
\sum_i(\delta_{j_i,\mathcal{N}+1/2})_i\epsilon_{proxy},
\end{gather}
where $H_0$ is given by Eq. (\ref{H0}) and the summation is for every valence particle $i$. This operator depends on the total angular momentum of each valence particle $j_i$. Let the summation be labeled:
\begin{equation}\label{H0pr}
H_{0,proxy}=\sum_i(\delta_{j_i,\mathcal{N}+1/2})_i\epsilon_{proxy},
\end{equation}
since it carries the effect of the proxy replacement of orbitals on the many--particle Hamiltonian.

The eigenvalues of $H_{0,proxy}$ have to be calculated in the $L$-projected (or $J$-projected for odd mass nuclei) \cite{Elliott3,Elliott4,Wilsdon} wave functions of the Elliott Model, which have $K,L,M$ as good quantum numbers with $K$ being the band label and $M$ is the projection of the angular momentum $L$ \cite{Elliott3}. The $L$-projected wave functions read \cite{Elliott3}:
\begin{gather}\label{Psi}
\Psi(KLM)={P(KLM)\over a(K,L)}\Phi,
\end{gather}
where $a(K,L)$ are the expansion coefficients of the Slater determinant $\Phi$, which consists by the cartesian single particle states $\ket{n_z,n_x,n_y,m_s}$, in the $L$-projected wave function:
\begin{equation}\label{PhiP}
\Phi=\sum_{K,L}a(K,L)\Psi(KLK),
\end{equation}
and $P(KLM)$ is the projection operator with explicit form given in the Appendix of Ref. \cite{Elliott3}. The matrix elements of the projection operator are labeled by \cite{Elliott3}:
\begin{equation}\label{P}
A(KLK')=\braket{\Phi|P(KLK')|\Phi},
\end{equation}
which for the special case of $K'=K$ become \cite{Elliott3}:
\begin{equation}\label{A}
A(KLK)=|a(K,L)|^2.
\end{equation}
Analytic formulae for the $a(K,L)$ coefficients are given in Table 2A of Ref. \cite{Vergados1968}. Finally following Eqs. (\ref{Psi})-(\ref{A}) the diagonal matrix elements of $H_{0,proxy}$ are:
\begin{gather}
\braket{\Psi(KLK)|H_{0,proxy}|\Psi(KLK)}=\nonumber\\
|a(K,L)|^2\braket{\Phi|H_{0,proxy}|\Phi}.\label{braket}
\end{gather}

A sample calculation of the above quantity for $\ce{^{16}O}$ will be given in section \ref{O}. In this sample calculation it will be demonstrated, how easy it is, to do calculations in the cartesian single--particle basis $\ket{n_z,n_x,n_y,m_s}$ (which suits to deformed nuclei) and then apply the $L$-projection technique \cite{Elliott3}, within the proxy-SU(3) Model. The coincidence of the calculation of the energy of the $0_2^+$ state (without any fitting) with the data for $\ce{^{16}O}$ will be outstanding.

A SO--like shell, with $\mathcal{N}$ quanta after the replacement of the intruder orbitals by their de Shalit--Goldhaber partners \cite{deShalit,proxy4}, has a $U(\Omega)$ symmetry, where $\Omega={(\mathcal{N}+1)(\mathcal{N}+2)\over 2}$. The proxy-SU(3) irreps $(\lambda,\mu)$ are then calculated by the code of Ref. \cite{code}. The various particle distributions in the valence Shell Model space result to a variety of SU(3) irreps \cite{proxy5,Assimakis}. Among them the highest weight irrep has been proven, to describe the low energy nuclear properties \cite{proxy2,pseudohw,proxy5}, since this one corresponds to the most symmetric spatial wave function \cite{proxy5} and in addition predicts the dominance of the prolate over the oblate shape \cite{proxy2}. The highest weight SU(3) irreps for the HO magic numbers and the SO--like magic numbers using the Elliott and the proxy-SU(3) symmetry respectively are listed in Tables 1, 2.

\begin{table*}[htb]

\caption{The highest weight SU(3) irreps for the spin-orbit (SO) like magic numbers 2, 6, 14, 28, 50, 82, 126 according to the proxy-SU(3) symmetry and for the harmonic oscillator (HO) magic numbers 2, 8, 20, 40, 70, 112, 168 according to the Elliott SU(3) symmetry. The results have been obtained by the code of Ref. \cite{code} and have been presented in \cite{Assimakis}. Examples about the calculation of the irreps have been given in Refs. \cite{Rila2018,proxy5,Book}. An analytic formula for the calculation of the highest weight irreps is given in \cite{Kota2018}.}\label{irrepsa}

\bigskip

\begin{tabular}{c c c c c c}

\hline

Particle Number & $(\lambda, \mu)_{SO}$&$(\lambda, \mu)_{HO}$& Particle Number & $(\lambda, \mu)_{SO}$&$(\lambda, \mu)_{HO}$\\
2 & (0, 0) & (0, 0) & 1 & (0, 0) & (0, 0)\\
4 & (0, 0) & (2, 0) & 3 & (0, 0) & (1, 0)\\
6 & (0, 0) & (0, 2) & 5 & (0, 0) & (1, 1)\\
8 & (2, 0) & (0, 0) & 7 & (1, 0) & (0, 1)\\
10 & (0, 2) & (4, 0) & 9 & (1, 1) & (2, 0) \\
12 & (0, 0) & (4, 2) & 11 & (0, 1) & (4, 1)\\
14 & (0, 0) & (6, 0) & 13& (0, 0) & (5, 1)\\
16 & (4, 0) & (2, 4) & 15 & (2, 0) & (4, 2)\\
18 & (4, 2) & (0, 4) & 17 & (4, 1) & (1, 4)\\
20 & (6, 0) & (0, 0) & 19 & (5, 1) & (0, 2)\\
22 & (2, 4) & (6, 0) & 21 & (4, 2) & (3, 0)\\
24 & (0, 4) & (8, 2) & 23 & (1, 4) & (7, 1)\\
26 & (0, 0) & (12, 0) & 25 & (0, 2) & (10, 1)\\
28 & (0, 0) & (10, 4) & 27 & (0, 0) & (11, 2)\\
30 & (6, 0) & (10, 4) & 29 & (3, 0) & (10, 4)\\
32 & (8, 2) & (12, 0) & 31 & (7, 1) & (11, 2)\\
34 & (12, 0) & (6, 6) & 33 & (10, 1) & (9, 3)\\
36 & (10, 4) & (2, 8) & 35 & (11, 2) & (4, 7)\\
38 & (10, 4) & (0, 6) & 37 & (10, 4) & (1, 7)\\
40 & (12, 0) & (0, 0) & 39 & (11, 2) & (0, 3)\\
42 & (6, 6) & (8, 0) & 41 & (9, 3) & (4, 0)\\
44 & (2, 8) & (12, 2) & 43 & (4, 7) & (10, 1)\\
46 & (0, 6) & (18, 0) & 45 & (1, 7) & (15, 1)\\
48 & (0, 0) & (18, 4) & 47 & (0, 3) & (18, 2)\\
50 & (0, 0) & (20, 4) & 49 & (0, 0) & (19, 4)\\
52 & (8, 0) & (24, 0) & 51 & (4, 0) & (22, 2)\\
54 & (12, 2) & (20, 6) & 53 & (10, 1) & (22, 3)\\
56 & (18, 0) & (18, 8) & 55 & (15, 1) & (19, 7)\\
58 & (18, 4) & (18, 6) & 57 & (18, 2) & (18, 7)\\
60 & (20, 4) & (20, 0) & 59 & (19, 4) & (19, 3)\\
62 & (24, 0) & (12, 8) & 61 & (22, 2) & (16, 4)\\
64 & (20, 6) & (6, 12) & 63 & (22, 3) & (9, 10)\\
66 & (18, 8) & (2, 12) & 65 & (19, 7) & (4, 12)\\
68 & (18, 6) & (0, 8) & 67 & (18, 7) & (1, 10)\\
70 & (20, 0) & (0, 0) & 69 & (19, 3) & (0, 4)\\
72&  (12, 8) & (10, 0) & 71 & (16, 4) & (5, 0)\\
74& (6, 12) & (16, 2) & 73 & (9, 10) & (13, 1) \\
76& (2, 12) & (24, 0) & 75 & (4, 12) & (20, 1)\\
78& (0, 8) & (26, 4) & 77 & (1, 10) & (25, 2)\\
80& (0, 0) & (30, 4) & 79 & (0, 4) & (28, 4)\\
82& (0, 0) & (36, 0) & 81 & (0, 0) & (33, 2)\\

\hline

\end{tabular}

\end{table*}

\setcounter{table}{0}

\begin{table*}[htb]

\caption{(continued)}

\bigskip

\begin{tabular}{c c c c c c}

\hline

Particle Number & $(\lambda, \mu)_{SO}$&$(\lambda, \mu)_{HO}$& Particle Number & $(\lambda, \mu)_{SO}$&$(\lambda, \mu)_{HO}$\\

84& (10, 0) & (34, 6) & 83 & (5, 0) & (35, 3)\\
86& (16, 2) & (34, 8) & 85 & (13, 1) & (34, 7)\\
88& (24, 0) & (36, 6) & 87 & (20, 1) & (35, 7)\\
90& (26, 4) & (40, 0) & 89 & (25, 2) & (38, 3)\\
92& (30, 4) & (34, 8) & 91 & (28, 4) & (37, 4)\\
94& (36, 0) & (30, 12) & 93 & (33, 2)& (32, 10)\\
96& (34, 6) & (28, 12) & 95 & (35, 3) & (29, 12)\\
98& (34, 8) & (28, 8) & 97 & (34, 7) & (28, 10)\\
100& (36, 6) & (30, 0) & 99 & (35, 7) & (29, 4)\\
102& (40, 0) & (20, 10) & 101 & (38, 3) & (25, 5)\\
104& (34, 8) & (12, 16) & 103 & (37, 4) & (16, 13)\\
106& (30, 12) & (6, 18) & 105 & (32, 10) & (9, 17)\\
108& (28, 12) & (2, 16) & 107 & (29, 12) & (4, 17)\\
110& (28, 8) & (0, 10) & 109 & (28, 10) & (1, 13)\\
112& (30, 0) & (0, 0) & 111 & (29, 4) & (0, 5)\\
114 & (20, 10) & (12, 0) & 113 & (25, 5) & (6, 0)\\
116 & (12, 16) & (20, 2) & 115 & (16, 13) & (16, 1)\\
118 & (6, 18) & (30, 0) & 117 & (9, 17) & (25, 1) \\
120 & (2, 16) & (34, 4) & 119 & (4, 17) & (32, 2)\\
122 & (0, 10) & (40, 4) & 121 & (1, 13) & (37, 4)\\
124 & (0, 0) & (48, 0) & 123 & (0, 5) & (44, 2)\\
126 & (0, 0) & (48, 6) & 125 & (0, 0) & (48, 3)\\

\hline

\end{tabular}

\end{table*}


\begin{table*}[htb]

\caption{The continuation of Table \ref{irrepsa} but for the spin-orbit (SO) like magic numbers 126, 184 using the proxy-SU(3) symmetry \cite{proxy1,proxy2,proxy3,proxy4,proxy5} and for the harmonic oscillator magic numbers 112, 168, 240 using the Elliott SU(3) symmetry \cite{Elliott1,Elliott2}. The highest weight irreps $(\lambda,\mu)$ have been calculated with the analytic formula of Ref. \cite{Kota2018}.}\label{irrepsb}

\bigskip

\begin{tabular}{c c c c c c}

\hline

Particle Number & $(\lambda, \mu)_{SO}$&$(\lambda, \mu)_{HO}$& Particle Number & $(\lambda, \mu)_{SO}$&$(\lambda, \mu)_{HO}$\\
128 & (12, 0) & (50, 8) & 127 & (6, 0) & (49, 7)\\
130 & (20, 2) & (54, 6) & 129 & (16, 1) & (52, 7)\\
132 & (30, 0) & (60, 0) & 131 & (25, 1) & (57, 3)\\
134 & (34, 4) & (56, 8) & 133 & (32, 2) & (58, 4)\\
136 & (40, 4) & (54, 12) & 135 & (37, 4) & (55, 10)\\
138 & (48, 0) & (54, 12) & 137 & (44, 2) & (54, 12)\\
140 & (48, 6) & (56, 8) & 139 & (48, 3) & (55, 10)\\
142 & (50, 8) & (60, 0) & 141 & (49, 7) & (58, 4)\\
144 & (54, 6) & (52, 10) & 143 & (52, 7) & (56, 5)\\
146 & (60, 0) & (46, 16) & 145 & (57, 3) & (49, 13)\\
148 & (56, 8) & (42, 18) & 147 & (58, 4) & (44, 17)\\
150 & (54, 12) & (40, 16) & 149 & (55, 10) & (41, 17)\\
152 & (54, 12) & (40, 10) & 151 & (54, 12) & (40, 13)\\
154 & (56, 8) & (42, 0) & 153 & (55, 10) & (41, 5)\\
156 & (60, 0) & (30, 12) & 155 & (58, 4) & (36, 6)\\
158 & (52, 10) & (20, 20) & 157 & (56, 5) & (25, 16)\\
160 & (46, 16) & (12, 24) & 159 & (49, 13) & (16, 22)\\
162 & (42, 18) & (6, 24) & 161 & (44, 17) & (9, 24)\\
164 & (40, 16) & (2, 20) & 163 & (41, 17) & (4, 22)\\
166 & (40, 10) & (0, 12) & 165 & (40, 13) & (1, 16)\\
168 & (42, 0) & (0, 0) & 167 & (41, 5) & (0, 6)\\
170 & (30, 12) & (14, 0) & 169 & (36, 6) & (7, 0)\\
172 & (20, 20) & (24, 2) & 171 & (25, 16) & (19, 1)\\
174 & (12, 24) & (36, 0) & 173 & (16, 22) & (30, 1)\\
176 & (6, 24) & (42, 4) & 175 & (9, 24) & (39, 2)\\
178 & (2, 20) & (50, 4) & 177 & (4, 22) & (46, 4)\\
180 & (0, 12) & (60, 0) & 179 & (1, 16) & (55, 2)\\
182 & (0, 0) & (62, 6) & 181 & (0, 6) & (61, 3)\\
184 & (0, 0) & (66, 8) & 183 & (0, 0) & (64, 7)\\

\hline

\end{tabular}

\end{table*}

Consequently a 3D-HO valence nuclear shell, which is created by magic numbers 2, 8, 20, 40, 70, 112, 168, is described by the Elliott SU(3) irreps $(\lambda,\mu)_{HO}$, while a SO shell, with magic numbers 6, 14, 28, 50, 82, 126, 182  by the irreps $(\lambda,\mu)_{SO}$ of the proxy-SU(3) symmetry. The eigenvalues of the $C_2$ operator, as derived by the two types of irreps, correspond to the deformation of the particle configuration within the two types of magic numbers (see Eq. (\ref{beta})), namely the HO and the SO--like magic numbers. In the following we will suggest, that the phenomenon of shape coexistence emerges from an interchange among the two types of magic numbers.

\section{Particle excitations}\label{excitations}

The state--of--the--art mechanism for shape coexistence is the particle--hole (p--h) excitation mechanism, which is reviewed in section IIA of Ref. \cite{Heyde2011}. The p--h excitation mechanism began with a suggestion of Morinaga in Ref. \cite{Morinaga1956} for the excited $0_2^+$ state of $\ce{^{16}O}$. He suggested, that this state could be understood as a 4 particle--4 hole excitation from the $p$ shell to the $sd$ shell. This 4p--4h excitation can also be understood as an excitation of an alpha particle across the $p$ shell \cite{Cseh2019}.

In the spherical Shell Model approach of shape coexistence, the Hamiltonian is the summation for every valence nucleon of all the single--particle energies plus the two--body nucleon--nucleon interactions (see Eq. (1) of Ref. \cite{Heyde2011}).  The energy gaps \cite{Otsuka2001} in this approach play a key role, since a reduction of the original valence space may lead to particle configurations with a different deformation than the initial one. Thus while moving away from closed shells, in areas where deformation arises, the knowledge of the effective single--particle energies, which determine the shell gaps, is of major importance for the particle--hole excitations.

In section \ref{Elliott} we have argued, that the intrinsic single--particle basis for the valence shell of deformed nuclei within the Elliott SU(3) symmetry is the cartesian basis $\ket{n_z,n_x,n_y,m_s}$, because such orbitals are eigenstates of the dominant $q_{0,i}q_{0,i'}$ interaction \cite{Martinou2019} and due to the fact, that rotational nuclear bands emerge from these states \cite{Elliott3,Elliott4}. In addition in section \ref{Elliott} in Eqs. (\ref{e0}), (\ref{Nc}), (\ref{Ns}) we have given the formula to calculate the single--particle energies of the Hamiltonian $h_0$ (as defined in Eq. (\ref{h0})) in the two types of bases $\ket{n,l,j,m_j}$, $\ket{n_z,n_x,n_y,m_s}$ in a 3D-HO valence shell among the HO magic numbers 2, 8, 20, 40, 70, 112, 168, which has the Elliott SU(3) symmetry. Furthermore in section \ref{proxy} we have presented the proxy-SU(3) symmetry, which is valid in the SO--like valence shells constituted by normal and intruder parity orbitals, among the magic numbers 6, 14, 28, 50, 82, 126, 184 (see Table 7 of Ref. \cite{proxy4}). In such shells the single particle Hamiltonian $h_{0,SO}$ of Eq. (\ref{h0pr}) is valid, which is {\it not} having the cartesian states $\ket{n_z,n_x,n_y,m_s}$ as eigenstates and as a result only the expectation values of Eq. (\ref{singlecpr}) can be calculated.

In this section we will present the neutron single--particle energies of a nucleus with 7 neutrons in the Elliott and the proxy-SU(3) scheme. Specifically we will clarify, which orbitals are occupied, if a) the valence neutrons obey the HO magic numbers and b) if they follow the SO--like magic numbers. Afterwards we will present the single--particle energies of these nucleons. Through a comparison of the single--particle energies it will become clear, that the particle configuration, which follows the SO shell 6-14, has excited single-particle energies, when compared with the particle configuration, which follows the HO shell 2-8. Thus {\it for certain nucleon numbers the transition of a particle from the HO shell to the neighboring SO shell could be interpreted as 
a particle excitation.}

Let us give an example, which suits to the case of the halo nucleus $\ce{^{11}_4Be_7}$ \cite{Geithner1999, Kondo2010}. If the 7 neutrons of this nucleus follow the HO magic numbers 2, 8, then the core consists of 2 neutrons, which occupy the $1s^{1/2}_{\pm 1/2}$ orbitals. The orbitals of a closed core can be described either in the spherical or in the cartesian basis, since both scenarios contribute the same in the energy and in the nuclear wave function. For simplicity we choose, to describe the closed core with the spherical basis. The valence shell among magic numbers 2-8, consists of the cartesian orbitals with $\mathcal{N}=1$ quanta (see Eq. (\ref{order})):
\begin{gather}\label{c1}
\ket{n_z,n_x,n_y,m_s}:\ket{1,0,0,\pm {1\over 2}}, \ket{0,1,0,\pm {1\over 2}}, \ket{0,0,1,\pm {1\over 2}}.
\end{gather}
Therefore the 5 valence neutrons occupy the orbitals:
\begin{gather}\label{o1}
\ket{n_z,n_x,n_y,m_s}:\ket{1,0,0,\pm {1\over 2}}, \ket{0,1,0,\pm {1\over 2}}, \ket{0,0,1,+ {1\over 2}}.
\end{gather}
The eigenvalues in units $\hbar\omega$ of the Hamiltonian $h_0$ of Eq. (\ref{h0}), as given by Eqs. (\ref{e0}), (\ref{Nc}), (\ref{Ns}) for the $1s^{1/2}_{\pm 1/2}$ and for the orbitals of Eq. (\ref{o1}) are ${3\over 2}, {5\over 2},{5\over 2}, {5\over 2}$ respectively.

If the 7 neutrons of this nucleus follow the SO magic numbers 6-14, then the core consists of the orbitals $1s^{1/2}_{\pm 1/2}$, $1p^{3/2}_{\pm 1/2}$, $1p^{3/2}_{\pm 3/2}$. The valence SO--like shell 6-14 consists of the orbitals $1p^{1/2}_{\pm 1/2}$, $1d^{5/2}_{\pm 1/2}$, $1d^{5/2}_{\pm 3/2}$, $1d^{5/2}_{\pm 5/2}$. The proxy valence shell, which results after the action of the unitary transformation of Eq. (\ref{unitary}) on the $1d^{5/2}_{m_j}$ orbitals \cite{proxy4}, consists of the orbitals  $1p^{1/2}_{\pm 1/2}$, $1p^{3/2}_{\pm 1/2}$, $1p^{3/2}_{\pm 3/2}$, with $\mathcal{N}=1$ quanta. Thus the proxy shell among magic numbers 6-12 \cite{proxy4} consists of the cartesian orbitals of Eq. (\ref{c1}). The one valence nucleon of the 6-12 proxy shell according to the occupancy order of Eq.  (\ref{order}) occupies the: 
\begin{gather}\label{o2}
\ket{n_z,n_x,n_y,m_s}:\ket{1,0,0,+{1\over 2}}
\end{gather}
orbital.

The single--particle energies $\epsilon_0$ of the Hamiltonian $h_0$ for the occupied orbitals $1s^{1/2}_{\pm 1/2}$, $1p^{3/2}_{\pm 1/2}$, $1p^{3/2}_{\pm 3/2}$ of the core in units $\hbar\omega$ are given by Eqs. (\ref{e0}), (\ref{Nc}), (\ref{Ns}) and result to be ${3\over 2}$, ${5\over 2}$, ${5\over 2}$ respectively. The expectation value of the occupied state of Eq. (\ref{o2}) is given by Eq. (\ref{singlecpr}). The transformation coefficients for the specific orbital are given in Eq. (\ref{ex}) to be:
\begin{gather}\label{coef}
\braket{n,l,j,m_j|n_z,n_x,n_y,m_s}:\\
\braket{0,1,{1\over 2},{1\over 2}|1,0,0,{1\over 2}}=-{1\over\sqrt{3}},\label{coef1}\\
\braket{0,1,{3\over 2},{1\over 2}|1,0,0,{1\over 2}}=\sqrt{2\over 3}.\label{coef2}
\end{gather}
Consequently the expectation value as given by Eqs. (\ref{singlecpr}), (\ref{coef1}), (\ref{coef2}) is: 
\begin{eqnarray}\label{expect1}
\langle\epsilon_0(\ket{1,0,0,{1\over 2}})\rangle_{SO}=\nonumber\\
{1\over 3}\epsilon_0(\ket{1p^{1/2}_{1/2}})+{2\over 3}\epsilon_0(\ket{1p^{3/2}_{1/2}})_{SO},
\end{eqnarray}
where according to Eqs. (\ref{e0}), (\ref{Ns}):
\begin{equation}\label{e012}
\epsilon_0(\ket{1p^{1/2}_{1/2}})=\left(\mathcal{N}+{3\over 2}\right)\hbar\omega
={5\over 2}\hbar\omega,
\end{equation}
and according to Eqs. (\ref{e0pr}), (\ref{Ns}):
\begin{equation}\label{e032}
\epsilon_0(\ket{1p^{3/2}_{1/2}})_{SO}=\left(\mathcal{N}+{3\over 2}\right)\hbar\omega+\epsilon_{proxy}={5\over 2}\hbar\omega+\epsilon_{proxy}.
\end{equation}
The $+\epsilon_{proxy}$ term in the above eigenvalue is the result of the replacement of the intruder orbitals by their proxies, which applies in the proxy SU(3) symmetry \cite{proxy1,proxy4}. Finally the expectation value of Eq. (\ref{expect1}) using Eqs. (\ref{e012}), (\ref{e032}) is:
\begin{equation}\label{expBe}
\langle\epsilon_0(\ket{1,0,0,{1\over 2}})\rangle_{SO}={5 \over 2}\hbar\omega+{2\epsilon_{proxy}\over 3}.
\end{equation}

For the ground state of $\ce{^{11}Be}$ we suppose, that the protons follow the HO magic numbers, while the neutrons follow the SO--like magic numbers, giving $(\lambda,\mu)_\varpi=(2,0)$ for protons and $(\lambda,\mu)_\nu=(1,0)$ for neutrons (see Table \ref{irrepsa}). The overall nuclear irrep is $(\lambda,\mu)=(3,0)$, as will be explained in section \ref{Be}. Using this irrep and Eqs. (\ref{C2})-(\ref{scale}), (\ref{def1}), (\ref{def2}) the deformation parameter of the ground state band of $\ce{^{11}Be}$ is predicted to be $\varepsilon=0.36$, which from Eq. (\ref{eproxy}) gives $\epsilon_{proxy}=0.76\hbar\omega$. Thus Eq. (\ref{expBe}) becomes:
\begin{equation}
\langle\epsilon_0(\ket{1,0,0,{1\over 2}})\rangle_{SO}\approx 3\hbar\omega.
\end{equation}

The results of this example are summarized in Table \ref{Beex}, where it becomes obvious, that the neutron configuration in the 6-14 SO--like magic numbers has excited single-particle energies comparing with those in the HO 2-8 magic numbers.
\begin{table*}
\caption{The list of the occupied orbitals for the neutron configuration of the halo nucleus $\ce{^{11}_4Be_{7}}$ for two sets of magic numbers, namely the harmonic oscillator magic numbers (HO) and the spin-orbit like (SO) magic numbers. The single--particle energies $\epsilon_0$ for the original orbitals have been calculated by Eq. (\ref{e0}), while the expectation value $\langle\epsilon_0(\ket{n_z,n_x,n_y,m_s})\rangle_{SO}$ for the proxy orbital $\ket{n_z=1,n_x=0,n_y=0,m_s={1\over 2}}$ has been calculated by Eq. (\ref{singlecpr}). See section \ref{excitations} for further discussion. In this example when the neutron configuration flips from the HO shell to the SO--like, one neutron is excited from the 2-8 HO shell to the 6-14 SO--like shell. }\label{Beex}
\begin{center}
\begin{tabular}{c c c| c c}
& Harmonic oscillator (HO) magic numbers && Spin-orbit (SO) like magic numbers \\
Neutrons & Orbitals & $\epsilon_0$ & Orbitals & $\langle\epsilon_0\rangle$\\
$1^{st}$ & $\ket{1s^{1/2}_{1/2}}$ & 1.5 &  $\ket{1s^{1/2}_{1/2}}$ & 1.5\\
$2^{nd}$ & $\ket{1s^{1/2}_{-1/2}}$ & 1.5 &  $\ket{1s^{1/2}_{-1/2}}$ & 1.5\\
$3^{rd}$ & $\ket{n_z=1,n_x=0,n_y=0,m_s={1\over 2}}$ & 2.5 &  $\ket{1p^{3/2}_{3/2}}$ & 2.5\\
$4^{th}$ & $\ket{n_z=1,n_x=0,n_y=0,m_s=-{1\over 2}}$ & 2.5 &  $\ket{1p^{3/2}_{1/2}}$ & 2.5\\
$5^{th}$ & $\ket{n_z=0,n_x=1,n_y=0,m_s={1\over 2}}$ & 2.5 &  $\ket{1p^{3/2}_{-1/2}}$ & 2.5\\
$6^{th}$ & $\ket{n_z=0,n_x=1,n_y=0,m_s=-{1\over 2}}$ & 2.5 &  $\ket{1p^{3/2}_{-3/2}}$ & 2.5\\
$7^{th}$ & $\ket{n_z=0,n_x=0,n_y=1,m_s={1\over 2}}$ & 2.5 & $\ket{n_z=1,n_x=0,n_y=0,m_s={1\over 2}}$  & 3\\
\end{tabular}
\end{center}
\end{table*}

Now a generalization can be made. A proxy SO--like shell among magic numbers 6-12, 14-26, 28-48, 50-80, 82-124, 126-184 (Table 7 of Ref. \cite{proxy4}) has excited single-particle energies comparing with those of a HO shell among magic numbers 2-8, 8-20, 20-40, 40-70, 70-112, 112-168 respectively. Thus one could say that, for certain proton or neutron numbers, where both types of shells are open and the single-particle states of the SO shell are excited comparing with those of the HO shell, then the transition from the HO shell to the relative SO--like shell could be interpreted as {\it particle excitations}. The particle numbers, for which this is possible, are presented in Table \ref{phexc}.
\begin{table*}
\caption{In the first column are presented the proton or neutron numbers, for which both types of shells, namely the HO and the SO--like, are open. For these nucleon numbers the SO--like shell of the the third column is excited, when compared to the HO shell of the second column. See section \ref{excitations} for further discussion.}\label{phexc}
\begin{center}
\begin{tabular}{c c c}
Nucleon number & HO shell & excited SO shell\\
6-8 & 2-8 & 6-14\\
14-20 & 8-20 & 14-28\\
28-40 & 20-40 & 28-50\\
50-70 & 40-70 & 50-82\\
82-112 & 70-112 & 82-126\\
126-168 & 112-168 & 126-184\\
\end{tabular}
\end{center}
\end{table*}

In this section we focused on the case of $\ce{^{11}Be}$, because this was the simplest example one could present. Through this example it became evident, that for certain nucleon numbers, which have been presented in Table \ref{phexc}, a transition from the HO shell to the open SO shell is equivalent to particle excitations. In section \ref{Be} we will present some predictions about $\ce{^{11}Be}$ within the dual--shell mechanism.

Returning to the case of $\ce{^{16}O}$, for which the study of shape coexistence began at 1956 \cite{Morinaga1956}, we could say, that a change of the proton and neutron configuration from the HO shell among magic numbers 2-8 to the SO shell among magic numbers 6-14 is equivalent to a 4 particle excitation \cite{Rowe2006}. Specifically if the 8 protons and the 8 neutrons of $\ce{^{16}O}$ inhabit the SO--like shells, then the 6-14 valence shell consists of 2 valence protons and 2 valence neutrons. Thus the on hand mechanism, suggests indeed an excitation of an alpha particle across the $p$ shell \cite{Cseh2019} to the mixed parity $1p^{1/2},1d^{5/2}$ SO--like shell.

In the HO scheme the proton ($\varpi$) and the neutron irrep ($\nu$) irreps for $\ce{^{16}O}$ are $(\lambda_\varpi,\nu_\varpi)_{HO}$ =$(0,0)$, $(\lambda_\nu,\mu_\nu)_{HO}$ =$(0,0)$ (see Table \ref{irrepsa}). The result of the outer product $(\lambda,\mu)_\varpi\otimes(\lambda,\mu)_\nu$  \cite{Harvey,Alex2011,Coleman1964,Troltenier1996} is:
\begin{equation}
(\lambda,\mu)_{HO}=(\lambda_\varpi+\lambda_\nu, \mu_\varpi+\mu_\nu)_{HO}=(0,0).
\end{equation}
In the SO--like configuration for $\ce{^{16}O}$  (see Table \ref{irrepsa}):
\begin{gather}
(\lambda_\varpi,\nu_\varpi)_{SO}=(2,0),\label{piproxy}\\
 (\lambda_\nu,\mu_\nu)_{SO}=(2,0),\label{nuproxy}
\end{gather}
giving:
\begin{equation}
(\lambda,\mu)_{SO}=(\lambda_\varpi+\lambda_\nu, \mu_\varpi+\mu_\nu)_{SO}=(4,0).
\end{equation}

Thus following Eq. (\ref{beta}) with the scaling factor of Eq. (\ref{scale}) $S_\varpi+S_\nu=12$ the dual--shell mechanism predicts a less deformed shape with $\beta=0.12$ coming from the HO configuration and a more deformed shape with $\beta=0.386$ from the SO--like particle configuration. The more deformed shape originating from the SO--like shell generates in the dual--shell mechanism the ground state band of $\ce{^{16}O}$ with experimental deformation variable $\beta=0.364$ \cite{Ni}, while the less deformed shape originating from the HO shell matches with the $0_2^+$ of $\ce{^{16}O}$ at 6.049 MeV \cite{Tilley1993}.  The transition $B(E2)\uparrow$ from the $0_1^+$ to the $2_1^+$ is used in Ref. \cite{Ni}, to derive the experimental deformation variable $\beta=0.364$. The more deformed shape with irrep $(4,0)$, predicts a ground state band with levels $L^+: 0_1^+, 2_1^+, 4_1^+$ according to Eq. (23) of Ref. \cite{Elliott2}. This sequence of states appears in the data \cite{Tilley1993} at $0$ MeV, $6.9$ MeV and $10.4$ MeV respectively. The observed $E(2)$ transitions among them could support, that they consist the ground state band. Their high energy values, could be predicted in future research by the additon of the pairing interaction into the Elliott SU(3) Hamiltonian, as in Ref. \cite{Bahri1995}, since in $N=Z$ nuclei with isospin $T=0$ \cite{Tilley1993} the pairing interaction is important. The energy of the $0_2^+$ state of $\ce{^{16}O}$ will be calculated in section \ref{O} within the dual--shell mechanism for shape coexistence. It will be explained in sections \ref{two} and \ref{islands}, how it is possible, that the SO--like shell, although it has excited single--particle energies comparing with those of the HO shell, reproduces the ground state band. Actually the answer in this paradox, will deliver the condition, which predicts the islands of shape coexistence on the nuclear map.

Another important consequence of the dual--shell mechanism is, that when the particles are placed in the SO--like shell, {\it no holes} are considered. For instance, when the particles are in the 6-14 SO--like shell, the previously filled orbitals $1s^{1/2},1p^{3/2}$ create a {\it closed core}, if one uses the proxy-SU(3) symmetry \cite{proxy4} (see Table I of Ref. \cite{Martinou2019}) and thus no hole irreps emerge. 

This is actually a difference of the dual--shell mechanism with the particle--hole mechanism as realized in the Symplectic Model \cite{Rowe2006,Rosensteel1979,Rosensteel1980,Rowe1980}. Both approaches consider particle excitations from the valence HO shell, but in the Symplectic Model, since only pure parity HO shells are treated, the excitation of one particle from the 2-8 shell to the 8-20 shell generates one hole in the 2-8 shell. 

Specifically for $\ce{^{16}O}$ the particle-hole excitation mechanism predicts, that for the protons a 2p--2h excitation is described by a particle $(\lambda,\mu)$ irrep in the 8-20 shell $(2,0)\otimes (2,0)=(4,0)$ and a hole irrep in the 2-8 shell $(0,1)\otimes (0,1)=(0,2)$ \cite{Rowe2006}. The same stands for the 2p--2h neutron configuration of $\ce{^{16}O}$ \cite{Rowe2006}. Finally the particle--hole excitation mechanism predicts a 4p--4h (2 protons and 2 neutrons) excitation from the 2-8 to the 8-20 shell. The proton and neutron irreps within the particle--hole excitation mechanism for $\ce{^{16}O}$ are \cite{Rowe2006}:
\begin{gather}
(4,0)\otimes(0,2)=(4,2)\mbox{ for 2 proton excitations},\label{pisymp}\\
(4,0)\otimes(0,2)=(4,2)\mbox{ for 2 neutron excitations},\label{nusymp}
\end{gather}
giving an overall nuclear irrep for protons and neutrons together:
\begin{equation}
(4,2)\otimes(4,2)=(8,4), 
\end{equation}
as is indicated in Table I of Ref. \cite{Rowe2006}.

The coexistence of the 0p--0h state with the 4p--4h state for this example results to the shape coexistence in $\ce{^{16}O}$ within the particle-hole excitation mechanism \cite{Rowe2006}. The 0p--0h configuration is set to be the ground state at 0 MeV, while the 4p--4h configuration represents the excited $0_2^+$ state at 4.9 MeV (see Table I of \cite{Rowe2006}), which is said to be close to the data at 6.049 MeV \cite{Tilley1993} for this state of $\ce{^{16}O}$.

\section{The dissolution of magic numbers}\label{mn}

 The SO--like shells are being created by a significant spin--orbit interaction, as discussed in section \ref{Elliott}. The coexistence of the SO--like shell with the HO shell becomes possible, because the significant single--particle energy gaps at the major magic numbers dissolve due to the deformation. When the open HO shell coexists with the open SO--like shell shape coexistence becomes possible. Each of the two types of active shells corresponds to a low--lying energy band. The resulting coexisting bands are characterized by different deformation parameters $(\beta, \gamma)$ of the Bohr and Mottelson Model \cite{BohrII}, which can be calculated by the dual--shell mechanism.

Specifically the fading out of the major magic numbers can be understood simply by looking at a Nilsson diagram \cite{Tables} (which can be found in section ``The Nilsson Model" of Ref. \cite{Casten}, or in Fig. 1 of Ref. \cite{Kota1998}), where the single--particle energies are plotted versus the Nilsson deformation parameter $\varepsilon$ or $\delta$ respectively \cite{Nilsson1}. For a spherical nucleus with $\varepsilon=0$ the nucleon numbers 28, 50, 82 are definitely magic numbers, which means, that there are large energy gaps in the single--particle energies above these nucleon numbers. But as deformation evolves ($\varepsilon>0$) the single particle energies of the Nilsson orbitals are affected by the deformation (see section 1.3 of Ref. \cite{Book} and Ref. \cite{Assimakis2019} for further understanding) and as a result at moderate deformation parameter $\varepsilon\approx 0.2$ the large energy gaps at 28, 50, 82 have already faded out. Consequently the Nilsson diagrams alone tattle, that {\it in deformed nuclei there are no major magic numbers}. 

While the Nilsson Model suffices for a qualitative explanation of the fading out of the major magic numbers at moderate deformations, the same can be observed within the microscopic approach of Energy Density Functionals (EDF) derived from an effective interaction, in the context of the self consistent mean field theory. More specifically, constraining the quadrupole moment of the nuclear mean field one is able to examine the change of the calculated single--particle energies with axial deformation. Examples of Nilsson diagrams produced with this method within the Relativistic Mean Field (RMF) have been reported in Ref. \cite{Prassa2013} and more recently in Ref. \cite{Kostas,Karakatsanis2020}. For instance in Figs. 5.4-5.13 of Ref. \cite{Kostas} the respective diagrams for several isotopes of $\ce{Hf}$, $\ce{Os}$ and $\ce{No}$ are shown and the disappearance of major gaps in the spectrum is clear.

The advantage of the RMF theory \cite{Boguta1977,Serot1984,Reinhard1989,Ring1996,Meng2006} is, that it is a no core theory, {\it i.e.}, all the nuclear shells participate in the calculation. Consequently shell merging is a build-in tool in the RMF theory. Another important aspect of shape coexistence is, that the excited coexisting states have short lifetimes, which means, that they are not stable states or isomeric states. Therefore we have not searched for shape coexistence in the various excited local minima of the energy surface as a function of the deformation variable $\beta$, which can be predicted by the EDF. We simply focused on the {\it ground state predictions} of the theory and we observed there the peculiarities of the single--particle energies, which will enable shape coexistence to appear. 

In addition, EDFs are extremely useful for calculating properties of isotopes far from the valley of stability, where there is limited experimental input and the large number of nucleons prohibits Shell Model calculations. In the suggested dual--shell mechanism we are interested in the energy gaps at the SO--like magic numbers and at the HO magic numbers and we are especially curious about the situation in heavy isotopes. Hence, we have used the functional DDME2 of Ref. \cite{Lalazissis2005} within the numerical implementation of the RMF model in the code of Ref. \cite{Niksic2014} to calculate the single--particle energies $\epsilon$ of the occupied Nilsson orbitals $K[Nn_z\Lambda]$ \cite{Nilsson1}. For an even--even nucleus $\ce{^A_ZX_N}$, with $Z,N$ being the protron and neutron number respectively, the energy gap at the $i^{th}$ nucleon is the difference:
\begin{equation}\label{gap}
\epsilon_{gap}=\epsilon_{i+2}-\epsilon_{i},
\end{equation}
where $i=1,2,...,N-2$ for neutron gaps, while $i=1,2,...,Z-2$ for proton gaps.

\begin{figure}
\begin{center}
\includegraphics[width=85mm]{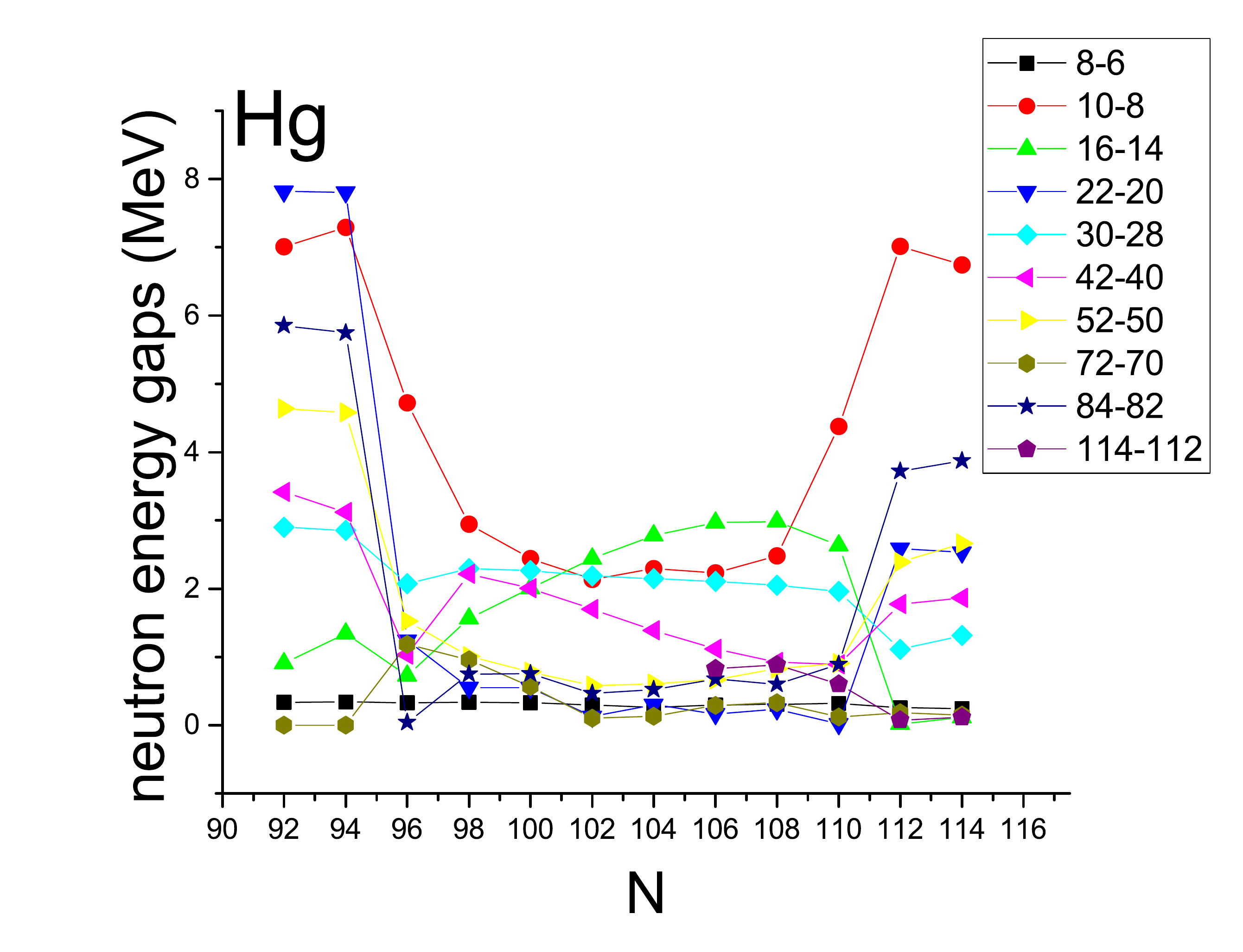}
\caption{Relativistic Mean Field calculations \cite{Lalazissis2005,Niksic2014} of the neutron energy gaps as defined in Eq. (\ref{gap}) for the $\ce{Hg}$ isotopes versus the neutron number. The legend ``8-6" signifies the energy difference (see Eq. (\ref{gap})) among the $6^{th}$ and the $8^{th}$ neutron for each isotope, the ``10-8" is the energy gap among the $8^{th}$ and the $10^{th}$ neutron etc. A valley of small energy gaps appears among $96\le N\le 110$, where exactly shape coexistence appears experimentally (see Fig. 10 of Ref. \cite{Heyde2011} and Fig. \ref{Hgdata}). For these mercury isotopes the major magic numbers have been dissolved and so the open SO--like neutron shell among magic numbers 82-126 can coexist with the open HO neutron shell among magic numbers 70-112. }\label{gapsHg}
\end{center}
\end{figure}

In Figs.\ref{gapsHg} and \ref{gapsPb} we present the predictions of the RMF theory for the neutron single--particle energy gaps as defined in Eq. (\ref{gap}) for mercury and lead isotopes. In these plots we clearly see, that the energy gaps at the major magic numbers are mitigated and so the coexistence of two types of open neutron shells, namely the HO and the SO--like shell, is possible. Specifically a valley of reduced neutron magic numbers appears among $96\le N\le 110$, which matches exactly with the appearance of shape coexistence in the mercury isotopes (see Fig. \ref{Hgdata}, and Fig. 10 of Ref. \cite{Heyde2011}).

\begin{figure}
\begin{center}
\includegraphics[width=85mm]{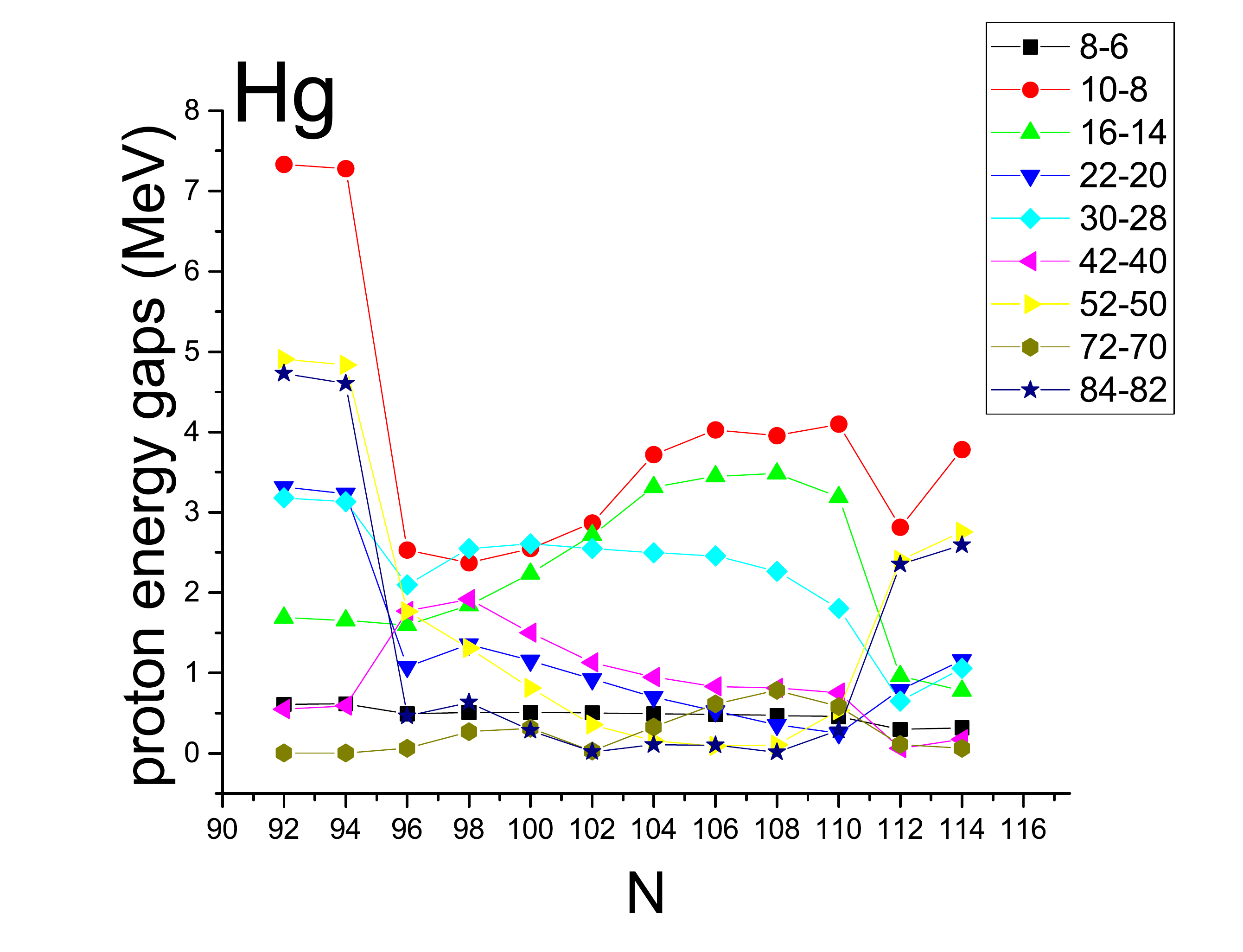}
\caption{The same as Fig. \ref{gapsHg} but for proton gaps. }\label{pgapsHg}
\end{center}
\end{figure}

\begin{figure}
\begin{center}
\includegraphics[width=85mm]{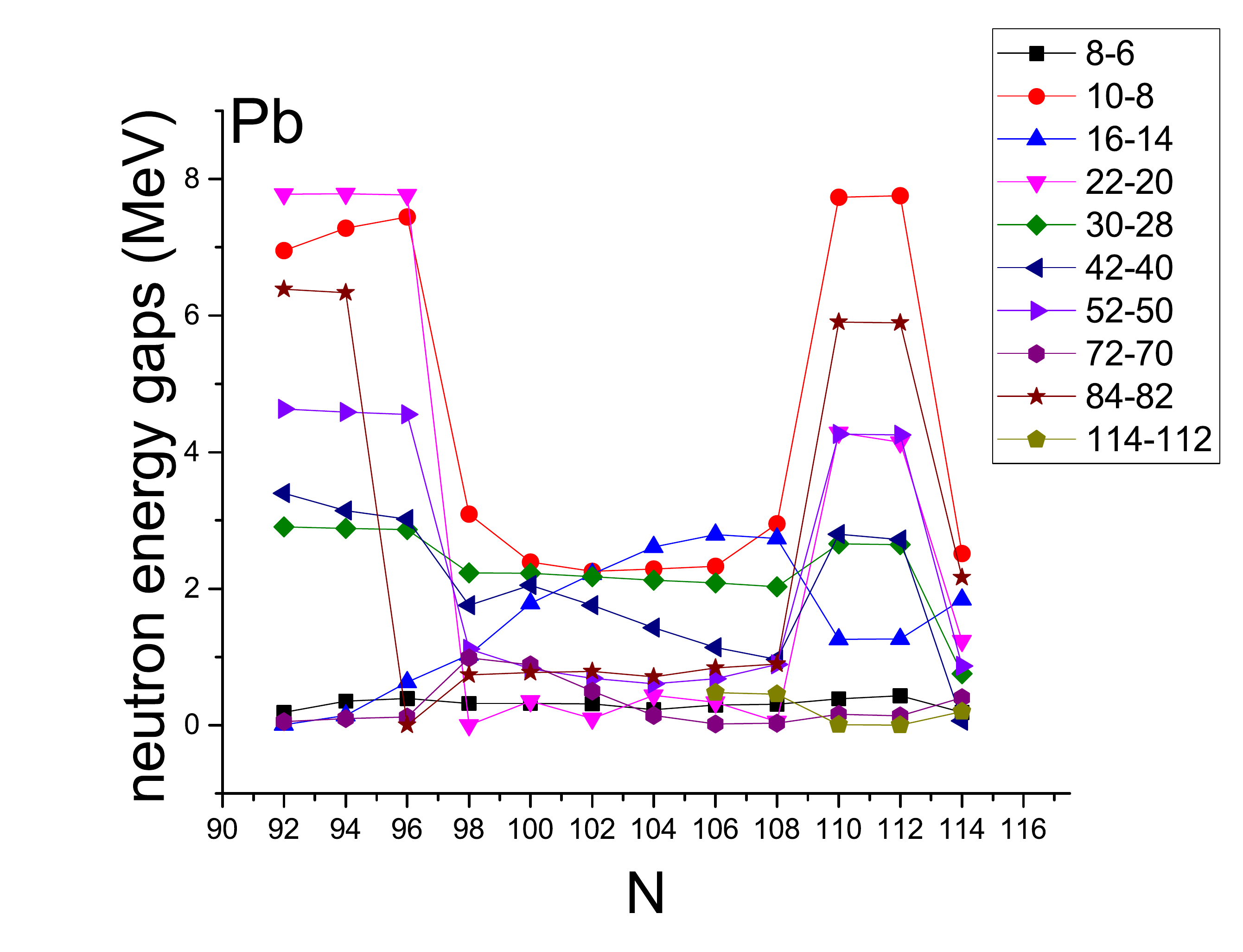}
\caption{The same as Fig. \ref{gapsHg} but for the lead isotopes.}\label{gapsPb}
\end{center}
\end{figure}

\begin{figure}
\begin{center}
\includegraphics[width=85mm]{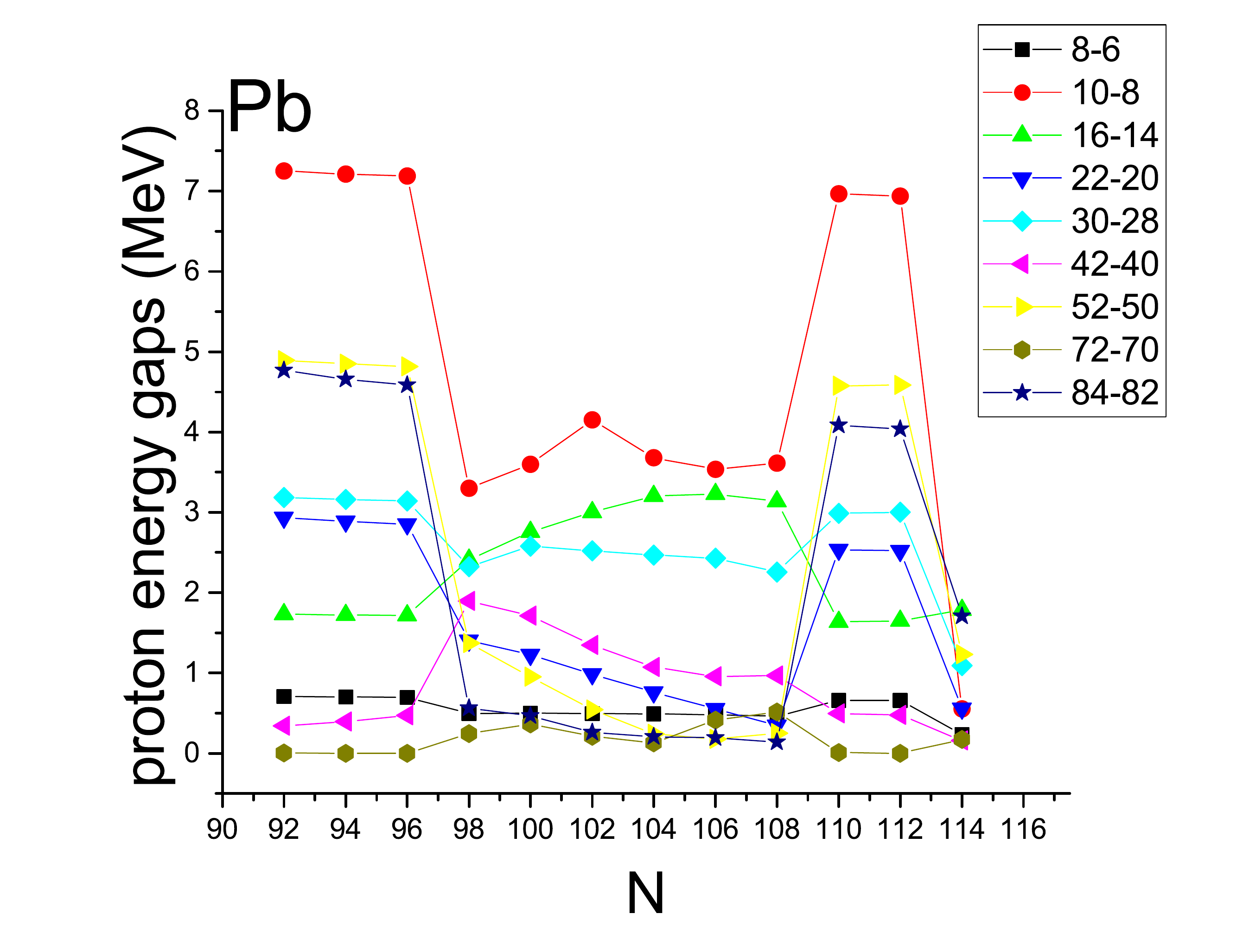}
\caption{The same as Fig. \ref{pgapsHg} but for the lead isotopes. }\label{pgapsPb}
\end{center}
\end{figure}

Figs. \ref{pgapsHg} and \ref{pgapsPb} are the analogs of Figs. \ref{gapsHg} and \ref{gapsPb} for the proton single--particle energy gaps in the $\ce{Hg}$ and $\ce{Pb}$ isotopes. We see, that in the case of proton gaps in Figs. \ref{pgapsHg} and \ref{pgapsPb} a valley appears in the central area of $96\leq N \leq 110$, similar to the one appearing in the neutron gaps in Figs. \ref{gapsHg} and \ref{gapsPb}. This finding clearly suggests that the protons in the $\ce{Hg}$ and $\ce{Pb}$ series of isotopes are not indifferent spectators of the gradual addition of neutrons. The protons are strongly influenced by the addition of neutrons and as a result a valley, in which the SO--like proton magic numbers collapse and become comparable to the HO magic numbers, is created. It is within this valley, that shape coexistence is observed experimentally, as seen, for example, in Fig. 10 of Ref. \cite{Heyde2011}. The role of the proton--neutron interaction in creating this valley calls for further investigation.  

Going into more detail, in Figs. \ref{holes} the proton Nilsson orbitals occurring in the DDME2 \cite{Lalazissis2005} calculation in the 50-82 shell and in the beginning of the next shell are depicted, drawn relatively to the Fermi energy $\epsilon_F$ in each isotope. In $\ce{Pb}$ one would expect, that all orbitals of the 50-82 shell (namely the  $1h^{11/2}$, $1g^{7/2}$, $2d^{5/2}$, $2d^{3/2}$, $3s^{1/2}$ orbitals) should be occupied, while all orbitals above 82 (the $1h^{9/2}$, 
$2f^{7/2}$ orbitals) should be empty. This is indeed the case for $N\leq 96$ and $N\geq 110$. But within the region $96\le N\le 110$ an interesting systematic effect is seen. The $1/2[400]$ and $3/2[402]$ Nilsson orbitals of  $2d^{3/2}$, as well as the $11/2[505]$ orbital of  $1h^{11/2}$, pop up above the Fermi energy, while the $1/2[541]$ and $3/2[532]$ orbitals of $1h^{9/2}$ are sunk below the Fermi energy, thus providing a clear picture of proton excitations. 

The present RMF results suggest, that the particle--hole mechanism for shape coexistence in heavy nuclei is fully justified microscopically within the $96\leq N \leq 110$ region, in which the present dual--shell mechanism also predicts the presence of shape coexistence, as will become evident in section \ref{islands}. The extra hint provided by the present dual--shell mechanism is, that for the $\ce{Hg},\ce{Pb}$ isotopes shape coexistence cannot exist outside the $96\leq N \leq 110$ region, as corroborated by the data (see Fig. 10 of Ref. \cite{Heyde2011}, for example). In other words, the present dual--shell mechanism is not disproving the particle-hole mechanism. In contrast, it corroborates it within the nuclear regions in which the particle--hole mechanism has been applied \cite{Heyde2011}.

\begin{figure}
\begin{center}
\includegraphics[width=85mm]{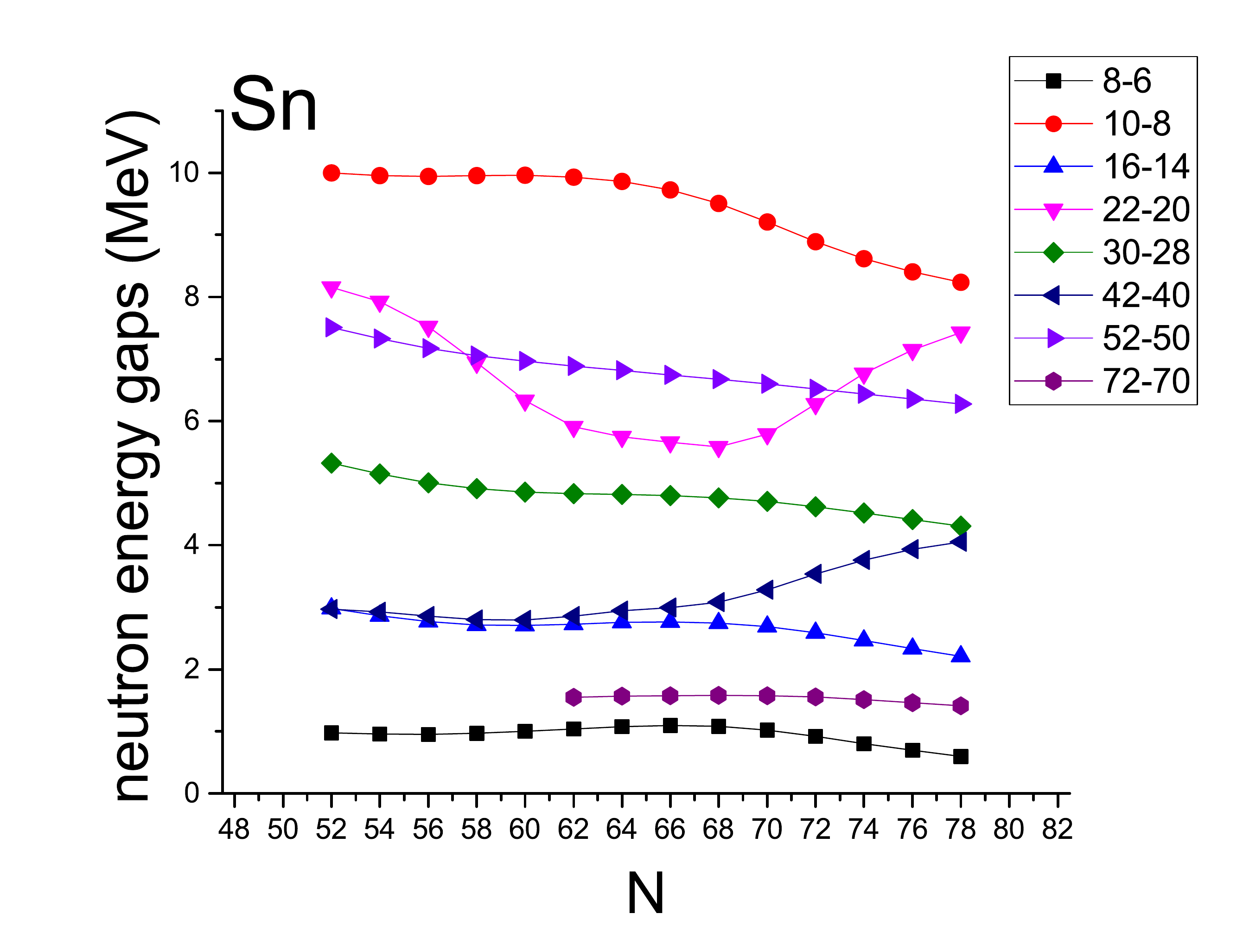}
\caption{The same as Fig. \ref{gapsHg} but for the tin isotopes, which in general are not very deformed. Despite that, the tins $\ce{^{110-120}Sn}$ manifest rotational energy bands and shape coexistence  (see Fig. 3.10 of Ref. \cite{Wood1992} and Fig. \ref{Sndata}). }\label{Sngapsn}
\end{center}
\end{figure}

\begin{figure*}

\subfigure[]{\includegraphics[width=65mm]{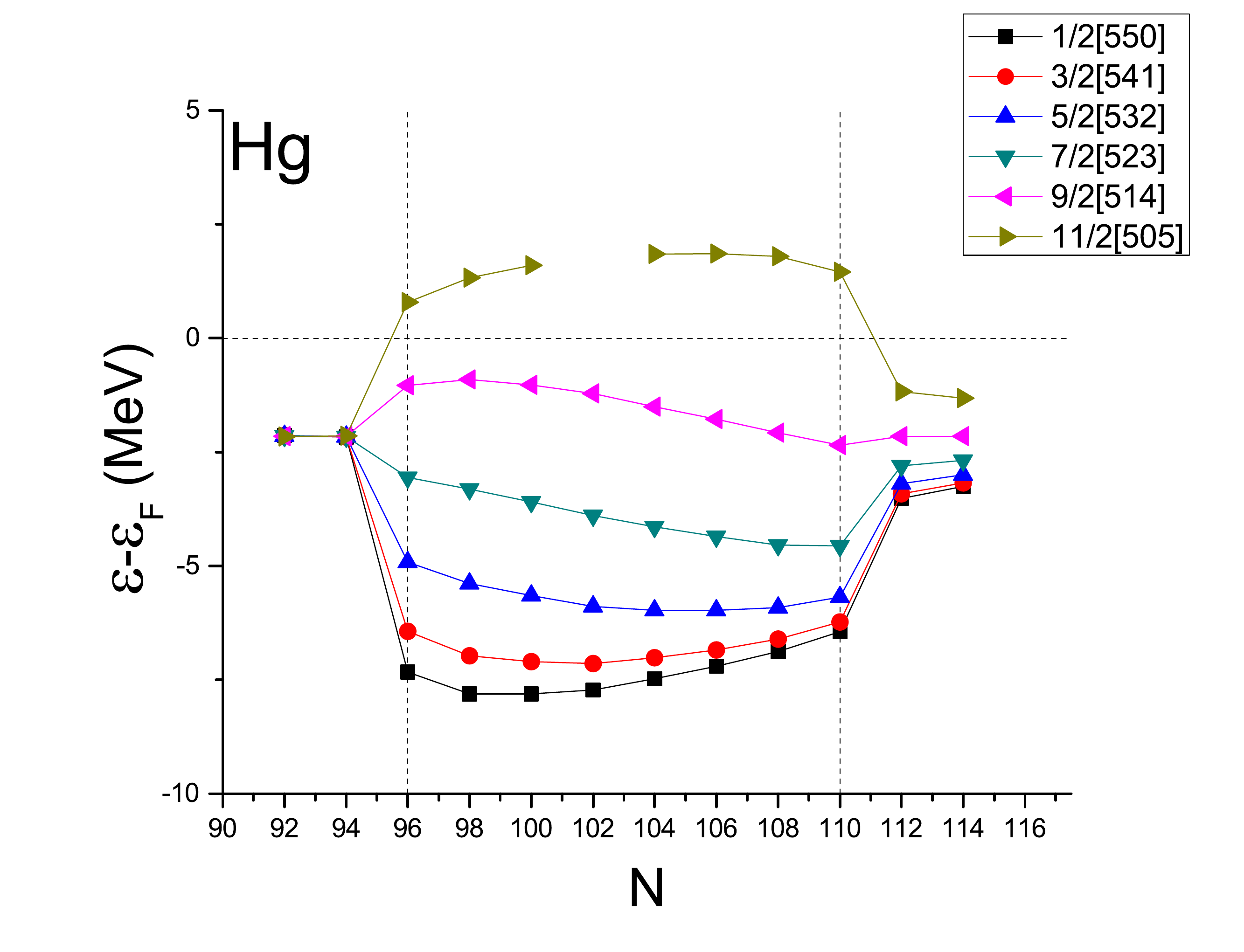}}\label{pexHg1}
\subfigure[]{\includegraphics[width=65mm]{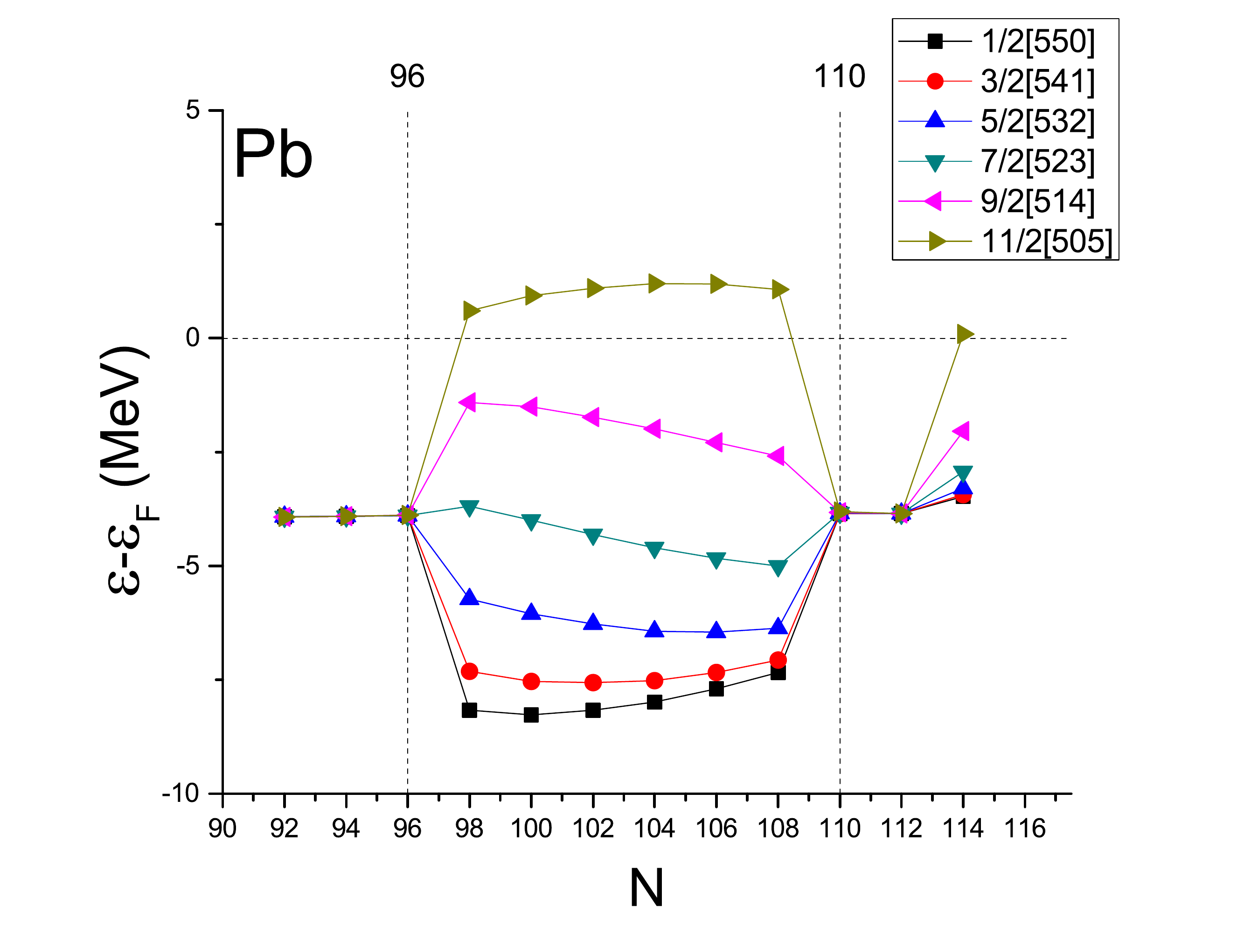}}\label{pexPb1}
\subfigure[]{\includegraphics[width=65mm]{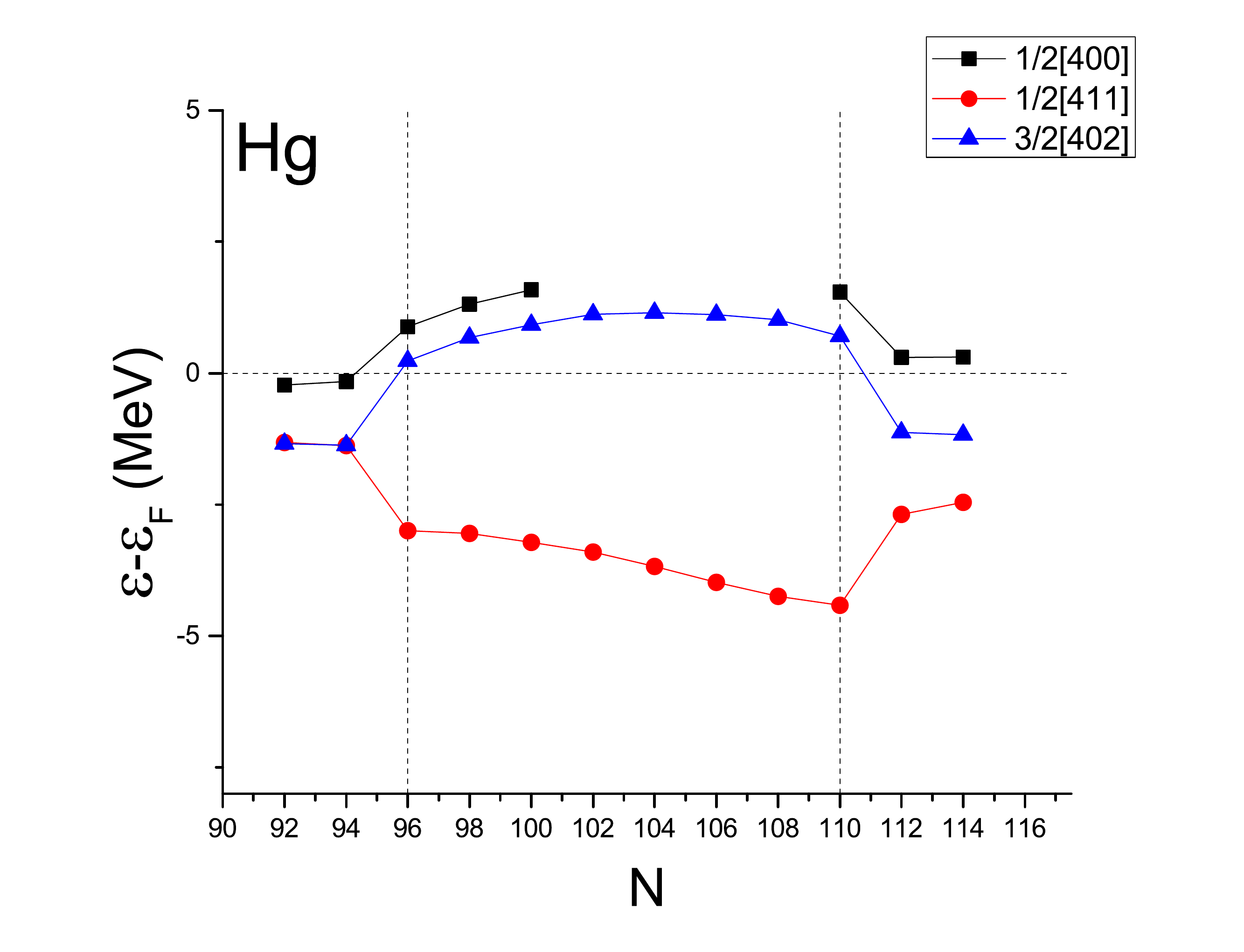}}\label{pexHg2}
\subfigure[]{\includegraphics[width=65mm]{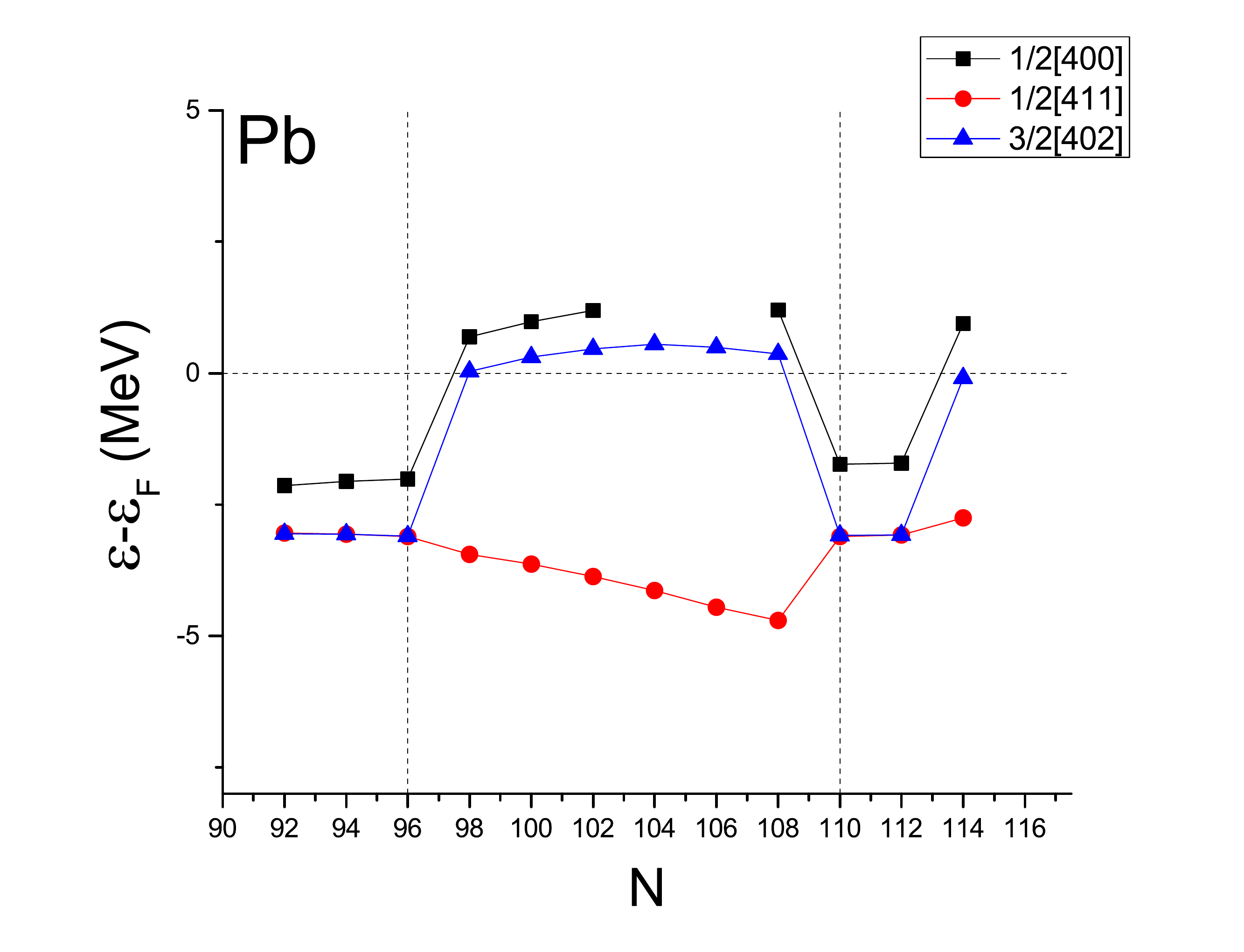}}\label{pexPb2}
\subfigure[]{\includegraphics[width=65mm]{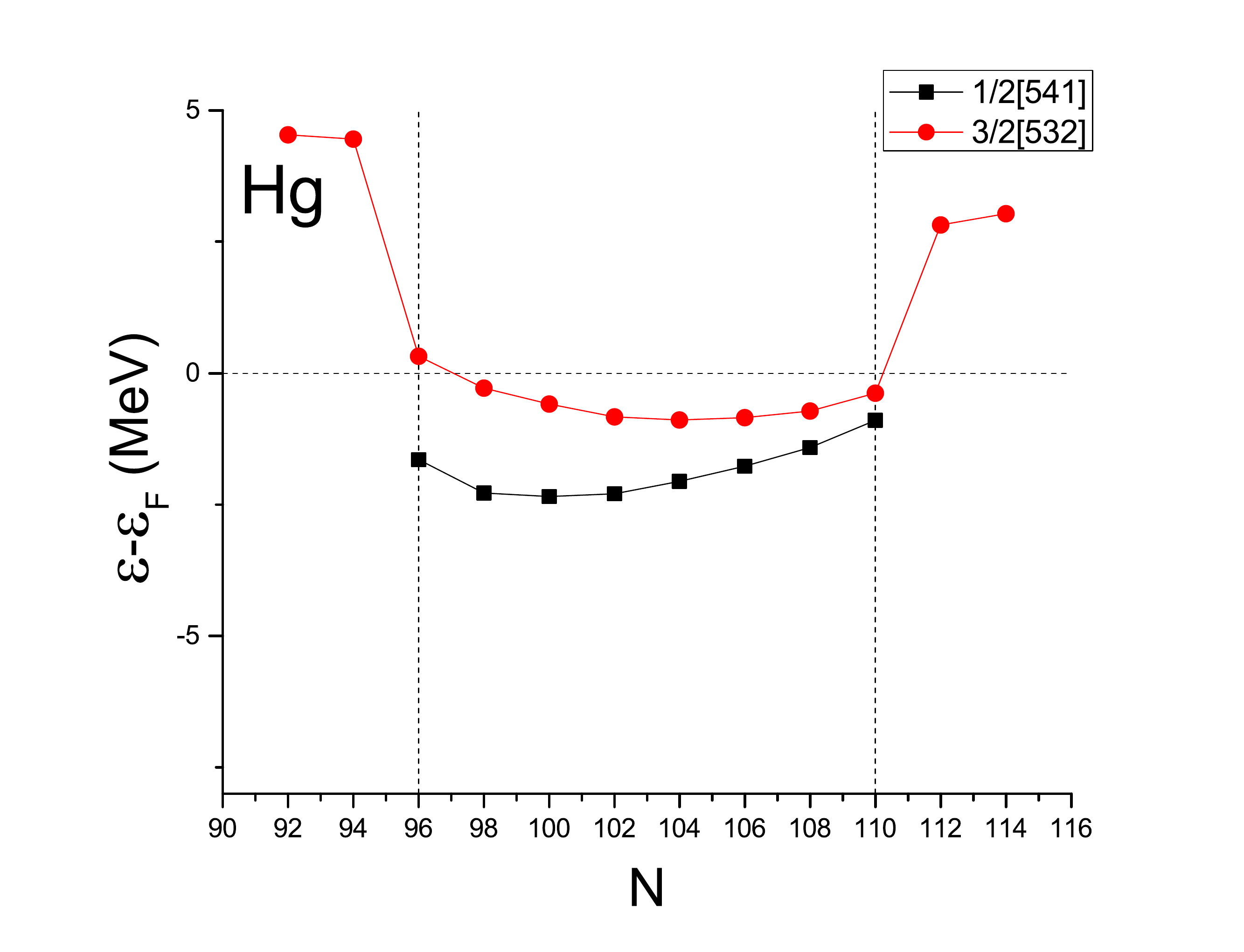}}\label{pexHg3}
\subfigure[]{\includegraphics[width=65mm]{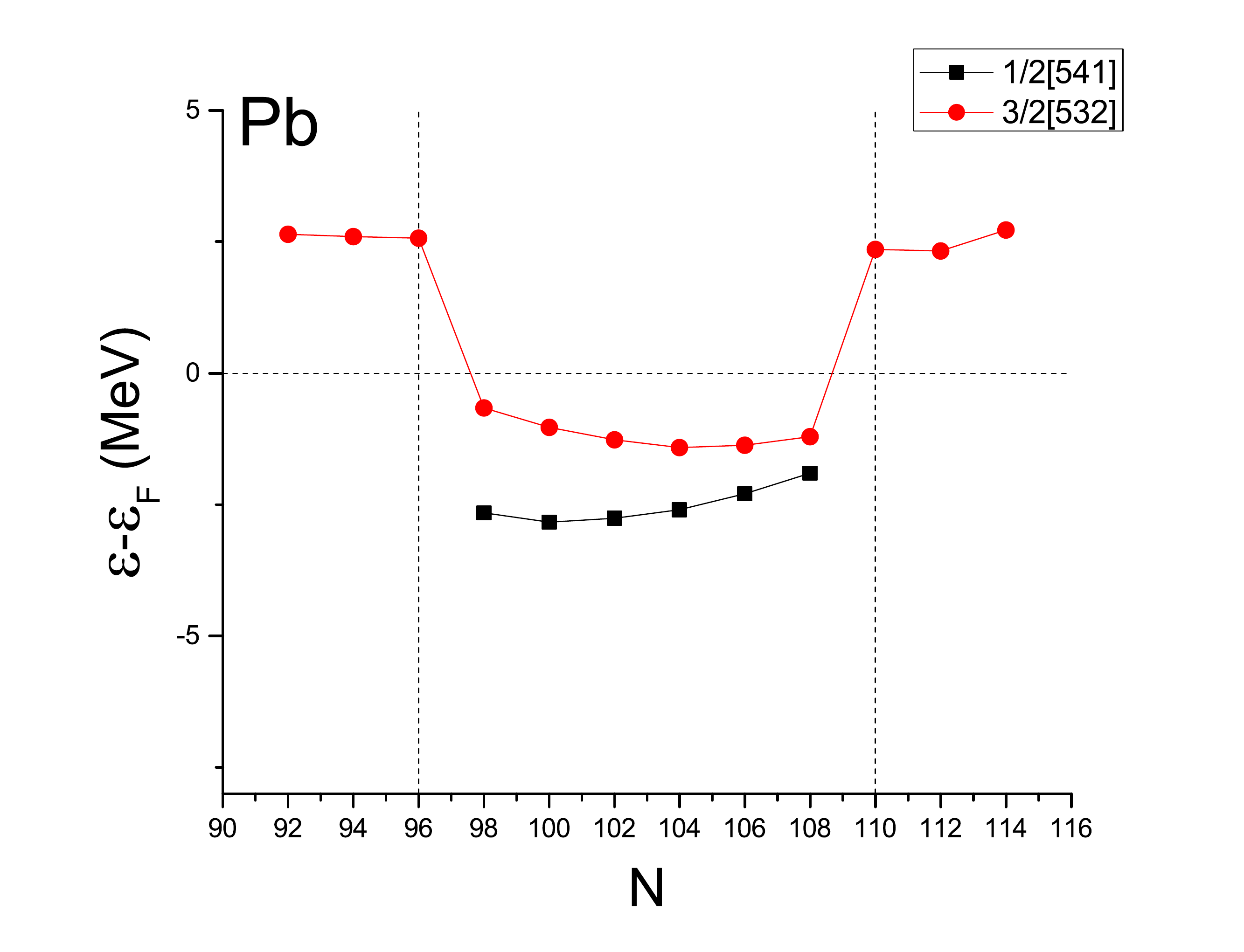}}\label{pexPb3}

\caption{Relativistic Mean Field calculations \cite{Lalazissis2005,Niksic2014} for the single particle energies $\epsilon$ for protons in the lead and mercury isotopes, reported relative to the proton Fermi energy $\epsilon_F$ in each isotope. We remark that the orbitals $1/2[541]$ and $3/2[532]$ of $2f^{7/2}$ of the 82-126 proton shell, which should be empty, are lying below the Fermi energy in the central region $98\leq N \leq 108$. On the contrary the $11/2[505]$, $1/2[400]$, $3/2[402]$, which were supposed to be filled, are empty. The large value of deformation may cause up to 4 proton excitations from the 50-82 SO--like shell to the 82-126 SO--like shell. See section \ref{mn} for further discussion.}\label{holes}

\end{figure*}

A more sensitive region of shape coexistence lies in the $\ce{Sr}, \ce{Zr}$ isotopes and in the lanthanides ($\ce{Nd}$, $\ce{Sm},...$, $\ce{Yb}$), where a sudden onset of deformation is observed \cite{Casten1985}. This sudden onset of deformation is attributed to the Federman--Pittel proton--neutron pairs \cite{Federman1977,Federman1979}. The onset of deformation due to the proton--neutron correlations leads to smaller proton energy gaps at: a) $Z=40$ in the $A\sim 100$ mass region, which includes the $\ce{Sr}, \ce{Zr}$ isotopes and at b) $Z=64$ in the $A\sim 150$ mass region, which corresponds to the $\ce{Nd}$, $\ce{Sm},...$, $\ce{Yb}$ isotopes \cite{Casten1985}. Consequently the protons of the $\ce{Sr}$, $\ce{Zr}$ may excite above the sub--shell closure at $Z=40$, while the protons of the lanthanides can excite above the proton sub--shell closure at $Z=64$  \cite{Heyde1985,Casten1985}. This is the standard particle--hole excitation mechanism for heavy nuclei. 

\begin{figure}
\begin{center}
\includegraphics[width=85mm]{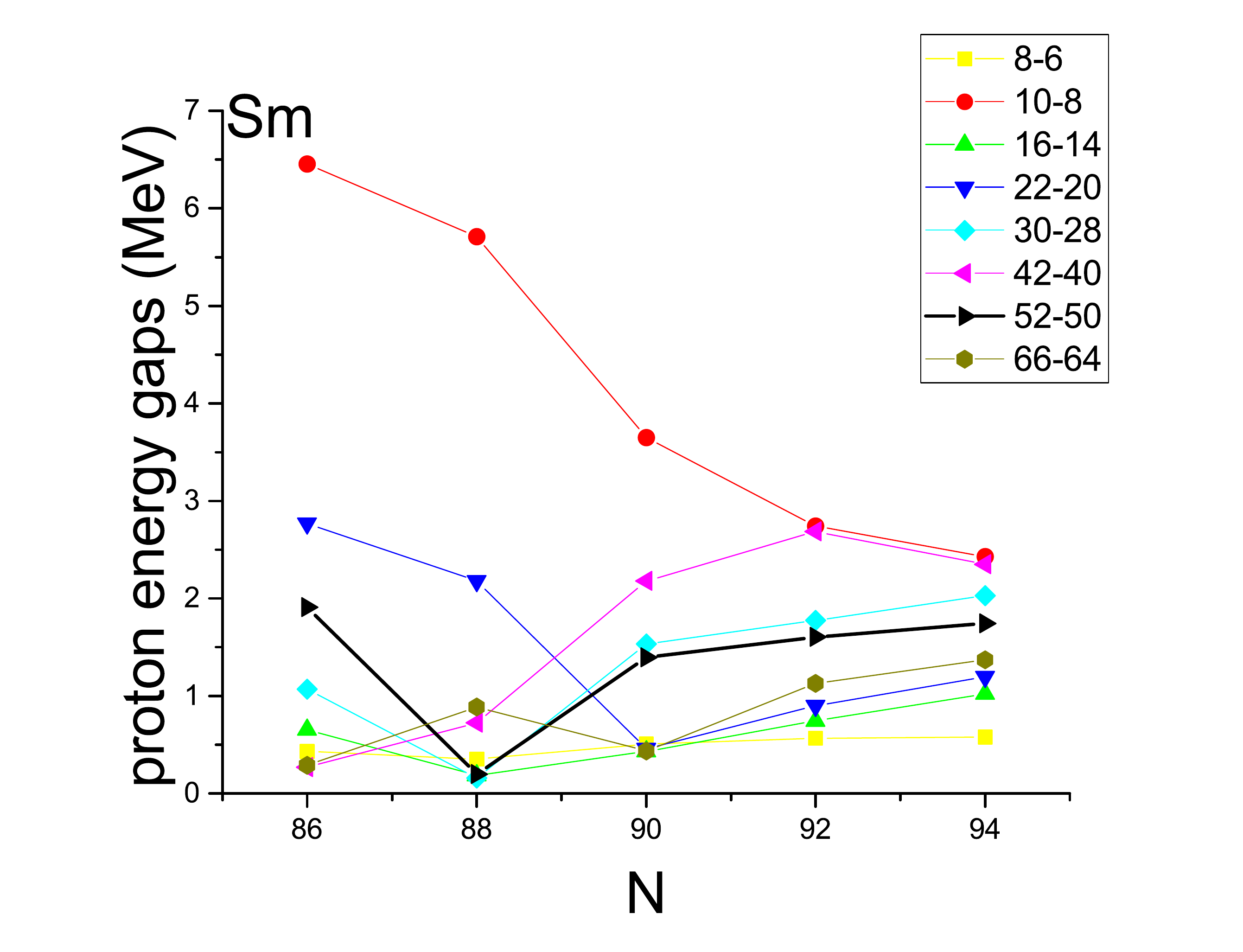}
\caption{The energy gaps for protons for the $\ce{Sm}$ isotopes as calculated by the RMF theory \cite{Lalazissis2005,Niksic2014}. For instance the legend ``52-50" indicates the energy gap among the $50^{th}$ and the $52^{nd}$ proton of the $\ce{Sm}$ isotopes. The large nuclear deformation, due to the Federman-Pittel proton--neutron pairs \cite{Federman1977,Federman1979}, is affecting {\it all} the energy gaps for protons, not only the gap at $Z=64$. Thus the one possibility is, that some of the protons of the $\ce{Sm}$ isotopes are able to excite above the sub-shell gap at $Z=64$ \cite{Ogawa1978}, as is explained in Ref. \cite{Casten1985,Heyde1985}. Another possibility is, that since the proton gap at among the $50^{th}$ and the $52^{nd}$ proton of the $\ce{Sm}$ isotopes is dissolved from the large deformation (especially around $N=88$), then the protons of the open 40-70 HO shell are active along with the protons of the 50-82 SO--like shell. }\label{gapsSm}
\end{center}
\end{figure}

\begin{figure}
\begin{center}
\includegraphics[width=85mm]{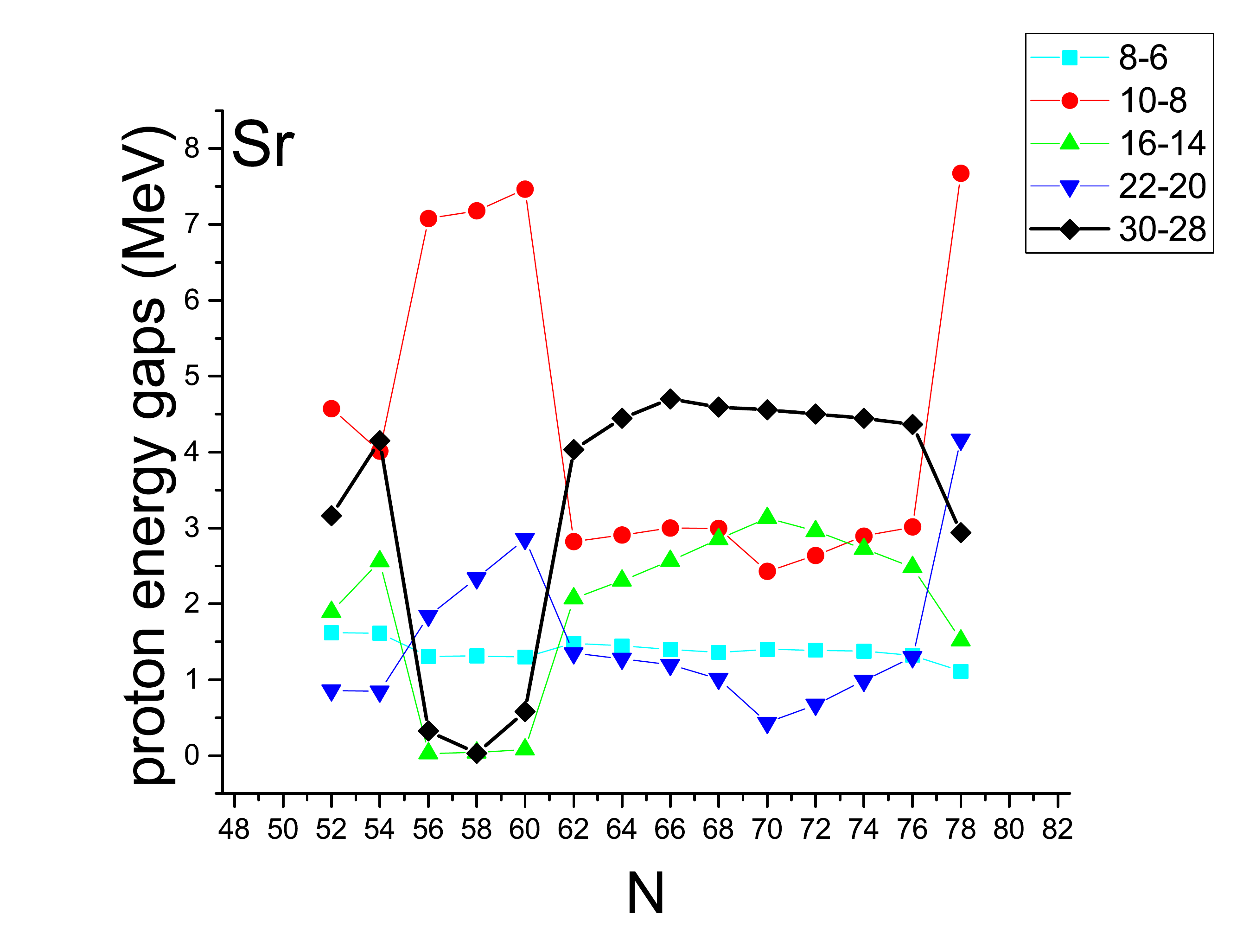}
\caption{The energy gaps for protons for the $\ce{Sr}$ isotopes as calculated by the RMF theory \cite{Lalazissis2005,Niksic2014}. For instance the legend ``28-30" indicates the energy gap among the $28^{th}$ and the $30^{th}$ proton of the $\ce{Sr}$ isotopes. The energy gap at the $28^{th}$ proton is mitigated before $N=60$. Thus the protons of the open 20-40 HO shell are active along with the protons of the 28-50 SO--like shell. }\label{gapsSr}
\end{center}
\end{figure}

But in the present dual--shell mechanism for shape coexistence we propose, that not only the proton gaps at $Z=40$ and $Z=64$ are reduced, due to the Federman--Pittel pairs, in the mass regions of $A\sim 100$ and $A\sim 150$ respectively, but {\it all} the proton gaps are affected by the onset of deformation. Calculations of the proton gaps with the RMF theory \cite{Lalazissis2005} are presented in Figs. \ref{gapsSm}, \ref{gapsSr}. In the $\ce{Sr}$ isotopes especially for $N=58$, as shown in Fig. \ref{gapsSr}, the proton gap among the $28^{th}$ and the $30^{th}$ protons is not major, thus a second possibility is now visible: that the protons of the 28-50 SO--like shell can interact with the protons of the 20-40 HO shell. It is a matter of fact, that for the $\ce{^{96}_{38}Sr_{58}}$ there is the largest known $E0$ transition for $A>56$ (see Fig. 27 of Ref. \cite{Heyde2011}). Similarly in the $\ce{Sm}$ isotopes, especially around $\ce{^{150}_{60}Sm_{88}}$, the energy gap among the $50^{th}$ and the $52^{nd}$ protons is mitigated by the onset of deformation, as seen in Fig. \ref{gapsSm}. Thus the active protons of the $\ce{Sm}$ isotopes lie among the SO--like shell 50-82 and the 40-70 HO shell. Similarly in the $N\approx 90$ lanthanides there are observed strong $E0$ transitions (Fig. 34 of \cite{Heyde2011} and Ref. \cite{Kibedi2005}).

To resume, the dual--shell mechanism about shape coexistence coincides with the particle--hole excitation mechanism in the light nuclei, as discussed in section \ref{excitations}, where the nucleons traditionally are in the HO shells., but it adds a new aspect in the standard particle--hole excitation mechanism \cite{Heyde1985} in the heavier nuclei, where the nucleons traditionally are in the SO--like shells. Specifically in the heavier nuclei the dual--shell mechanism supposes, that the protons (neutrons) of the SO--like valence shell 28-50, 50-82, 82-126, 126-184 merge with the protons (neutrons) of the still open HO shell  20-40, 40-70, 70-112, 12-168 and that this is possible only when the the proton (neutron) energy gaps at 28, 50, 82, 126 are dissolved by the large deformation.

In addition the effect of deformation on the single-particle energy gaps may cause the merging of the neutron HO shell with the relative neutron SO--like shell and this is what we call ``neutron induced" shape coexistence, but also may cause proton excitations in specific isotopes, as for example resulted from Figs. \ref{holes}. For the $Z\approx 40$ and $Z\approx 64$ regions we propose that the proton HO shell merges with the relative proton SO--like shell and this is what we call ``proton induced" shape coexistence. The shell merging is caused by the deformation, which mitigates the proton single-particle energy gaps, thus proton excitations are also possible.

It should be emphasized, however, that the present dual--shell mechanism is not implying that the particle--hole mechanism is not valid in heavy nuclei. It only implies that the particle--hole mechanism in heavy nuclei can take place only within the $N$ and $Z$ regions suggested by the dual--shell mechanism, given in the nuclear chart of Fig. \ref{map}.

\section{The dual--shell mechanism for shape coexistence}\label{mechanism}

With the understanding, that an important spin--orbit interaction is vital for the creation of the SO--like shells \cite{Bouten1967,Wilsdon}, we suggest, that the natural mechanism for shape coexistence involves the following steps:\\
a) The nucleus enters in a region with large $QQ$ proton or neutron interaction of the SO--like shell \cite{proxy2}, which usually occurs before the HO shell closure at proton or neutron number 8, 20, 40, 70, 112, 168 (see Eq. (\ref{QQ}) and Figs. \ref{isl1}-\ref{isl6}).\\
b) The large proton or neutron $QQ$ interaction of the SO shell along with the spin--orbit interaction dissolve the large single particle proton or neutron energy gaps as discussed in section \ref{mn}.\\
c) Therefore two types of open, valence proton or neutron shells are active for the proton or neutron numbers of Table \ref{phexc}, namely a HO shell and a SO--like shell. Shape coexistence can be the result of the coexistence of these two types of valence shells.

The above SU(3) mechanism for shape coexistence will be used in section \ref{paradigm} to explain the parity inversion in $\ce{Be}$ isotopes \cite{Geithner1999, Kondo2010}, the inversion of states in the $\ce{Mg}$ isotopes \cite{Himpe2008}, the reappearance of the magic number at $N=40$ in the $\ce{Ni}$ isotopes \cite{Nowacki2016}, the shape coexistence in the $\ce{Sn}$ \cite{Wood1992} and $\ce{Hg}$ \cite{Heyde2011} isotopes and the fission isomers at the $\ce{Pu}$ isotopes \cite{Thirolf2002}. Consequently the dual--shell mechanism is valid across all mass regions.

\section{The two low--lying nuclear bands}\label{two}

Within the dual--shell mechanism two low--lying energy bands are being predicted for each nucleus with shape coexistence. We will examine the simplified scenario, that one band is derived by the pure highest weight SU(3) irrep of the SO--like shell $(\lambda,\mu)_{SO}$ and the other one is derived by the pure highest weight SU(3) irrep of the HO shell $(\lambda,\mu)_{HO}$. The question is: which one is the ground state band and which one is the excited? As we shall see below, the answer lies in the symmetry of each nuclear wave function, which corresponds to each irrep.

In Ref. \cite{proxy5} it is exhibited, that a $(\lambda,\mu)$ irrep corresponds to a nuclear wave function, which has $\lambda+\mu$ symmetric quanta and $\mu$ quanta which are neither symmetric, nor antisymmetric. Thus the quantity $\lambda+\mu$ is a measure of symmetry. Also in Refs. \cite{Elliott1,Wilsdon,proxy5,Isacker1995,Isacker1997} it is stated, that the most symmetric irrep lies lower in energy. 

\begin{figure}
\begin{center}
\includegraphics[width=85mm]{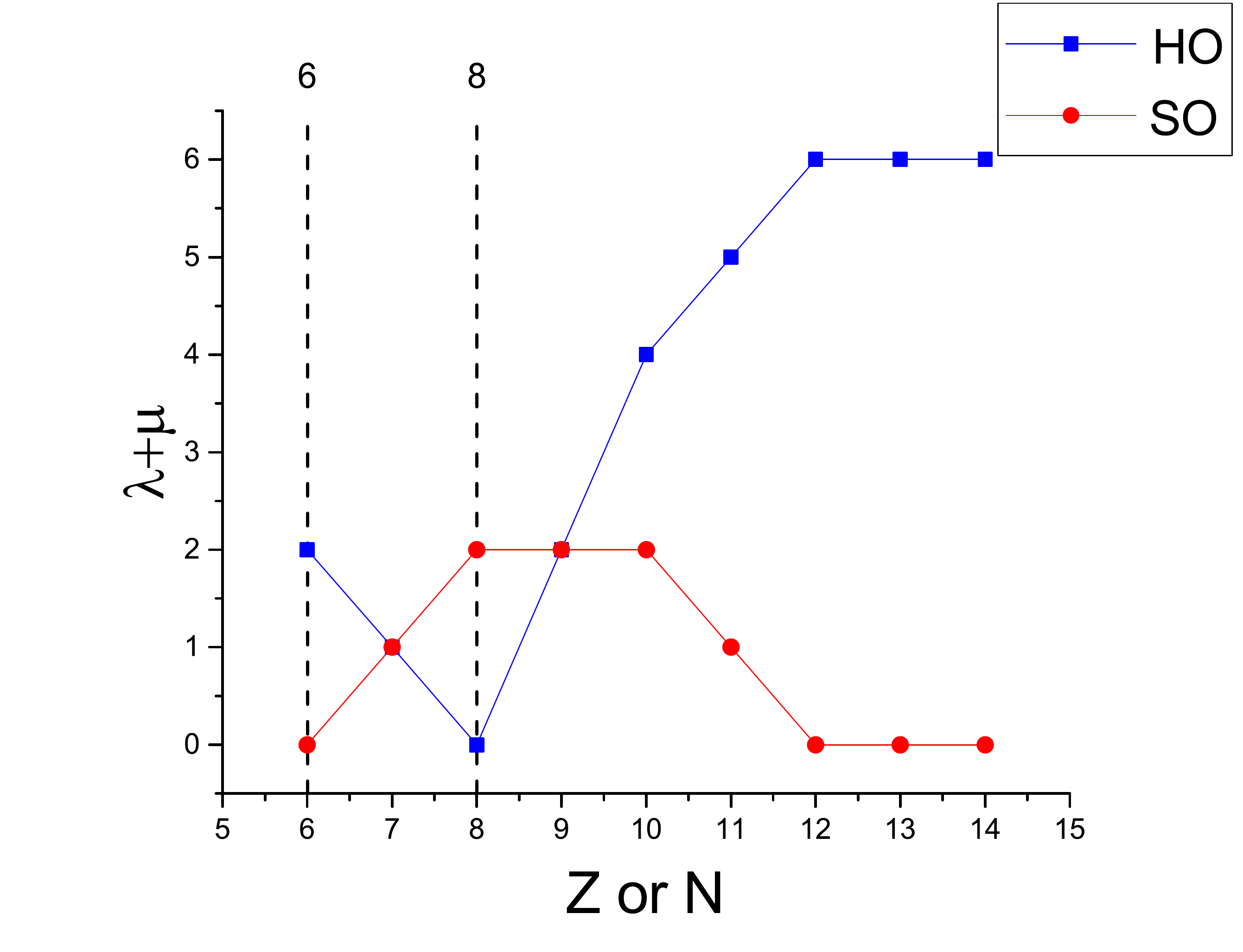}
\caption{The symmetric quanta $(\lambda+\mu)$ of the nuclear wave function for the two types of valence shells, namely the SO--like and the HO. Particle excitations from the 2-8 HO shell to the 6-14 SO--like shell are possible for proton or neutron numbers within the interval 6-8, where both types of shells are open (see Table \ref{phexc}). Just below 8 protons or neutrons the quadrupole deformation of the SO--like shell along with the spin--orbit interaction may decrease the major magic numbers, as discussed in section \ref{mn}. Just below 8 nucleons, the most symmetric irrep is the one of the SO--like shell: $(\lambda+\mu)_{SO}\ge(\lambda+\mu)_{HO}$. Thus the $(\lambda,\mu)_{SO}$ has to correspond to the ground state band and the less symmetric irrep $(\lambda,\mu)_{HO}$ to the excited band. }\label{lm1}
\end{center}
\end{figure}

\begin{figure}
\begin{center}
\includegraphics[width=85mm]{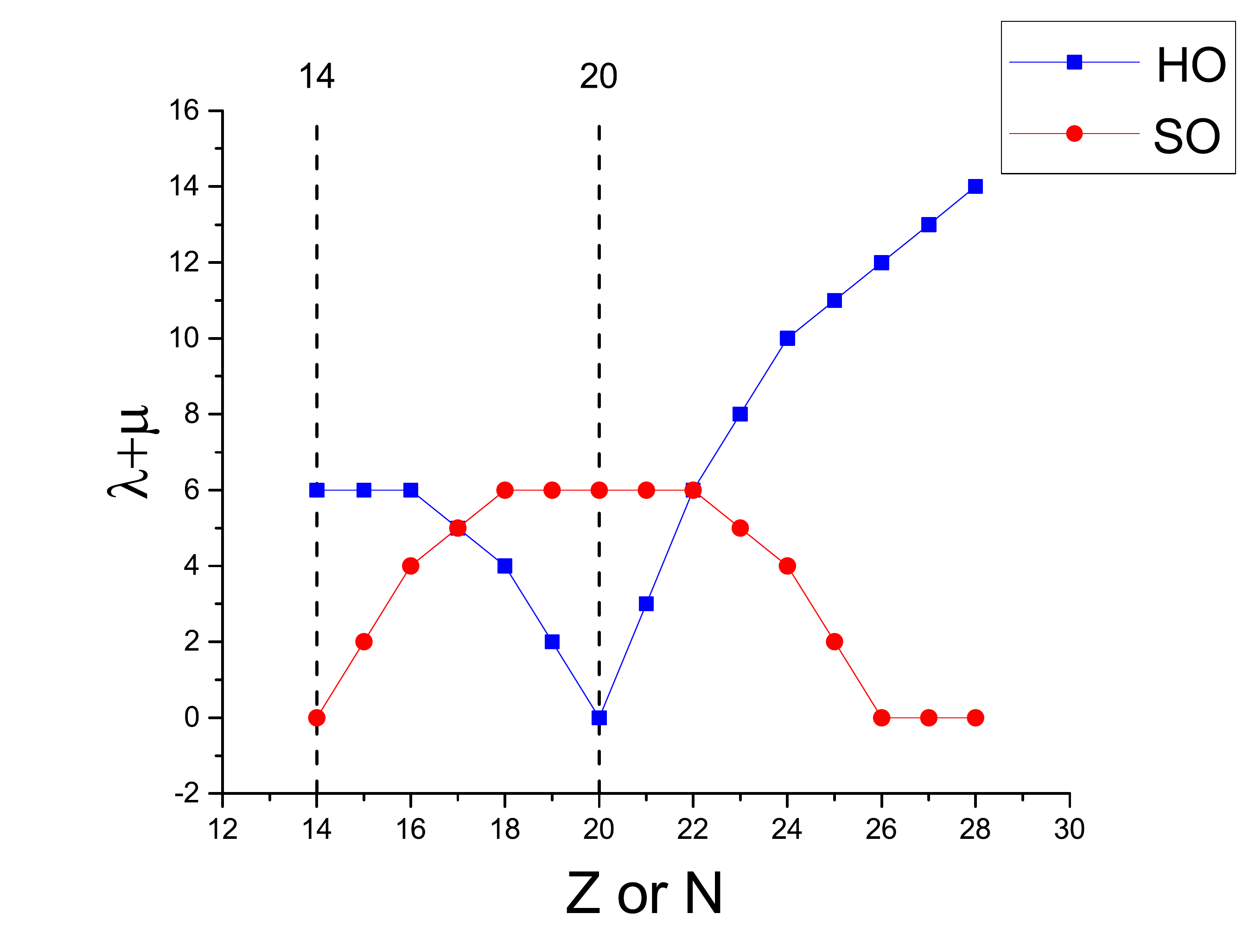}
\caption{The same as Fig. \ref{lm1} but for particle numbers among 14 and 28. Particle excitations from the 8-20 HO shell to the 14-28 SO--like shell are possible for proton or neutron numbers among 14-20, where both types of shells are open (see Table \ref{phexc}). Calculations of the spin--orbit interaction in the $sd$ shell \cite{Wilsdon,Bouten1967} have shown that in the second half of the 8-20 HO shell the strength of the spin-orbit force has grown very much. Thus the 14-28 SO--like shell emerges. Just below 20 protons or neutrons, the quadrupole deformation of the SO--like shell along with the spin--orbit interaction reduce the single--particle energy gaps, as discussed in section \ref{mn}. Just below 20 nucleons the most symmetric irrep is the one of the SO--like shell: $(\lambda+\mu)_{SO}\ge(\lambda+\mu)_{HO}$. Thus the $(\lambda,\mu)_{SO}$ has to derive the ground state band and the less symmetric irrep $(\lambda,\mu)_{HO}$ the excited band \cite{proxy5}.}\label{lm2}
\end{center}
\end{figure}

\begin{figure}
\begin{center}
\includegraphics[width=85mm]{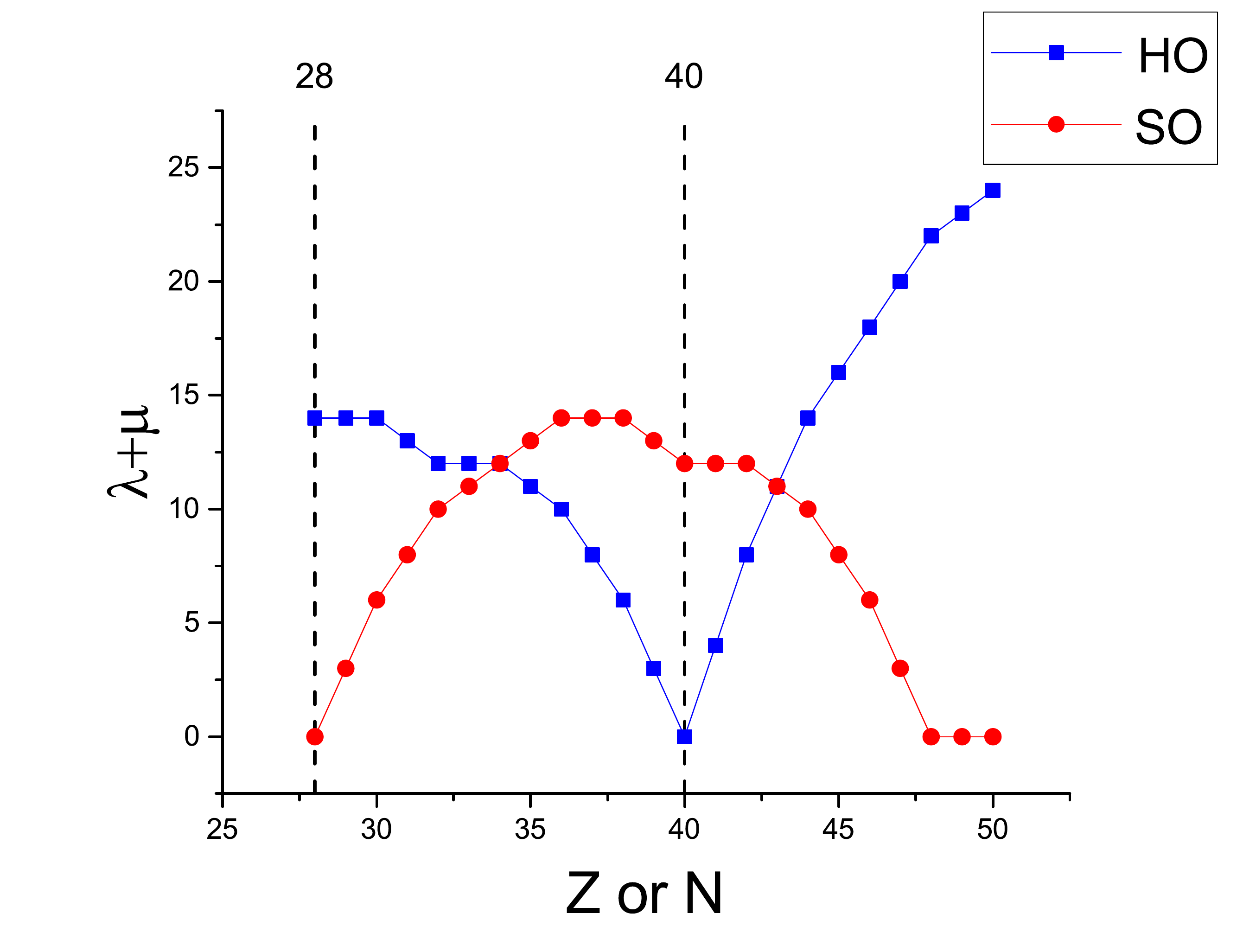}
\caption{The same as Fig. \ref{lm1}, but for nucleon numbers among 28 and 50. }\label{lm3}
\end{center}
\end{figure}

\begin{figure}
\begin{center}
\includegraphics[width=85mm]{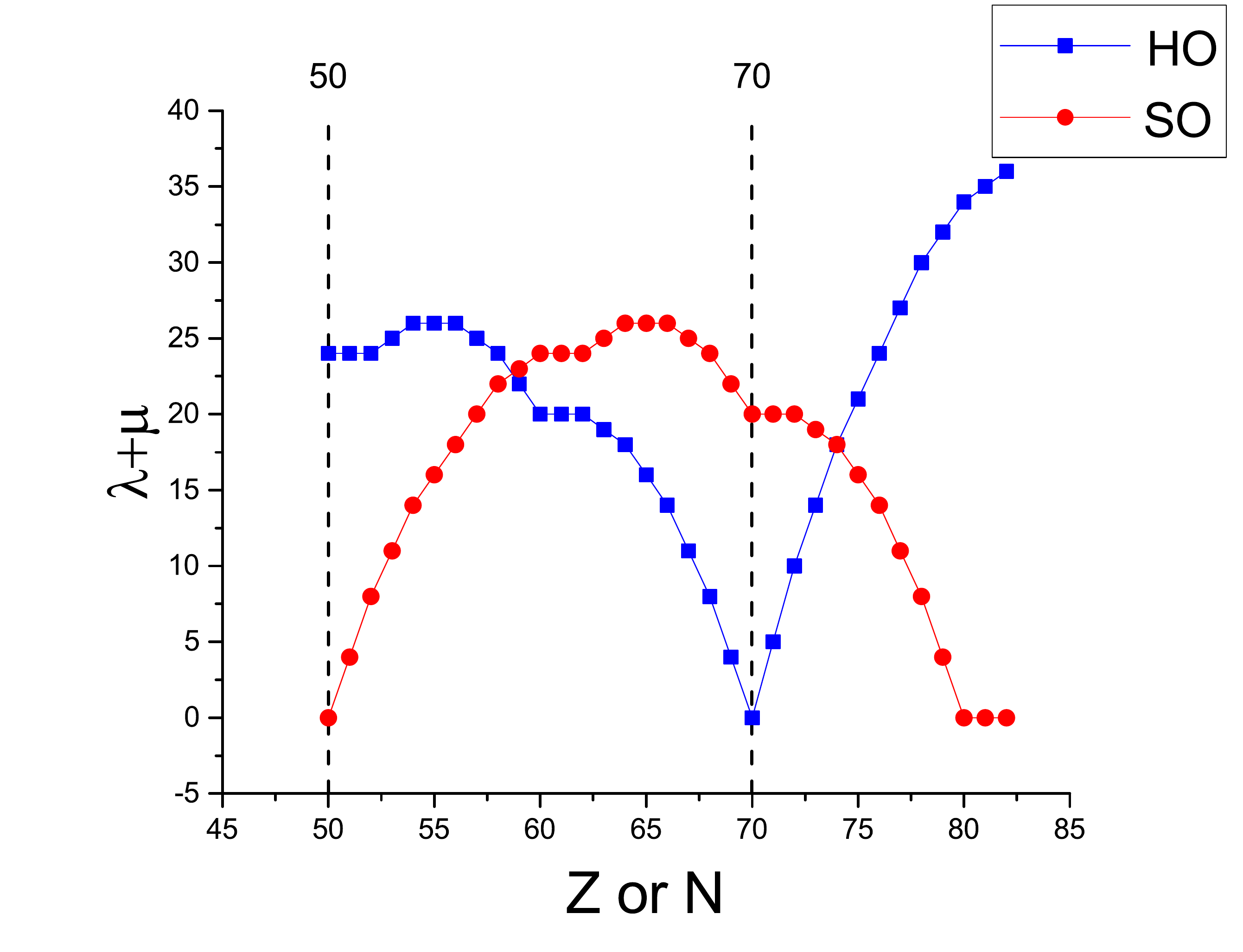}
\caption{The same as Fig. \ref{lm1}, but for nucleon numbers among 50 and 82. }\label{lm4}
\end{center}
\end{figure}

\begin{figure}
\begin{center}
\includegraphics[width=85mm]{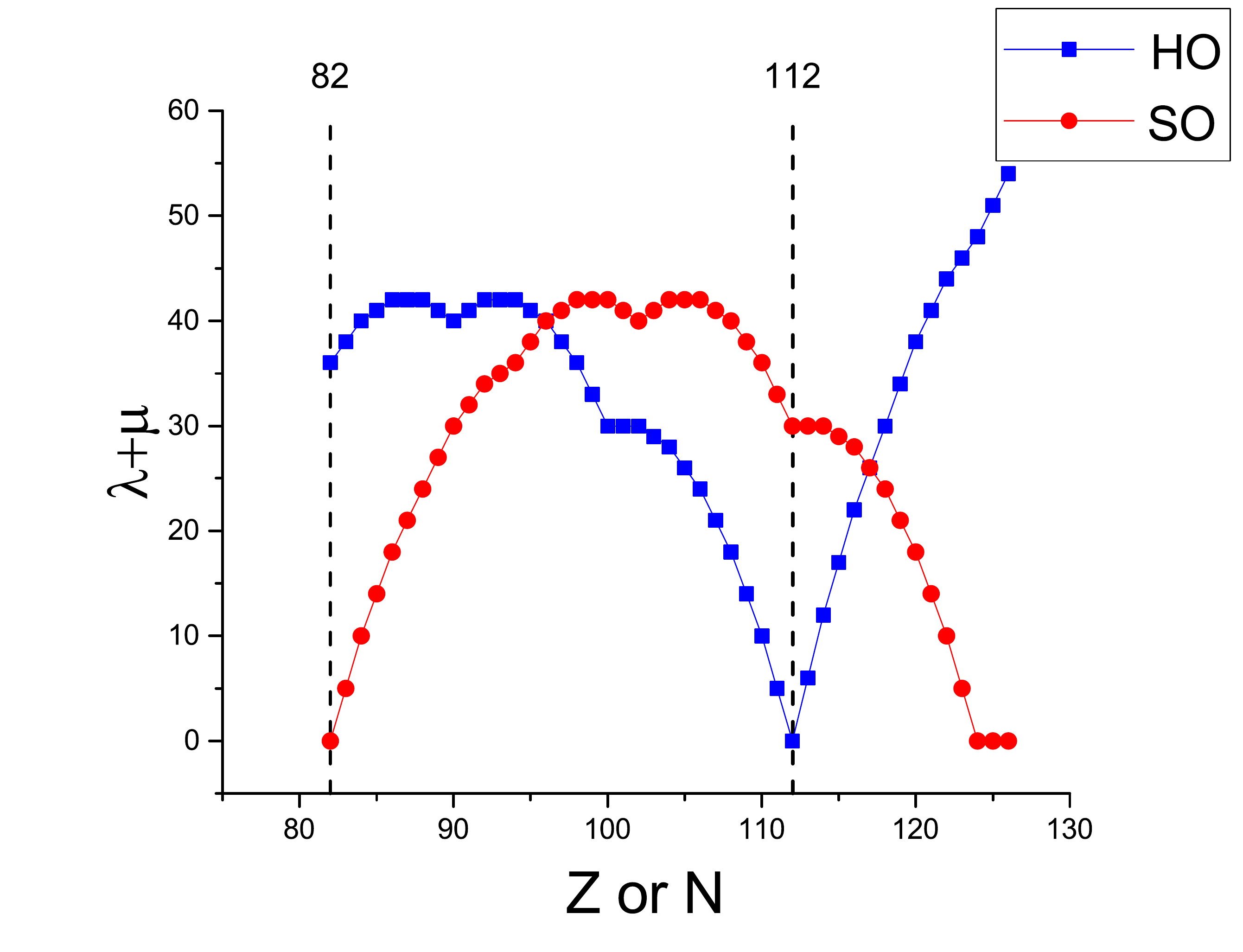}
\caption{The same as Fig. \ref{lm1}, but for nucleon numbers among 82 and 126.}\label{lm5}
\end{center}
\end{figure}

\begin{figure}
\begin{center}
\includegraphics[width=85mm]{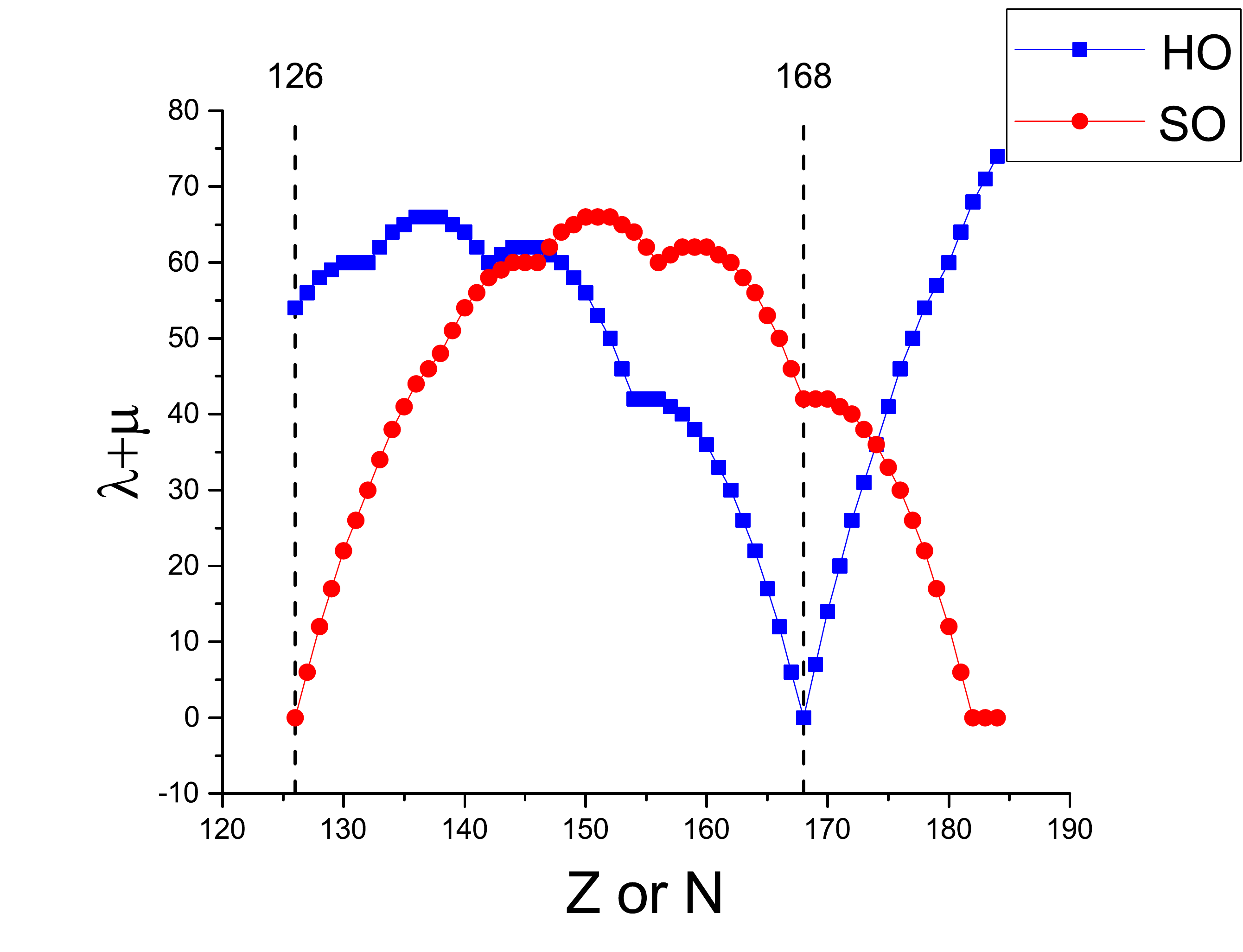}
\caption{The same as Fig. \ref{lm1}, but for nucleon numbers among 126 and 182. }\label{lm6}
\end{center}
\end{figure}

Shape coexistence has long been related with particle--hole excitations (see section \ref{excitations} and Ref. \cite{Heyde2011}). Table \ref{phexc} present the nucleon numbers, for which the SO--like shells have excited single--particle energies in comparison with those of the HO shells. We are further interested in the nucleon numbers, for which the $QQ$ interaction of the SO--like shell is intense. This interaction is necessary for the dissolution of the major magic numbers as discussed in section \ref{mn}. Thus we shall focus on the nucleon numbers before the HO shell closures 8, 20, 40, 70, 112, 168  (see Eq. (\ref{QQ}) and Figs. \ref{isl1}-\ref{isl6}). For these nucleon numbers we shall test, which irrep (the $(\lambda,\mu)_{SO}$ or the $(\lambda,\mu)_{HO}$) has more symmetric components $\lambda+\mu$. The results are presented in Figs. \ref{lm1}-\ref{lm6}.

From the Figs. \ref{lm1}-\ref{lm6} becomes evident, that before the HO shell closure the most symmetric pure SU(3) irrep is the $(\lambda,\mu)_{SO}$ and thus this irrep has to derive the ground state band. The less symmetric irrep is the $(\lambda,\mu)_{HO}$ and thus it has to derive the excited band of a nucleus with shape coexistence. This suffices to justify the inversion of states around $N=8$ and $N=20$, which is manifested in the $\ce{Be}$ and the $\ce{Mg}$ isotopes respectively. In these isotopes the neutrons normally are in a HO shell, but as soon as they enter into a region with shape coexistence, the ground state band is being derived by the SO--like shell and so inversion of states occurs. This subject will be further discussed in sections \ref{islands} and \ref{paradigm}.

\section{The islands of shape coexistence}\label{islands}

Shape coexistence is a phenomenon, which has been manifested in certain nuclei. A qualitative nuclear map with the islands of shape coexistence has been presented in Fig. 8 of Ref. \cite{Heyde2011}. This mechanism indicates, that shape coexistence may occur in certain islands on the nuclear map across all the mass regions. The key observation is, that the two types of valence shells start to coexist, whenever the proton or neutron configuration may change its valence shell from the SO--like to the HO and vice versa. 

In section \ref{Elliott} we have argued, that a strong spin--orbit interaction is necessary for the existence of the SO--like shells. Afterwards in section \ref{phexc} we displayed the nucleon numbers for which the SO--like shell has excited single--particle energies in comparison with the HO shell. In section \ref{mn} we argued, that a large $QQ$ interaction of the SO--like shell is necessary for the dissolution of the major magic numbers. All these phenomena together appear at nucleon numbers below the HO shell closure, which means below 8, 20, 40, 70, 112, 168 protons or neutrons. In these regions the most symmetric irrep, which lies lowest in energy, is the $(\lambda,\mu)_{SO}$, as discussed in section \ref{two}. The question now is, {\it how can it be possible, that the SO--like shell has excited single--particle energies and at the same time lies lower in energy?} The answer lies in the study of the $QQ$ interaction and will lead us to the condition, according to which, one can identify the islands of shape coexistence on the nuclear map. 

With the use of the Hamiltonian of Eq. (\ref{H0QQ}) the energy difference among the band--heads, that are being derived by the two valence shells, is the eigenvalue of:
\begin{eqnarray}\label{02}
H_{HO}-H_{SO}=(H_{0,HO}-H_{0,SO})\nonumber\\
+{\kappa \over 2}(QQ_{SO}-QQ_{HO}),
\end{eqnarray}
where $H_{0,SO}$ is given by Eq. (\ref{H0SO}) and $H_{0,HO}$ by Eq. (\ref{H0}).
The term
\begin{equation}\label{DN0H}
\Delta H_0=H_{0,HO}-H_{0,SO}
\end{equation}
is the difference between the single--particle terms, while the
\begin{equation}\label{DQQ}
\Delta (QQ)=QQ_{SO}-QQ_{HO}
\end{equation}
is the difference between the quadrupole--quadrupole interactions. If $N_{0,SO}$ is the eigenvalue of $H_{0,SO}$ of Eq. (\ref{H0SO}) and $N_{0,HO}$ of Eq. (\ref{N0}) is the eigenvalue of $H_{0,HO}$, then the eigenvalue of $\Delta H_0$ has a negative sign:
\begin{equation}\label{DN0}
N_{0,HO}-N_{0,SO}\le 0, 
\end{equation}
since the open SO--like shell 6-14, 14-28, 28-50, 50-82, 82-126, 126-184 has excited single--particle energies, when compared with those of the open harmonic oscillator shell 2-8, 8-20, 20-40, 40-70, 70-112, 112-168 respectively as discussed in section \ref{excitations}. Furthermore as discussed in section \ref{two} the excited band has to correspond to a particle configuration in the HO shell, which means, that:
\begin{equation}\label{DE}
E_{HO}-E_{SO}\ge 0,
\end{equation}
where $E_{HO}, E_{SO}$ are the eigenvalues of the Hamiltonians $H_{HO}, H_{SO}$ of Eq. (\ref{02}) respectively. One can satisfy both the conditions (\ref{DN0}) and (\ref{DE}), when:
\begin{equation}
QQ_{SO}\ge QQ_{HO},\label{cond1}
\end{equation}
and 
\begin{equation}
QQ_{SO}-QQ_{HO}\ge N_{0,SO}-N_{0,HO}\ge 0.\label{cond2}
\end{equation}
Consequently according to the inequality of Eq. (\ref{cond1}) the starting point of shape coexistence is at:
\begin{equation}\label{begin}
QQ_{SO}\approx QQ_{HO}.
\end{equation}
Afterwards the condition (\ref{cond2}) may narrow the range of shape coexistence on the nuclear map. 

In the Elliott Model \cite{Elliott1,Elliott2} the nuclear deformation is obtained from the SU(3) quantum numbers \cite{code,Castanos1988}. Therefore, the shape is determined by the distribution of the nucleons in the single--particle orbitals \cite{proxy5,Book}. In the ground state, due to the short--range and attractive nature of the nucleon--nucleon force, the particles result in space symmetric compact packing, giving rise to the highest weight SU(3) representations \cite{proxy5,Isacker1995}. In the HO and Elliott schemes the algebraic quadrupole nucleon--nucleon interaction defines the sequence of the orbitals, while in the spin--orbit scheme, described here by the proxy-SU(3) symmetry, the strong spin--orbit interaction \cite{Bouten1967,Wilsdon} plays an important role too, as discussed in sections \ref{Elliott} and \ref{excitations}. Therefore, in general, one has two different nucleon distributions, two different sets of SU(3) quantum numbers and two different deformations. Here, by Eq. (\ref{begin}) we put forward a conjecture, which says, that the shape coexistence appears in the special case, when the quadrupole deformation (given by the expectation of the second order Casimir operator of SU(3)) is in coincidence for the HO and SO schemes.

If the protons or neutrons lie within a harmonic oscillator shell, the $QQ_{HO}$ interaction of this shell can be calculated within the Elliott SU(3) symmetry. Especially for the ground state of an even-even nucleus with $L=0$ Eqs. (\ref{QQ}), (\ref{C2}) give:
\begin{eqnarray}
QQ_{HO}=4[\lambda^2+\mu^2+\lambda\mu+3(\lambda+\mu)]_{HO}\nonumber\\
=4C_{2,HO}.\nonumber\\
\end{eqnarray} 
Similarly for a SO--like shell, using the proxy-SU(3) symmetry one gets:
\begin{eqnarray}
QQ_{SO}=4[\lambda^2+\mu^2+\lambda\mu+3(\lambda+\mu)]_{SO}\nonumber\\
=4C_{2,SO}.
\end{eqnarray}

\begin{figure}
\begin{center}
\includegraphics[width=85mm]{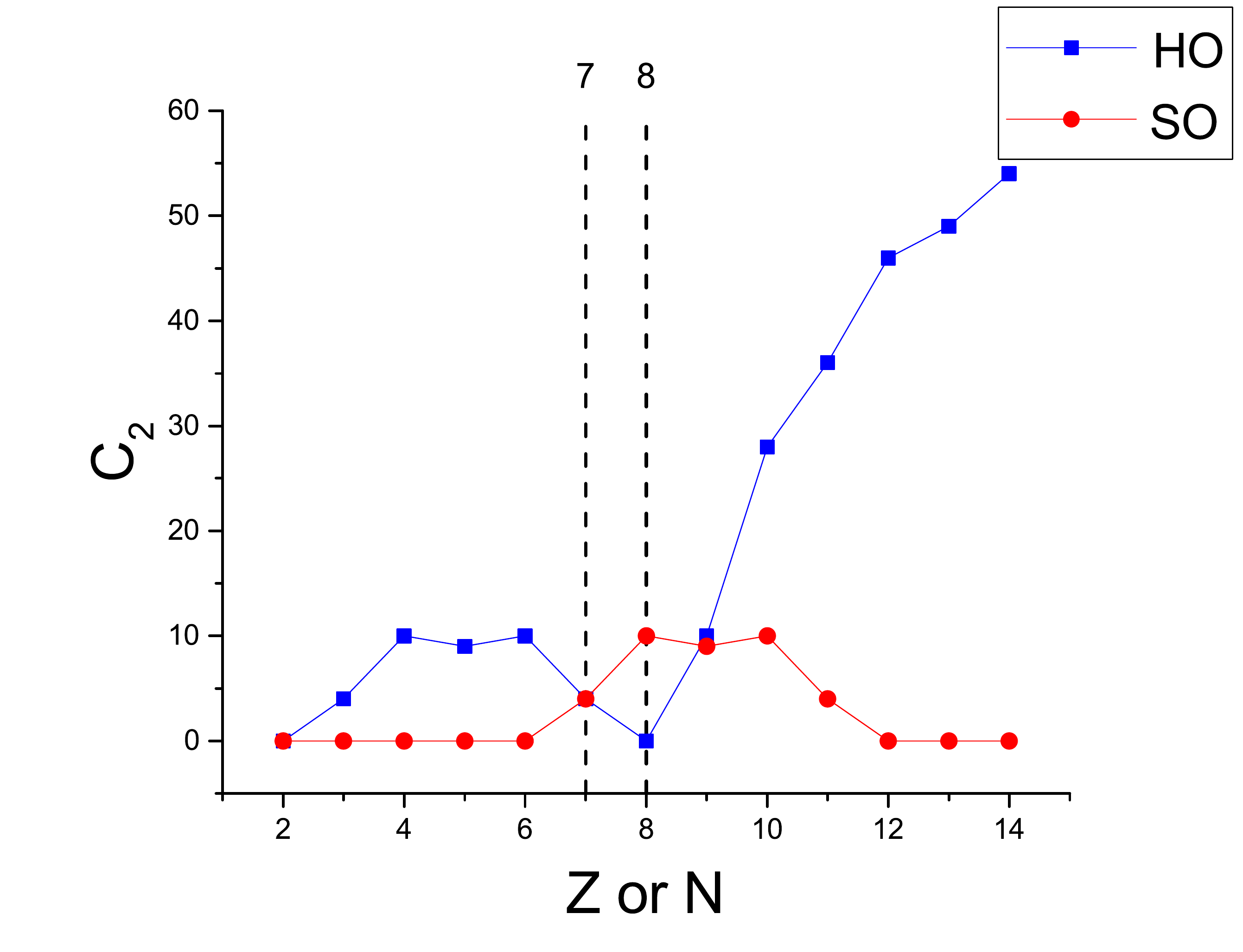}
\caption{ The eigenvalues of the second order Casimir operator of SU(3) versus the proton ($Z$) or neutron number ($N$). An island of shape coexistence is predicted within proton or neutron numbers $7-8$, where $C_{2,SO}\ge C_{2,HO}$. This island corresponds to the parity inversion in $\ce{^{11}Be}$ \cite{Geithner1999, Kondo2010}. See section \ref{paradigm} for further discussion. }\label{isl1}
\end{center}
\end{figure}

\begin{figure}
\begin{center}
\includegraphics[width=85mm]{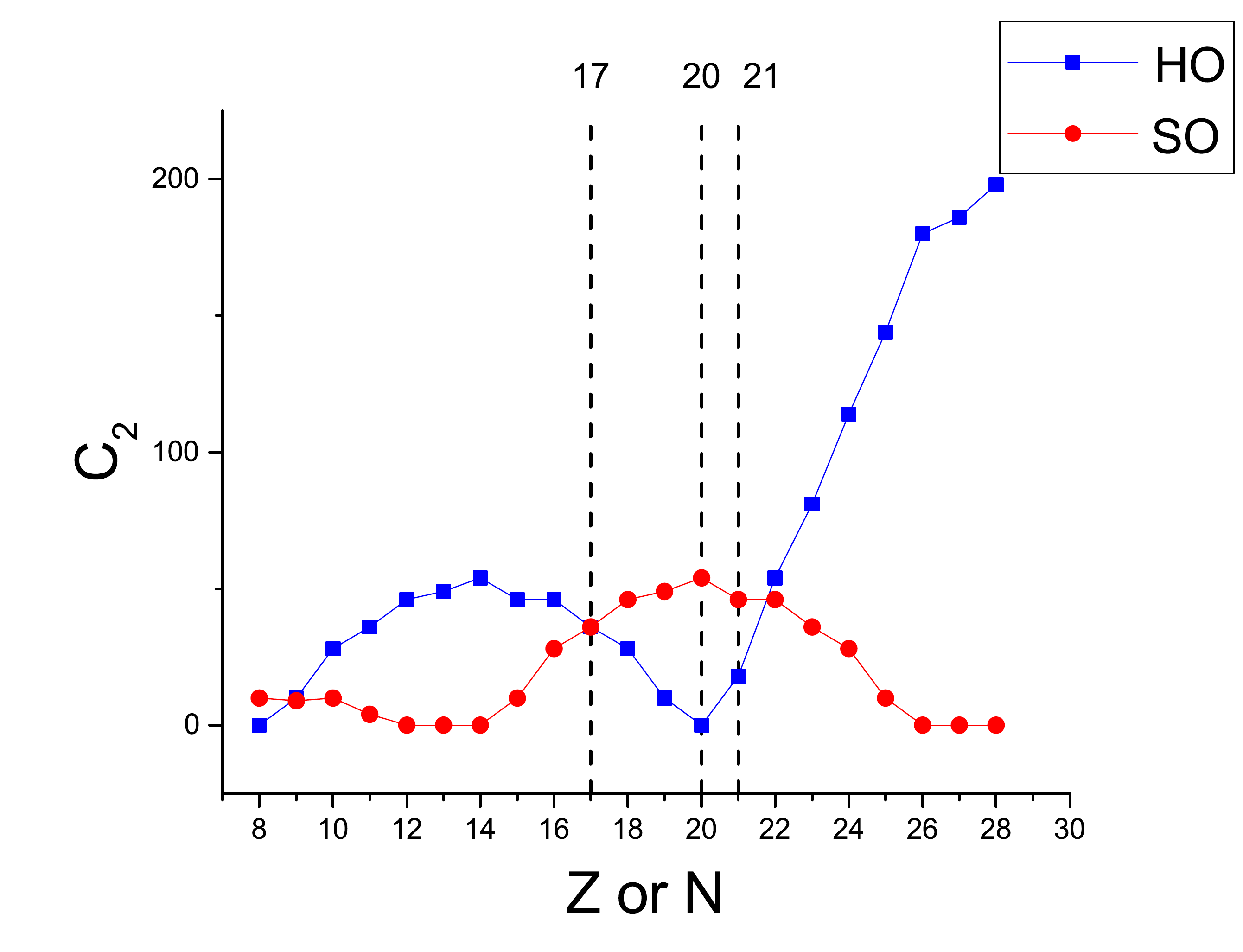}
\caption{The same as Fig. \ref{isl1}. An island of shape coexistence is predicted within proton or neutron numbers $17-20$. This island corresponds to the breaking of the magic number $N=20$ in the $\ce{Mg}$ isotopes \cite{Himpe2008}. See section \ref{paradigm} for further discussion. }\label{isl2}
\end{center}
\end{figure}

\begin{figure}
\begin{center}
\includegraphics[width=85mm]{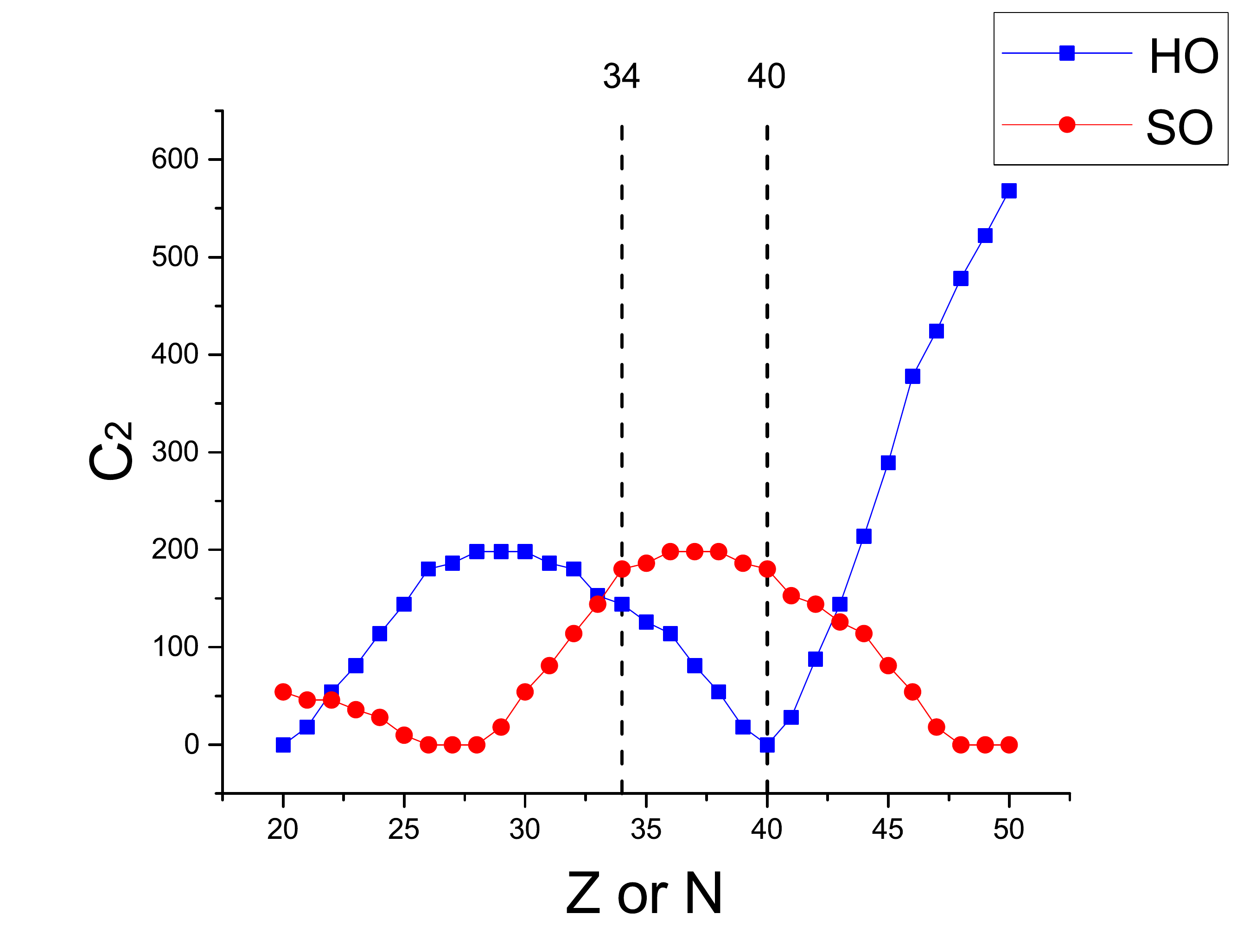}
\caption{The same as Fig. \ref{isl1}. An island of shape coexistence is predicted within proton or neutron numbers $34-40$. This island corresponds to the reappearance of the magic number $N=40$ at $\ce{Ni}$ isotopes \cite{Nowacki2016}. See section \ref{paradigm} for further discussion.}\label{isl3}
\end{center}
\end{figure}

\begin{figure}
\begin{center}
\includegraphics[width=85mm]{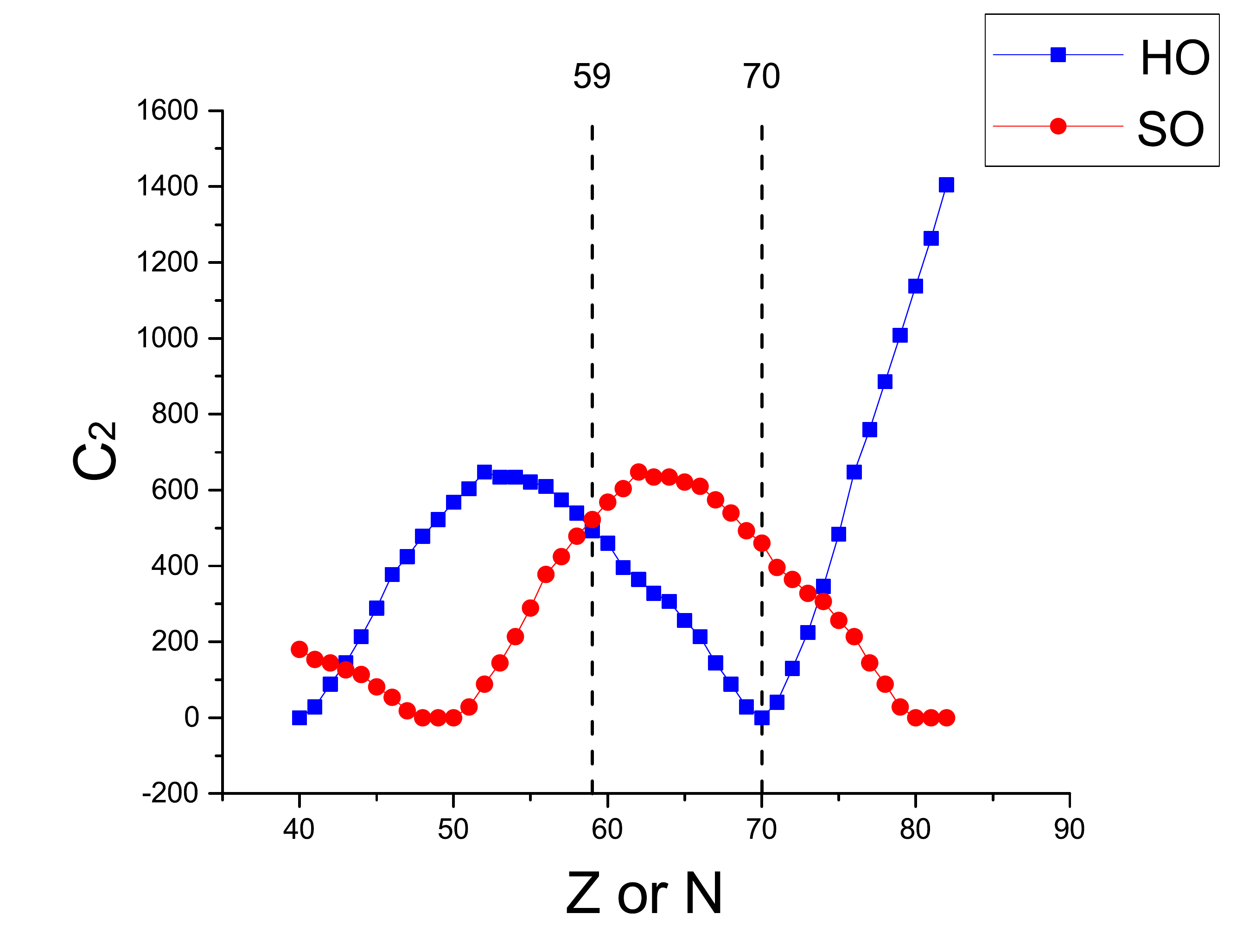}
\caption{The same as Fig. \ref{isl1}. An island of shape coexistence is predicted within proton or neutron numbers $59-70$. This island corresponds to the shape coexistence in the $\ce{Sn}$ isotopes \cite{Wood1992}. See section \ref{paradigm} for further discussion.}\label{isl4}
\end{center}
\end{figure}

\begin{figure}
\begin{center}
\includegraphics[width=85mm]{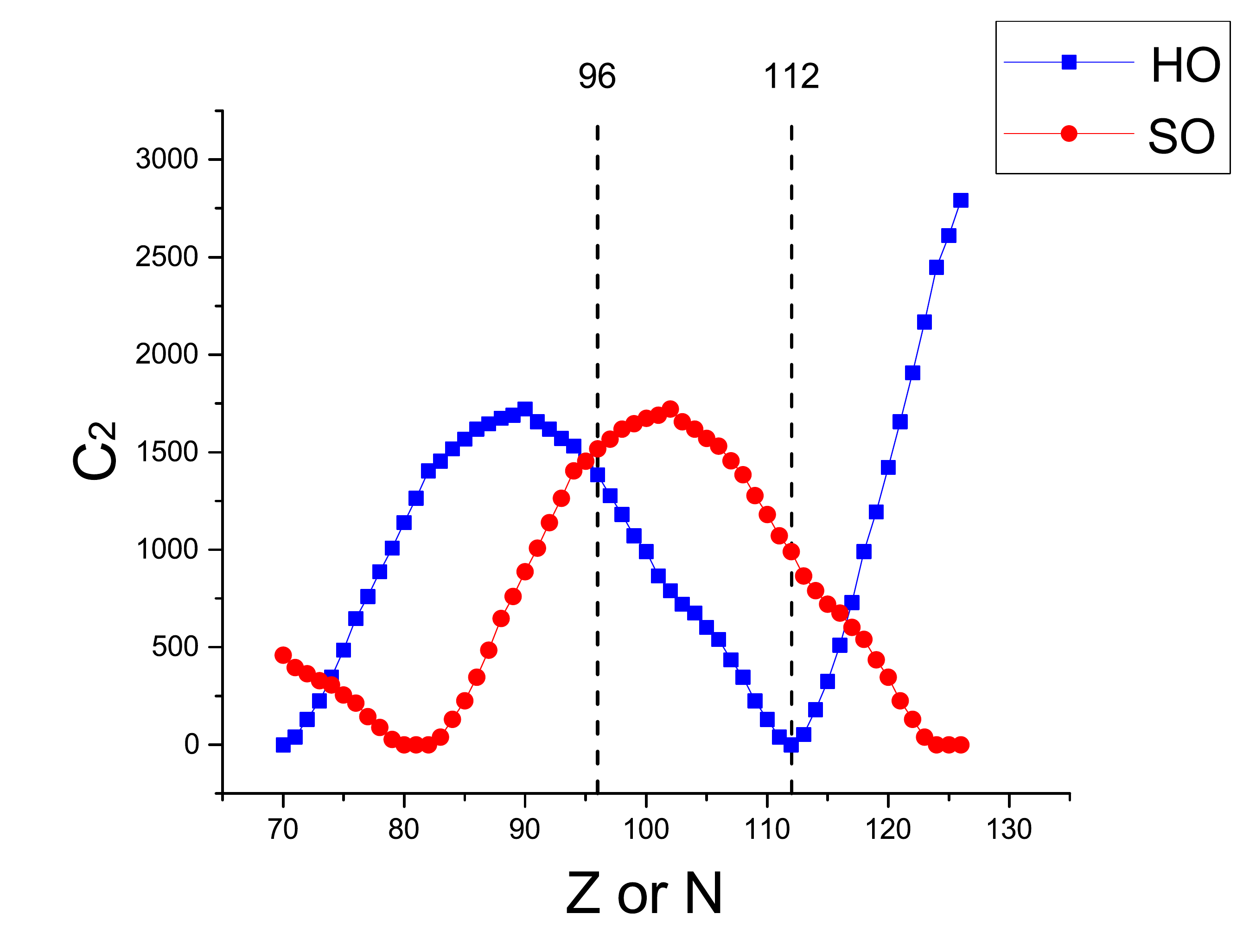}
\caption{The same as Fig. \ref{isl1}. An island of shape coexistence is predicted within proton or neutron numbers $96-112$. This island corresponds to the shape coexistence in the $\ce{Hg}$ isotopes \cite{Heyde2011}. See section \ref{paradigm} for further discussion.}\label{isl5}
\end{center}
\end{figure}

\begin{figure}
\begin{center}
\includegraphics[width=85mm]{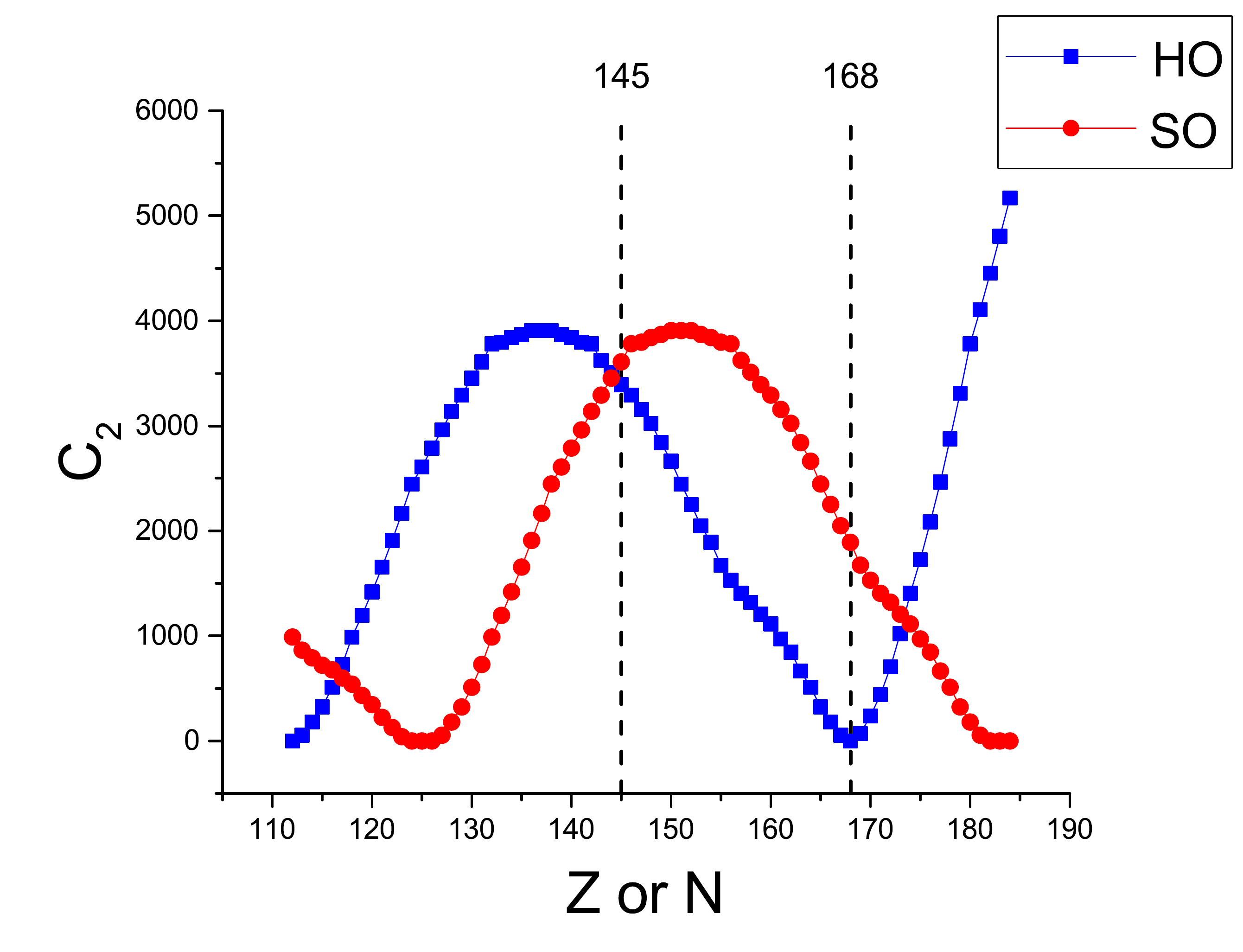}
\caption{The same as Fig. \ref{isl1}. An island of shape coexistence is predicted within proton or neutron numbers $145-168$. This island corresponds to the fission isomers in the $\ce{Pu}$ isotopes \cite{Thirolf2002}. See section \ref{paradigm} for further discussion.}\label{isl6}
\end{center}
\end{figure}

Figures \ref{isl1}-\ref{isl6} are plots of the eigenvalues of $C_2$ versus the proton or neutron number. In these figures it becomes obvious, that the condition of Eq. (\ref{begin}) is satisfied at proton or neutron numbers:
\begin{equation}\label{numbers1}
\mbox{\bf beginning }7, 17, 34, 59, 96, 145,
\end{equation}
which mark the beginning of shape coexistence across an isotopic or isotonic nuclear chain.
The ending of shape coexistence occurs at the harmonic oscillator shell closure
\begin{equation}
QQ_{HO}=0,
\end{equation}
which happens at proton or neutron numbers:
\begin{equation}\label{numbers2}
\mbox{\bf end } 8, 20, 40, 70, 112, 168.
\end{equation}
Figures about the $\beta,\gamma$ deformation variables using the two types of open shells are presented in Sec. 4.5 of Ref. \cite{MartinouThesis}.

Regions on the nuclear map (see Fig. \ref{map}) with proton or neutron numbers 7-8, 17-20, 34-40, 59-70, 96-112, 145-168 have a SO--like proton or neutron shell, which possesses a large value of $QQ_{SO}$ interaction, which in turn causes a fading out of the major single--particle energy gaps and thus two valence proton or neutron shells may coexist, namely the HO and the SO--like shell. Within these islands on the nuclear map it is valid the condition (\ref{cond1}).

Finally according to the dual--shell mechanism the islands of shape coexistence are expected to lie between proton or neutron numbers:
\begin{equation}\label{isl}
\mbox{\bf Z or N: } 7-8, 17-20, 34-40, 59-70, 96-112, 145-168.
\end{equation}
Using the above nucleon numbers one may draw a map with the islands of shape coexistence as in Fig. \ref{map}. Various nuclei in the colored areas of this map have experimentally observed shape coexistence \cite{Wood1992,Heyde2011}, while others do not have yet. In addition there is experimental evidence for shape coexistence in doubly magic nuclei \cite{Yang2016,Yang2016a}, which is not predicted by the map of Fig. {\ref{map}}. For the region of $\ce{Ge}$ \cite{Gottardo2016} the dual--shell mechanism aligns with the findings of \cite{Garcia2020}. Consequently the map of Fig. \ref{map} indicates, which nuclei have to be examined experimentally and theoretically for shape coexistence within the dual--shell mechanism. 

Returning to the question we addressed in the beginning of this section the answer is, that the particle configuration of the SO--like shell, although it has excited single--particle energies ($N_{0,SO}\ge N_{0,HO}$), lies lower in energy, because it also corresponds to a larger quadrupole-quadrupole interaction ($QQ_{SO}\ge QQ_{HO}$), which decreases the energy bearing a negative sign in the Hamiltonian of Eq. (\ref{H0QQ}). This fact etches the islands of shape coexistence on the nuclear map as in Fig. \ref{map} according to the nucleon numbers of Eq. (\ref{isl}). In section \ref{paradigm} we present, some emphatic cases of agreement between the predictions of the dual--shell mechanism for the islands of shape coexistence and the data.

The condition of Eq. (\ref{begin}) is satisfied beyond the numbers of Eq. (\ref{numbers1}) as seen in Figs. \ref{isl1}-\ref{isl6}, but in these nucleon numbers the $QQ_{SO}$ has a small value and thus major single--particle energy gaps separate the SO--like shell from the HO shell. Exceptions, in which the large deformation of the HO shell dominates and dissolves the major energy gaps, might appear \cite{Yang2016}. Such exceptions are warnings, that a detailed theoretical study has to be undertaken for each isotopic chain. 

This coexistence of the SO--like shell with the HO shell can explain, why inversion of states \cite{Otsuka2020} occurs around $N=8,20$. In these mass regions the neutrons traditionally occupy the HO shells 2-8 and 8-20. But as soon as the neutron number enters into an island of shape coexistence, as defined by Eq. (\ref{isl}), the neutrons flip from the HO shells 2-8, 8-20 to the SO--like shells 6-14, 14-28 respectively. The passing of the neutrons from the HO shell to the SO shells is accompanied by excitations in the single--particle energies. Therefore in light nuclei this SU(3) mechanism is similar with the particle-hole mechanism \cite{Heyde2011}. The term ``inversion" implies, that although the SO--like shell has excited single--particle energies, it lies lower in energy, due to the larger value of the $QQ$ interaction.

But in heavier nuclei, such as the $\ce{Sn}$ \cite{Wood1992} and $\ce{Hg}$ \cite{Heyde2011} isotopes are, the neutrons traditionally occupy the SO shells. In such cases the passing of the neutrons from the SO shells 50-82, 82-126 to the HO shells 40-70, 70-112 respectively cannot be interpreted as a particle excitation, since the SO shells have excited single--particle energies comparing with HO shells, as presented in section \ref{phexc}. 

In these mass regions a merging \cite{Vergados} of the proton (neutron) open SO--like shell with the open proton (neutron) HO shell can be treated through the outer product of the two SU(3) irreps:
\begin{equation}
(\lambda,\mu)_{SO}\otimes (\lambda,\mu)_{HO}
\end{equation}
For instance in the $\ce{Sn}$ isotopes a merging of the 50-82 shell with the 40-70 shell can be treated instead (see Fig. \ref{Sngapsn}). The idea of shell merging has already been applied in nuclei lighter than $\ce{Sn}$ in Ref. \cite{Nowacki2018}. The fact, that the irreps of a proton or neutron configuration in a SO--like and in a HO shell have to be coupled by their outer product \cite{Harvey,Alex2011,Coleman1964,Troltenier1996} is further reinforced by the experimentally observed monopole transition probabilities $B(E0)$ among the two bands of shape coexistence \cite{Heyde2011,Kibedi2005}. If the irreps of the two bands of shape coexistence were irrelevant, then the strong $B(E0)$s would not have been justified in the SU(3) symmetry. Furthermore in the Bohr-Hamiltonian treatment of shape coexistence of Refs. \cite{Budaca2018,Budaca2019,Buganu2018}, the finite barrier among the two minima of the potential guarantees, that $B(E0)s$ are allowed to occur among the coexisting shapes, while an infinite barrier in the double well potential would not allow tunneling among the two shapes.

The shell merging is accomplished within the SU(3) symmetry, by the outer product (coupling) of two irreps \cite{Alex2011,Harvey,Troltenier1996,Coleman1964}, which derives numerous irreps as a result. A physical criterion, based on the first principles of nuclear physics, has to be applied, in order to pick the right irrep, which describes the coexisting bands of a heavy nucleus with shape coexistence. If we consider, that the SO shell is the outer valence shell and the open HO shell is part of the inner core, then the irrep coupling represents the effect of the deformation  (see Eqs. (\ref{C2}), (\ref{beta})) induced by the inner core on the deformation induced by the valence shell. The coupling of a prolate irrep ($\lambda>\mu$) with an oblate irrep ($\lambda<\mu$) leads to a prolate--oblate shape coexistence. The coupling technique can be used for every nucleus with shape coexistence, in which the nucleons traditionally occupy the SO shells. The conclusions of section \ref{two} corresponded to the {\it pure} highest weight irreps for protons (neutrons) within the SO--like or the HO shell. It is yet unknown, which will be the coupled irreps and which one will be lowest in energy. This will be a subject of future research.

\section{The sudden onset of deformation}\label{onset}

It has long been understood, that the proton--neutron correlations enhance the deformation. The observation of an enhanced proton--neutron correlation was reported by de Shalit and Goldhaber in Ref. \cite{deShalit}, where the authors noticed, that when the last protons and the last neutrons occupy specific orbitals, then the $\beta$ transitions become slower. This stabilization of the nucleus was attributed to the proton--neutron interaction. By the way, many years later, the de Shalit--Goldhaber pairs lead to the unitary transformation, which is applied in the proxy-SU(3) symmetry \cite{proxy4}. Talmi in Ref. \cite{Talmi1962} stressed out, that the proton--neutron interaction is strong and attractive. Federman and Pittel in Refs. \cite{Federman1977,Federman1979} studied the transition from the spherical to the deformed shape, when the last protons and neutrons occupy specific orbitals, namely the Federman--Pittel pairs. The effect of the sudden onset of deformation on the phenomenon of shape coexistence has been pointed out by Heyde, Van Isacker, Casten and Wood in Ref. \cite{Heyde1985}. Systematics of nuclear observables, which are related with the onset of deformation, were presented in a simple way by Casten in Ref. \cite{Casten1985}, using the $N_pN_n$ scheme. So fruitful has been the study of the proton--neutron correlations, that triggered much work \cite{Dobaczewski1988,proxy1,Burcu2006,Burcu2010,Sofia2013,Duer2019} in the years to come. Recently the relation of shape transitions with shape coexistence has been investigated in Refs. \cite{Garcia2018,Garcia2019,garcianew}.

In this work we will discuss again the results of the plots of Ref. \cite{Casten1985}, having in mind the predictions of the present mechanism. We choose to concentrate on Ref. \cite{Casten1985}, because of the rich collection of experimental data presented in a simple way. The physical quantities used, are the ratio of the energy of the first $J=4^+$ state over the energy of the first $J=2^+$ state $E_{4_1^+}/E_{2_1^+}$,  as well as  $E_{2_1^+}$ itself. The ratio $E_{4_1^+}/E_{2_1^+}$ varies from $\sim 2$ for a vibrational nucleus, to $\sim 2.5$ for a $\gamma$-soft asymmetric rotor,  to $\sim 3.33$ for a deformed symmetric rotor \cite{Casten1985}. The energy $E_{2_1^+}$ varies from values near 1 MeV near closed shells to $\sim 100$ or $\sim 200$ keV for well-deformed rotational nuclei \cite{Casten1985}. In other words the higher is the ratio $E_{4_1^+}/E_{2_1^+}$, or the lower is $E_{2_1^+}$, the more deformed is the nucleus.

By considering Figs. 1 and 2 of Ref. \cite{Casten1985}, showing the $E_{4_1^+}/E_{2_1^+}$ratio versus the proton and neutron number respectively, we observe, that the sudden onset of deformation at $N=60$ is ``more sudden" for Sr and Zr ($Z=38, 40$), {\it i.e.}, when the proton number lies within the island of shape coexistence with $34\le Z\le 40$ (see present Fig. \ref{isl3}). Again, in Figs. 3 and 4 of Ref. \cite{Casten1985}, which present the energy  $E_{2_1^+}$ versus $N$ and $Z$ respectively, we observe a more violent onset of deformation for Sr and Zr ($Z=38,40$), which lie within an island of shape coexistence ($34\le Z\le 40$, as seen in present Fig. \ref{isl3}).

Similarly Figs. 12 and 13 of Ref. \cite{Casten1985}, showing the $E_{4_1^+}/E_{2_1^+}$ratio, indicate a more vivid onset of deformation at $N=90$ when the proton number ($Z=60,62,64,66,68,70$) lies within an island of shape coexistence ($59\le Z\le 70$, as seen in present Fig. \ref{isl4}). The same conclusion is drawn from Figs. 15 and 16 of Ref. \cite{Casten1985}, showing the energy  $E_{2_1^+}$. In addition,  Fig. 17 of Ref. \cite{Casten1985} indicates a sudden change in the trend of the energy $E_{2_2^+}$ after Ce ($Z=58$), {\it i.e.}, when the proton number enters the island of coexistence within $59\le Z \le 70$,  as seen in present Fig. \ref{isl4}.

On the contrary Figs. 6-11 of section 2.2 of Ref. \cite{Casten1985} indicate a smooth transition from the spherical to the deformed shape in the $A\simeq 130$ region. In the cases of the $\ce{Xe},\ce{Ba},\ce{Ce}$ isotopes ($Z=54,56,58)$ the proton number lies outside the islands of shape coexistence, which are predicted by the dual--shell mechanism (see expression (\ref{isl}) and present Fig. 25). 

In general we observe, that the onset of deformation is more vivid, when the proton number lies within an island of shape coexistence, which is predicted by the present mechanism. This more sudden transition from the spherical to the deformed shape could be related to the coexistence of three SU(3) irreps (two for protons $\varpi$ and one for neutrons $\nu$): 
\begin{equation}
(\lambda,\mu)_{\varpi,HO}\otimes (\lambda,\mu)_{\varpi,SO}\otimes (\lambda,\mu)_{\nu,SO},
\end{equation}
when the proton number lies within an island of shape coexistence, as was defined in Eq.  (\ref{isl}), or it could be related to four SU(3) irreps:
\begin{equation}
(\lambda,\mu)_{\varpi,HO}\otimes (\lambda,\mu)_{\varpi,SO}\otimes (\lambda,\mu)_{\nu,HO}\otimes  (\lambda,\mu)_{\nu,SO}
\end{equation}
when both the proton and neutron numbers lie within an island of shape coexistence. These possibilities call for further investigations. 

Another important experimental fingerprint of the proton induced shape coexistence is, that it is accompanied by large electrical monopole transitions $\rho^2(E0)$. As an example one may see Figs. 27, 31 and 34 of Ref. \cite{Heyde2011}. In general in the proton induced islands of shape coexistence, which are marked by the horizontal stripes on the map of Fig. 25, one has to look for $B(E0)s$, while in the neutron induced islands, which are marked by the vertical stripes on the map of Fig. 25, $B(E2)s$ among the coexisting bands are more common.

\section{Paradigmatic isotopes with shape coexistence}\label{paradigm}

In the next we will demonstrate some exceptional manifestations of shape coexistence, which are predicted by the dual--shell mechanism. The paradigms will begin from the $\ce{Be}$ and will end to the $\ce{Pu}$, a fact that highlights, that the dual--shell mechanism can be applied in all mass regions. The reported phenomena will be the parity inversion, the inversion of states, the come-back of the HO magic numbers, the shape coexistence and the fission isomers, all of them being just the various faces of shape coexistence in different mass regions.

\subsection{The $\ce{^{11}Be}$}\label{Be}

The first island of shape coexistence, which is presented in Fig. \ref{isl1}, appears in the light nuclei. A paradigm in this mass region is the $\ce{^{11}_4Be_{7}}$ halo nucleus, which is known for the phenomenon of parity inversion \cite{Esbensen1995}. The last unpaired neutron of $\ce{^{9}_4Be_{5}}$ lies in the $1p^{3/2}$ orbit and thus this isotope exhibits a ground state with negative parity \cite{Tilley2004}. Similarly the $\ce{^{11}_4Be_{7}}$ nucleus should possess a negative parity ground state, too. But magnetic moment measurements \cite{Geithner1999} have revealed that $\ce{^{11}_4Be_{7}}$ has a positive parity ground state. Furthermore, measurements of the nuclear charge radii in this isotopic chain \cite{Noertershaeuser2009} have established the $\ce{^{11}_4Be_{7}}$ nucleus as a halo nucleus. Spectroscopic factors led to the conclusion, that in $\ce{^{12}Be}$ the last neutron pair occupies partially ($2\over 3$) the $2s^{1/2},1d^{5/2}$ orbitals \cite{Navin2000}.

This parity inversion can be justified by the first island of shape coexistence of Fig. \ref{isl1}. The neutrons of the $\ce{Be}$ isotopes traditionally occupy the 2-8 harmonic oscillator shell, namely the $p$ shell. But as soon as the condition (\ref{begin}) is satisfied, right at $N=7$, the neutrons are excited to the SO--like shell 6-14, which consists of the $1p^{1/2}_{m_j},1d^{5/2}_{m_j}$ orbitals and thus a positive parity ground state becomes possible. The orbitals occupied by neutrons within the two sets of magic numbers have been presented in Table \ref{Beex} and the calculation of the single particle energies within the Elliott and the proxy-SU(3) symmetry has been analyzed in section \ref{excitations}.

At this point, it has to be clear, that the total angular momentum and the parity ($J^\pm$) of a state of an even-odd nucleus within the Elliott SU(3) symmetry, is {\it not} the $J^\pm$ of the last uncoupled neutron. The Elliott SU(3) model derives the $J$ quantum number with the rules of Ref. \cite{Elliott4}:
\begin{gather}
K_S=S, (S-1), ..., -S,\label{KS}\\
K=K_S+K_L\ge 0,\label{K}\\
K_L=\mu,\mu-2,...,-\mu,\label{KL}\\
J=K,K+1,...,\lambda+\mu+S,\label{J}
\end{gather}
with the exception that if $K_L=K_S=K=0$ then $J$ is even or odd with $J_{max}=\lambda+S$. In the above $S={1\over 2}$ is the nuclear spin of the odd mass nucleus, $K_S$ is the nuclear spin projection, $K_L$ is the projection of the orbital angular momentum of the nucleus and $K$ is the nuclear projection of the total angular momentum. As a result the $2s^{1/2}_{m_j}$ orbital is not necessary for the prediction of the $1/2^+$ ground state of  $\ce{^{11}_4Be_{7}}$ within the Elliott SU(3) model. Instead using a proton ($\varpi$) irrep $(\lambda_\varpi,\mu_\varpi)_{HO}=(2,0)$ and a neutron ($\nu$) irrep $(\lambda_\nu,\mu_\nu)_{SO}=(1,0)$ (see Table I), the total irrep of the  $\ce{^{11}_4Be_{7}}$ is $(\lambda,\mu)=(\lambda_\varpi+\lambda_\nu,\mu_\varpi+\mu_\nu)=(3,0)$ \cite{proxy2}, which gives \cite{Elliott4}:
\begin{gather}
S={1\over 2},\\
K_S={1\over 2},\\
K_L=0,\\
K=K_S+K_L={1\over 2},\\
J={1\over 2},{3\over 2},{5\over 2}.
\end{gather}
Consequently the $J$ of the ground state of this halo nucleus, with the one last neutron lying in the $1p^{1/2}_{m_j},1d^{5/2}_{m_j}$ shell, is $J={1\over 2}$ within the dual--shell mechanism. A positive parity becomes possible due to the participation in the SO--like shell of the $1d^{5/2}_{m_j}$ orbital.

\subsection{The $\ce{^{16}O}$}\label{O}

The case of $\ce{^{16}O}$ has already been discussed in section \ref{excitations}, where an irrep $(4,0)$ has been predicted from the SO--like shell 6-14 to derive the ground $0_1^+$ state, and an irrep (0,0) originated from the HO shell 2-8, matches with the excited $0_2^+$ state. In section \ref{islands} we had argued, that although the 6-14 shell is excited in comparison with the 2-8 shell, the first lies lower in energy, because it is characterized by larger $QQ$ interaction. In this section we will calculate the energy of the $0_2^+$ state for this nucleus using the on hand mechanism. The energy of the $0_2^+$ is the eigenvalue of Eq. (\ref{02}):
\begin{gather}\label{02O}
E_{0_2^+}=(N_{0,HO}-N_{0,SO})+{\kappa\over 2}(4C_{2,SO}-4C_{2,HO}),
\end{gather}
where Eq. (\ref{QQ}) has been used for $L=0$.

The $N_0$ of the HO shell is:
\begin{equation}
N_{0,HO}=N_0
\end{equation}
of Eq. (\ref{N0}), while the $N_{0,SO}$ is the eigenvalue of Eq. (\ref{H0SO}):
\begin{equation}\label{N0SO}
N_{0,SO}=N_0+\braket{\Psi|H_{0,proxy}|\Psi},
\end{equation}
where $\Psi$ is the $L$-projected Elliott wave function \cite{Elliott3}. Thus the eigenvalue of Eq. (\ref{DN0}) is:
\begin{equation}\label{DN0b}
\Delta N_0=-\braket{\Psi|H_{0,proxy}|\Psi}.
\end{equation}

The two protons and two neutrons of the 6-14 shell are placed in the cartesian state $\ket{n_z=1,n_x=0,n_y=0}$ (with opposite spin projections $m_s=\pm 1/2$ and isospin projections $m_t=+1/2,-1/2$ for protons and neutrons respectively), according to the order of Eq. (\ref{order}), possessing $(\lambda,\mu)=(4,0)$, as derived from Eqs. (\ref{lambda}), (\ref{mu}). The resulting states, which are given in Eq. (\ref{ex}), form the Slater determinant $\Phi$. Following the notation of section 2.6 of Ref. \cite{Wilsdon} the cartesian state $\ket{n_z,n_x,n_y}$ = $\ket{1,0,0}$ with $m_t=+1/2,m_s=-1/2$ for the first particle will be labeled by $\phi^{+-}(1)$. The Slater determinant for the 4 particle system is:
\begin{equation}
\Phi={1\over\sqrt{4!}} 
\begin{vmatrix}
\phi^{++}(1) & \phi^{+-}(1) & \phi^{-+}(1) & \phi^{--}(1) \\ 
\phi^{++}(2) & \phi^{+-}(2) & \phi^{-+}(2) & \phi^{--}(2) \\ 
\phi^{++}(3) & \phi^{+-}(3) & \phi^{-+}(3) & \phi^{--}(3) \\ 
\phi^{++}(4) & \phi^{+-}(4) & \phi^{-+}(4) & \phi^{--}(4)
\end{vmatrix}.
\end{equation}

 The $\delta_{j,\mathcal{N}+1/2}=\delta_{j,3/2}$ term, which is included in $H_{0,proxy}$ of Eq. (\ref{H0pr}), affects only the $1p^{3/2}_{\pm 1/2}$ component of the $\ket{n_z=1,n_x=0,n_y=0,m_s=\pm 1/2}$ states of Eq. (\ref{ex}). Thus for the 4--particle excitation of the $\ce{^{16}O}$:
\begin{equation}\label{Phi}
\braket{\Phi|H_{0,proxy}|\Phi}={8\over 3}\epsilon_{proxy}.
\end{equation}

The irrep $(\lambda,\mu)=(4,0)$ of the 4 particles in the SO--like shell generates only the $K=0$ band, following Eq. (\ref{KL}) with $S=K_S=0$ for an even--even nucleus. Consequently from Eqs. (\ref{braket}), (\ref{Phi}) for the ground state with $K=L=0$ we get:
\begin{equation}
\braket{\Psi|H_{0,proxy}|\Psi}=|a(0,0)|^2{8\over 3}\epsilon_{proxy}. 
\end{equation}
The coefficient $a(K,L)$ for a $\mu=0$ case, as this one is, can be taken from Table 2A of Ref. \cite{Vergados1968} to be $a(0,0)={1\over\sqrt{5}}$, while the value of $\hbar\omega$ is calculated from Eq. (\ref{hbar}) to be $\hbar\omega=16.27$ MeV. By substituting the experimental value of the deformation variable $\beta=0.364$ for $\ce{^{16}O}$ \cite{Ni} in Eqs. (\ref{def1}), (\ref{def2}) we derive, that the deformation parameter of the Nilsson asymptotic basis is $\varepsilon=0.36$, which leads to $\epsilon_{proxy}=0.76\hbar\omega$ using Eq. (\ref{eproxy}). The $\Delta N_0$ of Eq. (\ref{DN0b}) for the 4--particle excitation in $\ce{^{16}O}$ is:
\begin{equation}\label{DN0O}
\Delta N_0=-\braket{\Psi|H_{0,proxy}|\Psi}=-6.595 ~\mathrm{MeV}.
\end{equation}

The quadrupole difference $\Delta(QQ)$ of Eq. (\ref{DQQ}) among the SO--like irrep $(4,0)$ for the 4--particle excitation and the HO irrep $(0,0)$ for the 0--particle excitation reads \cite{Rowe2006}:
\begin{equation}
\Delta(QQ)={\kappa\over2}(4C_{2,SO}-4C_{2,HO})=2\kappa C_{2,SO},
\end{equation}
where $\kappa$ is given by Eq. (\ref{kappa}) and $C_{2,SO}=28$, $C_{2,HO}=0$ are given by Eq. (\ref{C2}). Using $N_0=36$ as derived from Eq. (\ref{N0}) for the 16 particles of this nucleus one gets that $2\kappa=0.452$ MeV and so
\begin{equation}\label{DQQO}
\Delta(QQ)=12.655~\mathrm{MeV}.
\end{equation}
Finally by substituting Eqs. (\ref{DN0O}), (\ref{DQQO}) into (\ref{02O}), the predicted energy for the $0_2^+$ state of $\ce{^{16}O}$ within the dual--shell mechanism for shape coexistence is:
\begin{equation}
E_{0_2^+}=6.06~\mathrm{MeV},
\end{equation}
which is in very good agreement with the data at $6.049$ MeV \cite{Tilley1993}.

In this sample calculation of the band-head of the coexisting band became obvious the competition between the $\Delta N_0$ and the $\Delta(QQ)$ terms. Actually this competition leads to the condition $\Delta(QQ)\ge-\Delta N_0\ge 0$, which derived the islands of shape coexistence in section \ref{islands}. It also became obvious, that the Elliott SU(3) and the proxy-SU(3) symmetry are capable of all types of calculations concerning the nuclear shape and that the agreement with the data emerges naturally using global parameters without any fitting.

\subsection{The $\ce{Mg}$ isotopes}\label{Mg}

\begin{figure}
\begin{center}
\includegraphics[width=85mm]{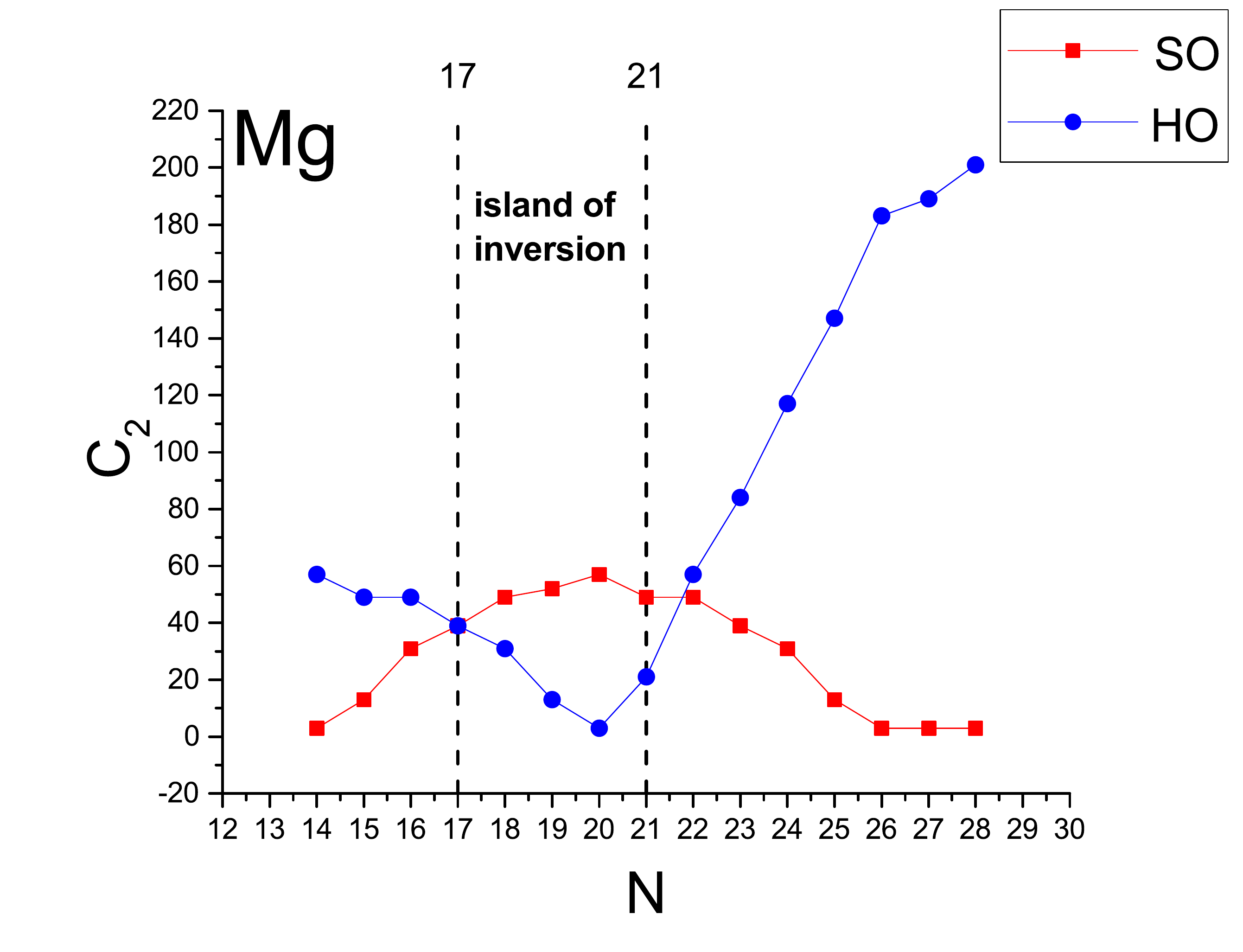}
\caption{The eigenvalues of the SU(3) Casimir operator $C_2$ for the $\ce{Mg}$ isotopes. The 12 protons of the $\ce{Mg}$ isotopes lie in the 6-14 SO--like shell \cite{Saxena2013} and thus posses $(\lambda_\varpi,\mu_\varpi)_{SO}=(0,0)$.  For neutron numbers $17\le N\le 20$, the neutron configuration which is derived from the SO--like neutron shell, has to correspond to the ground state band (see section \ref{two}), while the less deformed configuration, coming from the HO neutron shell, has to correspond to the excited band, if only pure SU(3) irreps are considered. Inversion of states is predicted in the $\ce{^{29-33}_{12}Mg_{17-21}}$ isotopes, which possess $QQ_{SO}\ge QQ_{HO}$ according to the condition of Eq. (\ref{cond1}), while shape coexistence is predicted in the $\ce{^{29-32}_{12}Mg_{17-20}}$ isotopes. These predictions align with the data of Refs. \cite{Himpe2008,Kowalska2008,Schwerdtfeger2009}. See section \ref{Mg} for further discussion.} \label{MgC2}
\end{center}
\end{figure}

In $\ce{Mg}$ isotopes the neutrons traditionally occupy the 8-20 harmonic oscillator shell. But as the dual--shell mechanism predicts, just after $N=17$ (see Fig. \ref{isl2}) the neutrons can flip from the 8-20 HO shell to the 14-28 SO shell. This procedure is able to explain the inversion of states \cite{Otsuka2020} in the $\ce{Mg}$ isotopes and the breaking of the magicity of $N=20$. According to Figs. \ref{isl2} and \ref{MgC2} shape coexistence and inversion of states in the $\ce{Mg}$ isotopes begin just after $N=17$, while shape coexistence ends at the harmonic oscillator shell closure at $N=20$, and inversion of states endures as far as $QQ_{SO}\ge QQ_{HO}$, which is valid till $N=21$ (see Fig. \ref{MgC2}), predictions which align with the experimental facts \cite{Himpe2008,Kowalska2008,Schwerdtfeger2009,Wimmer2010}. 

\subsection{The $\ce{Ni}$ isotopes}\label{Nikel}
\begin{figure}
\begin{center}
\includegraphics[width=85mm]{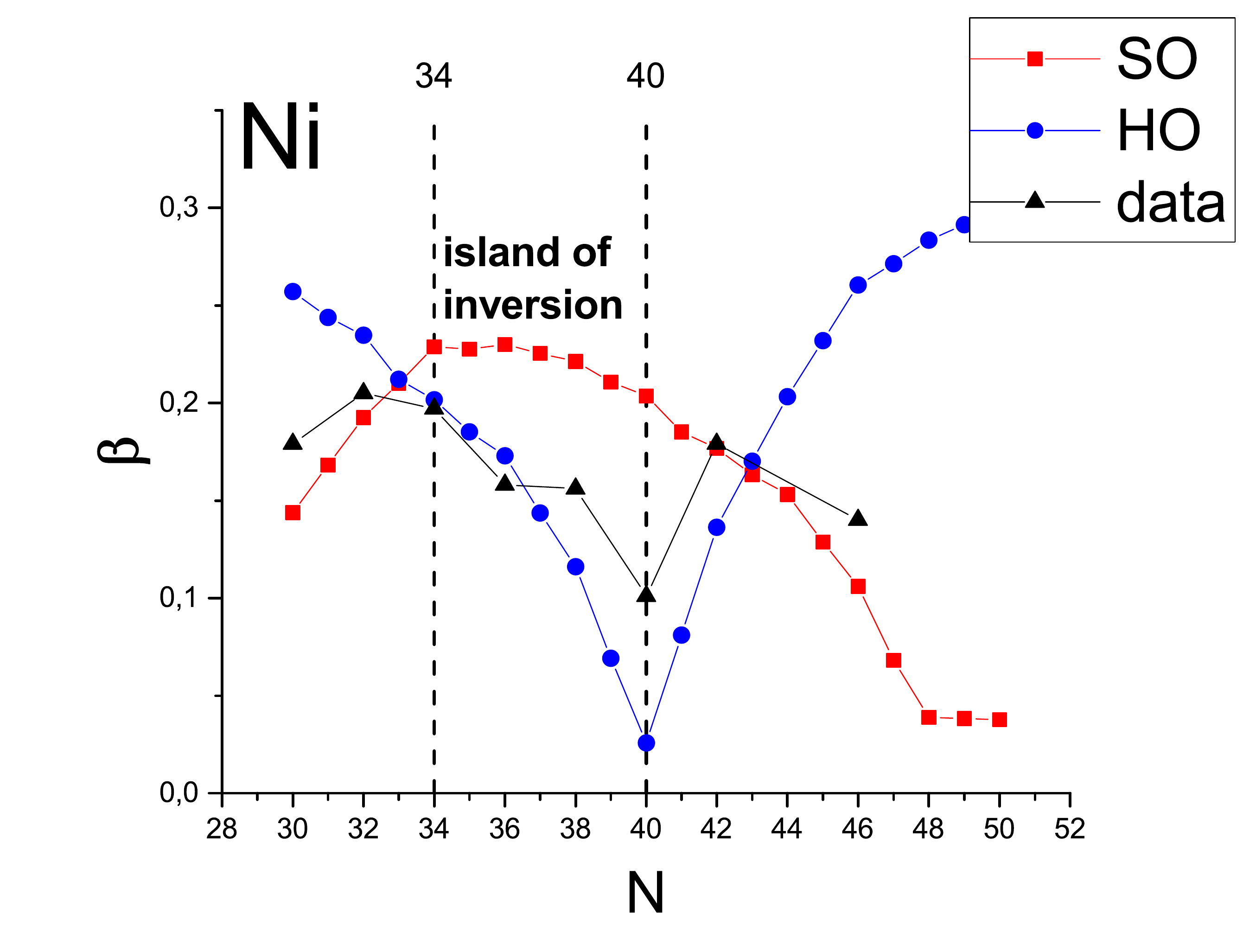}
\caption{The deformation parameter $\beta$ as calculated from Eq. (\ref{beta}) with the scaling factor of Eq. (\ref{scale}) \cite{proxy2} for the $\ce{Ni}$ isotopes for two valence neutron shells. The protons lie in the 28-50 SO--like shell, with proton irrep $(\lambda_\varpi, \mu_\varpi)=(0,0)$ (see Table \ref{irrepsa}). The data on the deformation $\beta$ \cite{Ni,Ni2} indicate, that just after $N=34$ the 20-40 HO shell becomes preferable for the neutron configuration. The neutrons are predicted to flip back from the 20-40 shell to the 28-50 shell at the harmonic oscillator shell closure, which occurs at neutron number $N=40$.  See section \ref{Nikel} for further discussion.}\label{Nibeta}
\end{center}
\end{figure}

The next island of inversion is indicated by Fig. \ref{isl3}, which supports, that the SO--like magic numbers are being competed by the harmonic oscillator magic number \cite{Guenaut2005} for isotopic chains with $N\ge 34$. This prediction is in agreement with the data on the deformation variable $\beta$ in the $\ce{Ni}$ isotopes as plotted in Fig. \ref{Nibeta}. 

\subsection{Heavier nuclei}\label{heavy}

\begin{figure}
\begin{center}
\includegraphics[width=85mm]{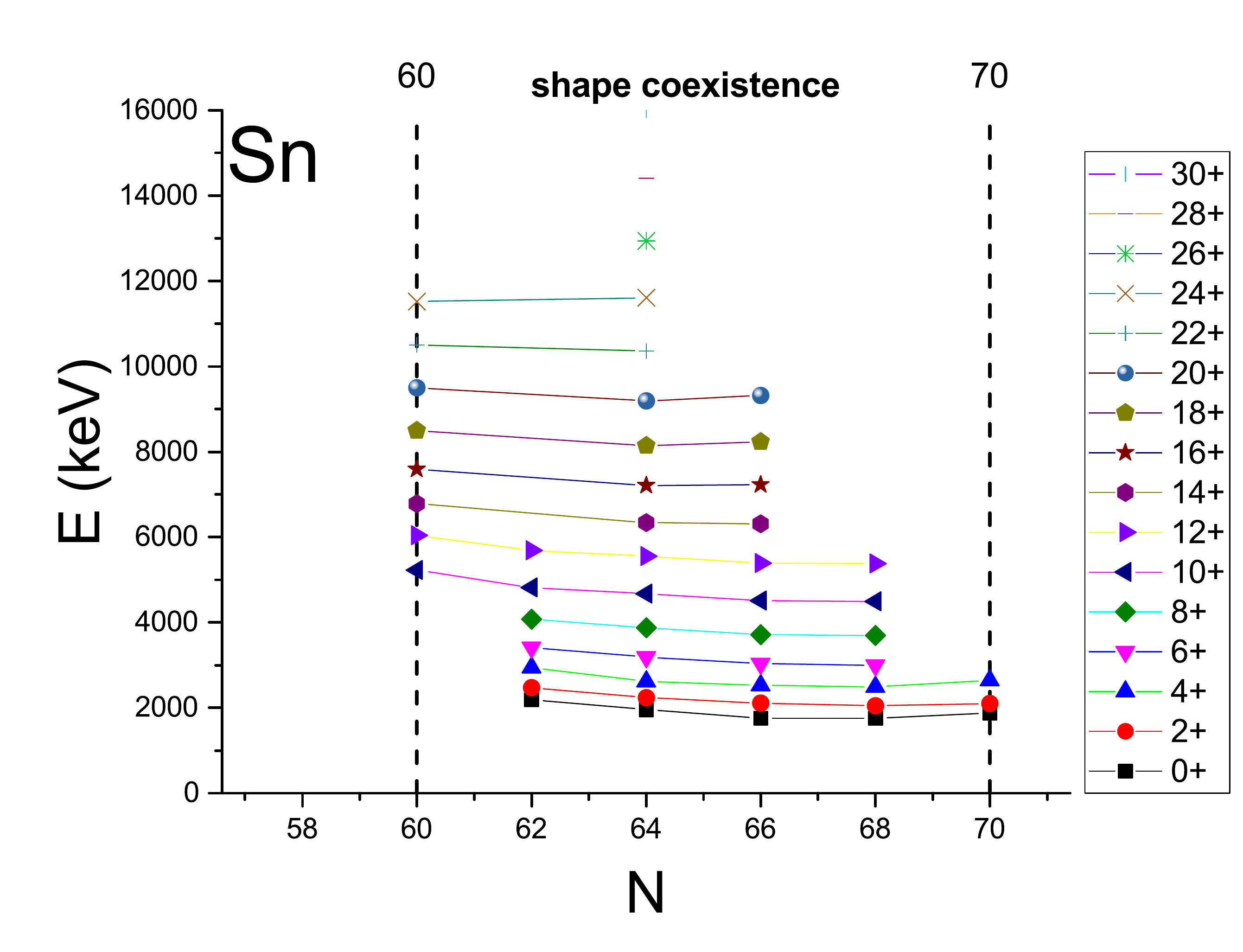}
\caption{One of the most striking manifestations of shape coexistence appears in the $\ce{Sn}$ isotopes. An excited $K=0^+$ band, which is attributed to shape coexistence (see Fig. 3.10 of Ref. \cite{Wood1992}), appears within neutron numbers $60\le N\le 70$. Energy levels have been obtained by \cite{Guerdal2012,Lalkovski2015,Blachot2012,Blachot2010,Kitao1995,KITAO2002}. See section \ref{heavy} for further discussion.}\label{Sndata}
\end{center}
\end{figure}

An island of shape coexistence is indicated by Fig. \ref{isl4}, which predicts the phenomenon among $59\le N \le 70$ for an isotopic chain. The most impressive example for this island of shape coexistence is the parabolic line of the excited $K=0^+$ bands in the $\ce{^{110-120}Sn}$ isotopes as presented in Fig. \ref{Sndata} and in Ref. \cite{Wood1992}.

\begin{figure}
\begin{center}
\includegraphics[width=85mm]{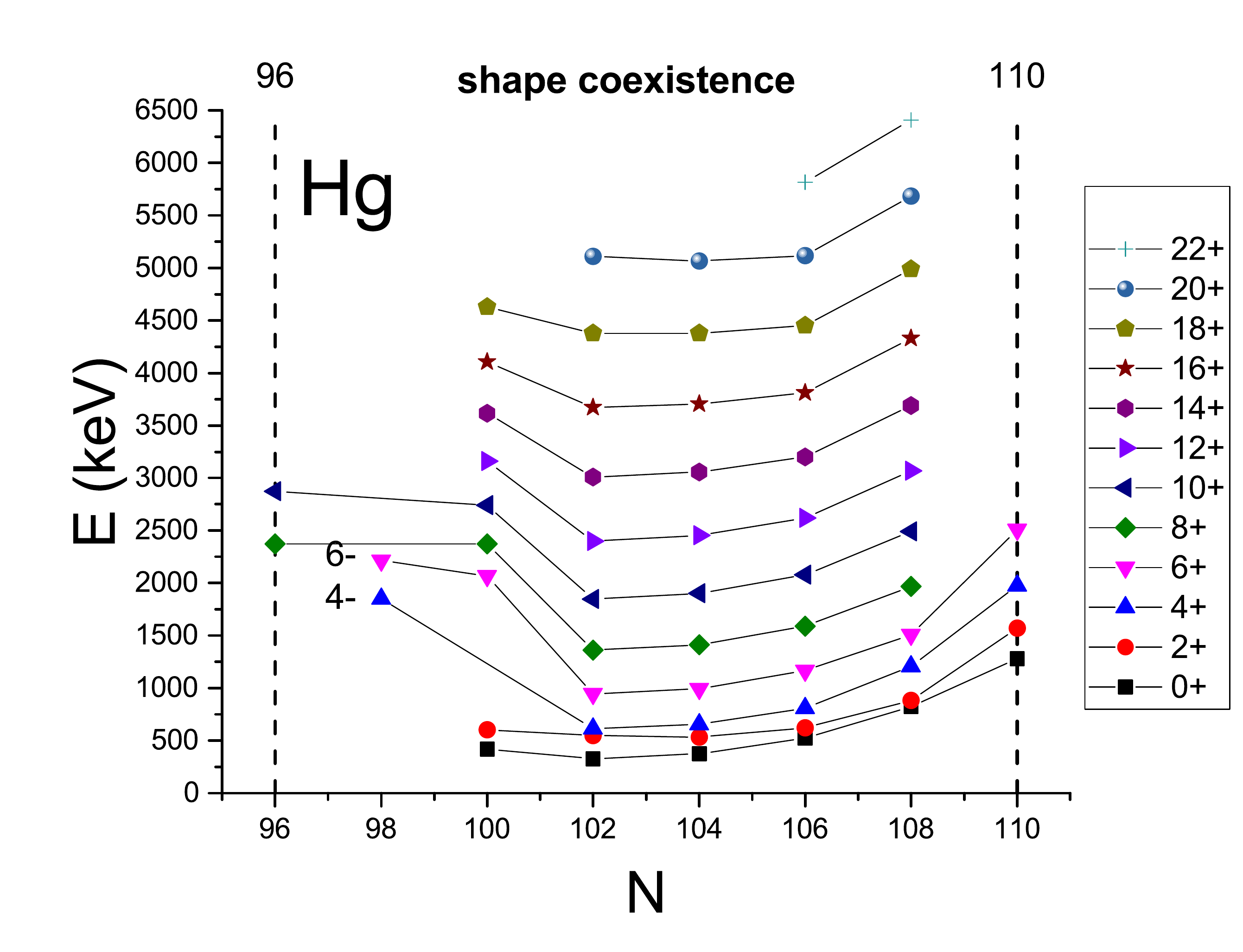}
\caption{A peerless manifestation of shape coexistence appears in the $\ce{Hg}$ isotopes with neutron numbers within $96\le N \le 110$ (see Fig. 10 of Ref. \cite{Heyde2011}). The data have been collected from \cite{Basunia2006,Achterberg2009,McCutchan2015,Singh2015,Baglin2010,BAGLIN2003,Kondev2018,SINGH2003}. See section \ref{heavy} for further discussion.}\label{Hgdata}
\end{center}
\end{figure}

\begin{turnpage}

\begin{figure} 
  
    \includegraphics[width=240mm]{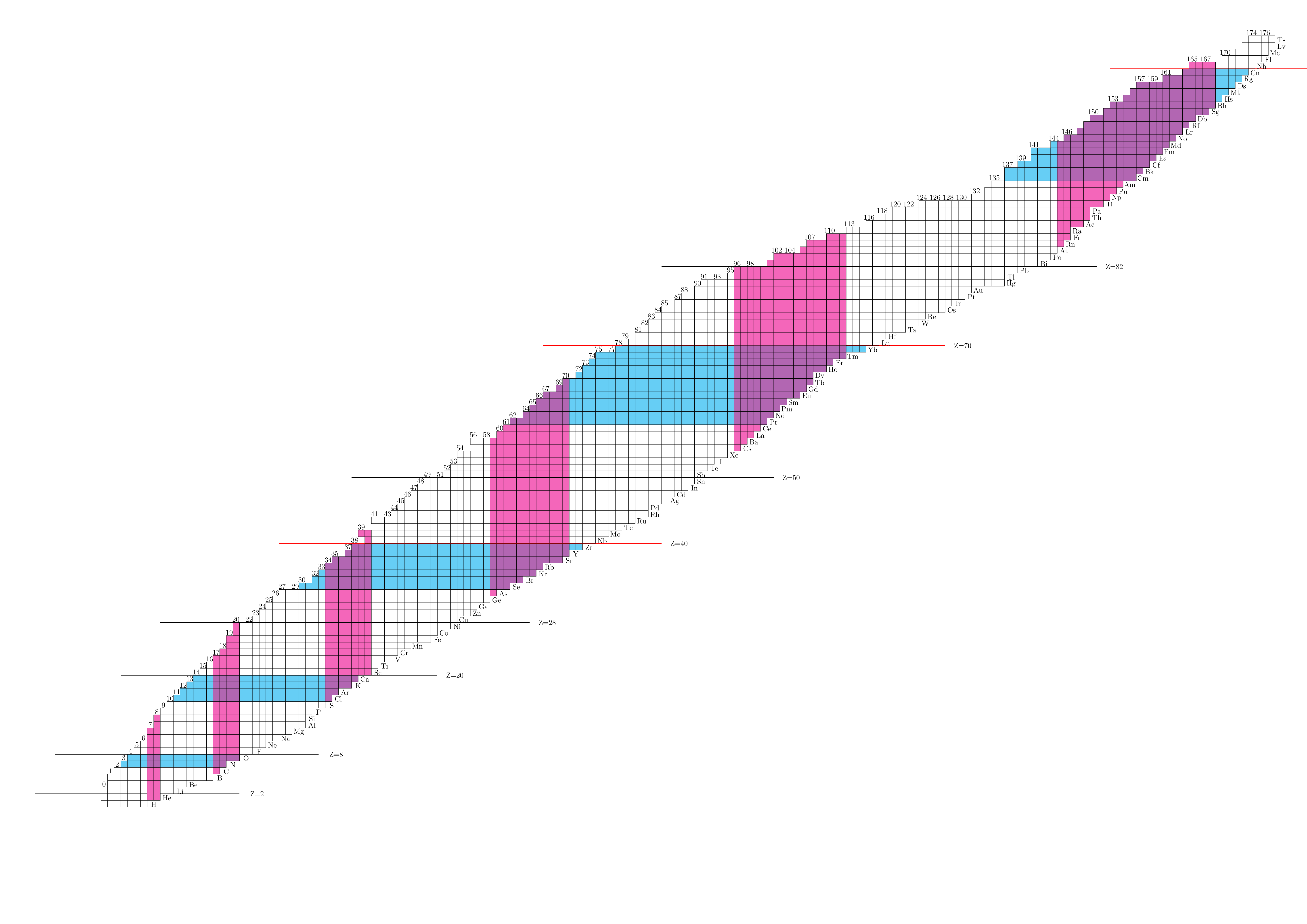}

    \caption{This map indicates, which nuclei have to be examined both theoretically and experimentally for manifesting shape coexistence according to the proposed mechanism. The colored regions possess proton or neutron number between $7-8$, $17-20$, $34-40$, $59-70$, $96-112$, $145-168$. The horizontal stripes correspond to the proton induced shape coexistence, while the vertical stripes correspond to the neutron induced shape coexistence. }

    \label{map}
    
\end{figure}

\end{turnpage}

Another astonishing example of shape coexistence lies in the $\ce{^{176-190}Hg}$ isotopes (see Fig. 10 of Ref. \cite{Heyde2011}), which matches with the predictions of Fig. \ref{isl5}. The parabolic line of the excited $K=0^+$ bands begins at $N=96$ and ends at $N=110$, just two neutrons below the harmonic oscillator shell closure at $N=112$. The data for the parabolic and excited $K=0^+$ band for these isotopes are plotted in Fig. \ref{Hgdata}.

Finally Fig. \ref{isl6} corresponds to the super--heavies and predicts a comeback of the harmonic oscillator shell after $N=145$. This region corresponds to the fission isomers centered around $\ce{^{240}_{94}Pu_{146}}$ \cite{Thirolf2002}.

\section{Discussion}\label{discussion}

In the present work a novel mechanism for the phenomenon of shape coexistence is introduced. Its main features are summarized here:\\
1) The proposed mechanism is based on the SU(3) symmetry. In light nuclei, up to the $sd$ shell, the SU(3) symmetry of the 3-dimensional (3D) isotropic harmonic oscillator (HO) is present, as used in the Elliott SU(3) model. Beyond the $sd$ shell, the recently introduced proxy-SU(3) symmetry is used. \\
2) The present mechanism is based on the interplay between the HO magic numbers 2, 8, 20, 40, 72, 112, 168, and the spin-orbit (SO) like magic numbers  6, 14, 28, 50, 82, 126, 184. The main element of the new mechanism are particle excitations occurring between the HO and SO sets of shells.\\
3) These particle excitations lead to the dissolution of the magic numbers and to the merging of the two types of shells. \\
4) The main novel prediction of the present mechanism is, that shape coexistence cannot appear everywhere on the nuclear chart, but only within specific regions, called islands of shape coexistence, the shores of which are determined through group theoretical arguments in a parameter independent way. \\
5) The present mechanism allows for a parameter-free prediction of the energy of the $0_2^+$ band being the shape coexistence partner of the ground state band. \\
6) The present mechanism suggests that shape coexistence, parity inversion, inversion of states, and fission isomers are just shades of the same phenomenon, {\it  i.e.}, the coexistence of the HO valence shell with the SO--like shell.

In relation to the widely accepted particle--hole mechanism of shape coexistence, the following comments apply:\\
1) The islands predicted by the present mechanism are fully compatible with the regions of the nuclear chart in which the particle--hole mechanism has been applied. No contradiction between the two mechanisms arises. \\
2) The novel message from the present mechanism is, that the particle--hole mechanism cannot occur all over the nuclear chart, but only within the specific regions predicted by the present mechanism based on parameter--free SU(3) symmetry arguments. \\
3) In light nuclei the two mechanisms coincide, since the present mechanism predicts particle excitations from the HO shell to the neighboring SO--like shell within certain islands on the nuclear map. In medium mass and heavy nuclei, the present mechanism is based on shell merging.

The proposed mechanism is supported by results for the single particle energies provided by Density Functional Theory (DFT) calculations. In particular:\\
1) In the $\ce{Pb}$ and $\ce{Hg}$ isotopes both the neutron and the proton energy gaps above the HO and the SO magic numbers collapse within the $96\leq N \leq 110$ region (as predicted by the present mechanism), leading to shell merging.   \\
2) Plotting the single particle energies relatively to the Fermi energy in the $\ce{Pb}$ and $\ce{Hg}$ isotopes, one sees within the $96\leq N \leq 110$ region (predicted by the present mechanism) that some of the proton orbitals belonging to the shell above the 82 magic number sink below the Fermi energy, while some of the proton orbitals belonging to the 50-82 shell pop up above the Fermi energy. These results corroborate the compatibility between the particle-hole mechanism and the present shell merging mechanism, giving in parallel the message that the particle-hole mechanism is applicable only within the islands predicted by the present approach. 

The proposed mechanism is rather general and applies to all mass regions. It should be noticed that SU(3) is used as a classification scheme and not as a dynamical symmetry, therefore the method is applicable over the whole nuclear chart and is not limited within regions of highly deformed nuclei. For instance the dual--shell mechanism predicts the tin isotopes with shape coexistence, despite that these isotopes are not very deformed. Furthermore, its predictions are parameter-independent.

 As a consequence the present work can be considered as a {\it first step} towards a unified understanding of shape coexistence and related phenomena. Detailed studies for different isotopes and mass regions are called for, which might reveal the need for further elaboration of the mechanism.  The calculation of nuclear observables within the dual--shell and the comparison of the results with the data and with the predictions of other nuclear models and mechanisms \cite{Lalazissis1999,Niksic2002,Rowe2020,Caurier2001,Caurier2007,Caurier2014,Budaca2018,Budaca2019,Otsuka2019} could provide a better understanding of the realm of nuclear structure. The proton--neutron interaction in the Elliott and the proxy-SU(3) schemes has to be studied in the future, since we expect, that it will give us a more clear view of the islands of shape coexistence on the nuclear chart. Also the pairing interaction has to be included in the Hamiltonian, so as to estimate accurately the energy levels of the coexisting bands.
 
 \section*{Acknowledgements}

Discussions with R. F. Casten, K. Blaum and J. Cseh improved considerably the manuscript. 
Financial support by the Greek State Scholarships Foundation (IKY) and the European Union within the MIS 5033021 action, by the Bulgarian National Science Fund (BNSF) under Contract No.KP-06-N48/1 and by the Tenure Track Pilot Programme of the Croatian Science Foundation and the \'{E}cole Polytechnique F\'ed\'erale de Lausanne and the Project No. TTP-2018-07-3554 Exotic Nuclear Structure and Dynamics with funds of the Croatian-Swiss Research Programme is gratefully acknowledged.


\end{document}